\documentclass{aa}  

\usepackage{graphicx}
\usepackage{txfonts}
\usepackage{float}
\usepackage{parskip}
\usepackage{adjustbox}
\usepackage[switch]{lineno}   
\usepackage{placeins}
\begin{document}

   \title{A study in scarlet - I. Photometric properties of a sample of Intermediate Luminosity Red Transients}


   \authorrunning{Valerin et al.}
   \author{G. Valerin\inst{1}\thanks{E--mail: giorgio.valerin@inaf.it},
A. Pastorello\inst{1},
A. Reguitti\inst{1,2}, 
S. Benetti\inst{1},
Y.-Z. Cai\inst{3,4,5}, 
T.-W. Chen\inst{6},          
D. Eappachen\inst{7,8}, 
N. Elias-Rosa\inst{1,9},
M. Fraser\inst{10}, 
A. Gangopadhyay\inst{11,12},  
E. Y. Hsiao\inst{13},            
D. A. Howell\inst{14,15},
C. Inserra\inst{16},            
L. Izzo\inst{17,18},
J. Jencson\inst{19},         
E. Kankare\inst{20},
R. Kotak\inst{20}, 
P. A. Mazzali\inst{21,22},         
K. Misra\inst{23}, 
G. Pignata\inst{24},
S. J. Prentice\inst{25},           
D. J. Sand\inst{26},         
S. J. Smartt\inst{27,28},
M. D. Stritzinger\inst{29},    
L. Tartaglia\inst{30},
S. Valenti\inst{31},
J. P. Anderson\inst{32,33},
J. E. Andrews\inst{26},       
R. C. Amaro\inst{26},                           
S. Brennan\inst {11},
F. Bufano\inst{34},          
E. Callis\inst {11},
E. Cappellaro\inst{1},     
R. Dastidar\inst{35,33},
M. Della Valle\inst{17,36},
A. Fiore\inst{1,37,38},          
M. D. Fulton\inst{28},
L. Galbany\inst{9,39},        
T. Heikkil\"{a}\inst{20},            
D. Hiramatsu\inst{13,14,40,41},            
E. Karamehmetoglu\inst{11,29},    
H. Kuncarayakti\inst{20,42},
G. Leloudas\inst{43},             
M. Lundquist\inst{26},               
C. McCully\inst{14},
T. E. Müller-Bravo\inst{9,39},
M. Nicholl\inst{28},                
P. Ochner\inst{1,44},                
E. Padilla Gonzalez\inst{14,15},    
E. Paraskeva\inst{45},
C. Pellegrino\inst{46},          
D. E. Reichart\inst{47},          
T. M. Reynolds\inst{20,48,49},      
R. Roy\inst{50},                                     
I. Salmaso\inst{1},  
M. Singh\inst{51},
M. Turatto\inst{1},              
L. Tomasella\inst{1},              
S. Wyatt\inst{26},
D. R. Young\inst{28}
          }

   \institute{Affiliations are listed after the acknowledgements
             }

   \date{Received ----; accepted ----}

 
  \abstract
   {}
   {We investigate the photometric characteristics of a sample of Intermediate Luminosity Red Transients (ILRTs), a class of elusive objects with peak luminosity between that of classical novae and standard supernovae. Our goal is to provide a stepping stone in the path to unveil the physical origin of such events, thanks to the analysis of the datasets collected.}
   {We present the multi-wavelength photometric follow-up of four ILRTs, namely NGC 300 2008OT-1, AT 2019abn, AT 2019ahd and AT 2019udc. Through the analysis and modelling of their spectral energy distribution and bolometric light curves we infer the physical parameters associated with these transients. }
   {All four objects display a single peaked light curve which ends in a linear decline in magnitudes at late phases. A flux excess with respect to a single black body emission is detected in the infrared domain for three objects in our sample, a few months after maximum. This feature, commonly found in ILRTs, is interpreted as a sign of dust formation. Mid infrared monitoring of NGC 300 2008OT-1 761 days after maximum allows us to infer the presence of $\sim$10$^{-3}$-10$^{-5}$ M$_{\odot}$ of dust, depending on the chemical composition and the grain size adopted. The late time decline of the bolometric light curves of the considered ILRTs is shallower than expected for \textsuperscript{56}Ni decay, hence requiring an additional powering mechanism. James Webb Space Telescope observations of AT 2019abn prove that the object has faded below its progenitor luminosity in the mid-infrared domain, five years after its peak. Together with the disappearance of NGC 300 2008OT-1 in Spitzer images seven years after its discovery, this supports the terminal explosion scenario for ILRTs. With a simple semi-analytical model we try to reproduce the observed bolometric light curves in the context of few M$_{\odot}$ of material ejected at few 10$^{3}$ km s$^{-1}$ and enshrouded in an optically thick circumstellar medium.}
   {}


   \maketitle
%
\section{Introduction}


It is well established that single stars with an initial mass below $\sim$8 M$_{\odot}$ will end their lives as white dwarfs, cooling down while supported by the electron degeneracy pressure in their cores. In contrast, stars with initial masses between $\sim$10 M$_{\odot}$ and 40 M$_{\odot}$ will complete all the nuclear burning cycles and will undergo a violent explosion as their core collapses \citep{WoosleyMassiveStars}.
This apparently simple distinction raises a complicated question: what is the exact initial mass limit that separates the two opposite fates?
Stars with a zero-age-main-sequence mass between 8 M$_{\odot}$ and 10 M$_{\odot}$ are expected to form a degenerate O-Ne-Mg core during their lifetime \citep{Nomoto1984}. Such stars are labelled Super-Asymptotic Giant Branch stars (SAGB), and the outcome of their evolution is uncertain. In fact, if the O-Ne-Mg core accretes enough material to approach the Chandrasekhar limit, the star will explode as an Electron Capture Supernova (ECSN), but if the core fails to reach this critical mass the star will end its evolution as an O-Ne-Mg white dwarf (e.g. \citealt{Miyaji1980,Nomoto1984,Jones2013,Moriya2014,Doherty2015, LimongiAGB}).
Whether such critical mass can be reached depends on the competing effects of mixing, convective overshooting and mass loss rates, which make the modelling of the core and its evolution a challenging endeavor \citep{Poelarends2008}.
An additional complication, as pointed out by \cite{Kozyreva2021}, is that even small changes in the initial mass and metallicity of the progenitor star may give rise to a Fe core-collapse supernova (SN) instead of an ECSN, overall showing similar observables.

While stellar evolution theory predicts the existence of ECSNe, finding their observational counterparts is still an open issue. Proving that a transient originates from the core-collapse of an O-Ne-Mg core, rather than from a classical Fe core collapse, is not trivial. However, there has been no shortage of attempts: Low Luminosity Supernovae Type IIP (LL SNe IIP) (e.g. \citealt{Spiro2014,Reguitti2018hwm,Valerin2020cxd}) and also some interacting transients (e.g. \citealt{2009knSmith,2018zdHiramatsu}) have been proposed as ECSN candidates.
In order to be a reasonable ECSN candidate, an object should fulfill the key expectations for the explosion following the collapse of an O-Ne-Mg core.
First of all, the energy released by an ECSN should be significantly lower ($\sim$ 10$^{50}$ erg) compared to classical SN explosions, therefore directing the investigation towards faint targets with low velocity ejecta \citep{Janka2008_Energy}.
Secondly, the nucleosynthesis following the collapse of an O-Ne-Mg core yields limited amounts of \textsuperscript{56}Ni (few 10$^{-3}$ M$_{\odot}$), placing constraints on the luminosity of the late time decline of the candidate \citep{Wanajo2009_56Ni}.
Finally, the progenitor star of a candidate should be compatible with a luminous ($\sim$10$^{5}$ L$_{\odot}$) SAGB star, since that is the only kind of star able to produce a degenerate O-Ne-Mg core massive enough to trigger an ECSN explosion \citep{Poelarends2008}.

Intermediate Luminosity Red Transients (ILRTs) are a class of objects, populating the luminosity gap between classical novae and standard SNe \citep{Pasto2019Gap}, which are appealing ECSN candidates.
Their physical origin is still debated, with some studies associating ILRTs to non--terminal eruptions of post--main sequence stars (e.g. \citealt{Hump2011}).
However, there are several indicators that favour the ECSN interpretation to explain the observed properties of these transients.
The low luminosity characterizing ILRTs, which show peak absolute magnitudes ranging between M$_{r}$ $\sim$ --12 mag and --15 mag, is consistent with the expected weak explosion originating from the collapse of an O-Ne-Mg core \citep{Pumo2009}. Likewise, the late time decline in luminosity points towards small \textsuperscript{56}Ni masses synthesized, fulfilling the condition presented by \cite{Wanajo2009_56Ni} (see also \citealt{Yongzhi2021ILRT}).
Furthermore, all of the progenitor stars associated with ILRTs have been consistent with a SAGB star, corroborating the ECSN scenario \citep{ThompsonProgenitor2009,Jencson2019abn}. Recent estimates have shown that the rate of ILRTs is a few percent of all the local CC SNe events ($\approx$8\% according to \citealt{Yongzhi2021ILRT}, or $\approx$1--5\%, according to \citealt{Karambelkar2023}), compatible with the theoretical expectations for ECSNe \citep{Poelarends2008,ThompsonProgenitor2009,Doherty2015}.
Finally, an important step towards understanding the nature of ILRTs was performed by \cite{Adams2016}, who showed that a few years after their maximum luminosity the remnants of the two ILRTs SN 2008S \citep{Botticella2008S} and NGC 300 2008OT-1 \citep{Bond2009,Berger2009,Hump2011} had become fainter in the mid-infrared (MIR) than their progenitor stars.
Extreme dust extinctions would be needed to obscure a surviving star, therefore favouring a genuine terminal explosion over a non-terminal outburst.

While the considerations presented so far are certainly encouraging, the discussion on ILRTs as ECSN candidates is still ongoing. This is, after all, a relatively young class of transients, only established around a fifteen years ago. Their rates are not particularly low but their faintness makes their discovery occasional and their follow-up challenging. Since only a handful of ILRTs have been accurately characterized so far, additional data is key to improve our understanding of this poorly studied class of objects.
In this paper we present and analyse original photometric data of four ILRTs: NGC 300 2008OT-1, AT 2019abn, AT 2019ahd and AT 2019udc. This work is the first part of a series of two papers: in the second installment (``A study in scarlet II. Spectroscopic properties of a sample of Intermediate Luminosity Red Transients", Valerin et al. 2024, hereafter ``Paper II") we will present and discuss the spectroscopic data collected for the same targets analysed here.
This paper is organized as follows: in Section \ref{DataRedu_sect} we discuss the methodology used to obtain and reduce the data, while in Section \ref{Photometry_sect} the photometric data and the host properties are presented. In Section \ref{reddeningEstimate} we discuss the reddening estimate, while in Section \ref{SED_evol_sect} we discuss the physical parameters obtained through modelling of the spectral energy distribution fits. In Section \ref{LATETIME} we study the late time behaviour of NGC 300 2008OT-1 and in Section \ref{TOYMODEL} we present a simple model to reproduce the light curves of ILRTs. Finally, in Section \ref{Conclusion_sect} we summarise the results obtained.

\begin{table*}
  \centering
  \resizebox{0.9\textwidth}{!}{%
  \begin{adjustbox}{tabular = lccccccc, center}
    &  &  &  &  &  &  &  \\
  \hline
    Transient & Host Galaxy & Type & Redshift & Distance [Mpc] & $\mu$ [mag] & Galactic A$_{V}$ [mag] & Local A$_{V}$ [mag] \\
    \hline
     AT 2019abn & M51 & Sa  & 1.54 (-) x10$^{-3}$  & 8.6 (0.1) & 29.67 (0.02) & 0.096 (0.006) & 2.34 (0.06) \\
     AT 2019ahd & NGC 3423 & SA(s)cd & 3.35 (0.01) x10$^{-3}$  & 11.1 (0.7) & 30.22 (0.14) & 0.079 (0.003) & 0.37 (0.03) \\
     AT 2019udc & NGC 0718 & SAB(s)a & 5.78 (0.03) x10$^{-3}$ & 19.9 (1.4) & 31.49 (0.15) & 0.100 (0.001) & 0.00 (0.00) \\
     NGC 300 OT & NGC 300 & SA(s)d & 4.80 (0.03) x10$^{-4}$  & 1.92 (0.14) & 26.43 (0.09) & 0.034 (0.001) & 0.78 (0.06) \\
    \hline
  \end{adjustbox}}
  \caption{Basic information on the galaxies hosting the transients in the sample. Morphological classifications are from \protect \cite{M51CLASSIFICATION} for M51 and from \protect \cite{DeVaucouleursBOOK} for the other galaxies.}
  \label{tab:1}
\end{table*}

\section{Data reduction} \label{DataRedu_sect}
The objects presented in this paper were followed with several instruments at different facilities reported in Table [\ref{Instrum}]. In particular, the majority of the private data was collected with the Nordic Optical Telescope (NOT) within the NOT Unbiased Transient Survey 2 (NUTS2) collaboration \citep{Holmbo2019ATEL}, with the Liverpool Telescope \citep{LiverpoolTelescope}, with the GROND imager \citep{GROND_imager}, within the ePESSTO+ collaboration \citep{PESSTO} as well as the Global Supernova Project (GSP, \citealt{AndyLCO}).
Images obtained were reduced through standard \textsc{iraf} tasks \citep{IRAF_Toby}, removing the overscan, correcting them for bias and flat field. When multiple exposures were taken the same night, we combined them to improve the signal to noise ratio (S/N). To measure the magnitudes of the transients observed, we used a dedicated, \textsc{python}-based pipeline called \textsc{ecsnoopy} \citep{snoopy}. \textsc{ecnoopy} is a collection of \textsc{python} scripts that call \textsc{iraf} standard tasks like \textsc{daophot} through \textsc{pyraf}, and it was designed for Point Spread Function (PSF) fitting of multi-wavelenght data acquired from different instruments and telescopes. The PSF model was built from the profiles of isolated, unsaturated stars in the field. The instrumental magnitude of the transient was then retrieved by fitting this PSF model and accounting for the background contribution around the target position through a low-order polynomial fit. The error on this procedure was obtained through artificially placed stars close to the target, with magnitudes similar to that inferred for the object. The dispersion of the artificial stars instrumental magnitudes was combined in quadrature with the PSF fitting error given by \textsc{daophot} to obtain the total error associated with that measure. Zero Point (ZP) and Colour Terms (CT) corrections were computed for each instrument by observing standard fields: SDSS \citep{sdssTrue} was used as reference for Sloan filters, the \cite{Landolt} catalogue was used for Johnson filters and the 2MASS \citep{2MASS} catalogue was used for Near Infrared (NIR) filters. 

For the "Asteroid Terrestrial--impact Last Alert System" (ATLAS) data \citep{TonryATLAS}, we combined the flux values obtained through forced photometry released from their archive\footnote{https://fallingstar--data.com/forcedphot/}, and converted the result into magnitudes as prescribed in the ATLAS webpage.
While reducing WISE \citep{WISE} and Spitzer \citep{Spitzer} data we adopted the ZP provided on their respective websites\footnote{https://wise2.ipac.caltech.edu/docs/release/prelim/expsup/sec4\_3g.html, https://irsa.ipac.caltech.edu/data/SPITZER/docs/irac/iracinstrumenthand book/14/}. We also performed PSF fitting measurements on publicly available images taken with the James Webb Space Telescope (JWST, \citealt{JamesWebb2006}). The pipeline-reduced JWST+NIRCam/MIRI `i2d' images (and Level-3 mosaics when available) were retrieved from the MAST archive\footnote{https://mast.stsci.edu/portal/Mashup/Clients/Mast/Portal.html}.
It is worth noticing that in the NIR and Mid Infrared (MIR) we assumed negligible CT, so we only computed the ZP correction. 
Images from the $Swift$ UV/Optical Telescope \cite{Swift2005} were reduced using standard \textsc{heasoft} tasks, and the magnitudes were retrieved through aperture photometry, applying the aperture correction reported in the $Swift$
website\footnote{https://swift.gsfc.nasa.gov/analysis/uvot\textunderscore digest/apercor.html} when needed.
In order to account for non-photometric nights, we selected a series of stars in the field of each observed transient: measuring the average magnitude variation of the reference stars, we computed the ZP correction for each night in each band. Applying ZP and CT corrections to the instrumental magnitudes of our targets, we obtained the apparent magnitudes which are reported in this paper. We adopted the AB magnitudes system for $u$, $g$, $r$, $i$, $z$, $cyan$, $orange$ bands and Vega magnitudes for $UVW2$, $UVM2$, $UVW1$, $U$, $B$, $V$, $R$, $I$, $J$, $H$, $K$, $W1$, $W2$, $[3.6]$ $\mu m$, $[4.5]$ $\mu m$ bands.
We resorted to template subtraction at late epochs, when the transients were too faint to be detected otherwise. The template subtraction procedure was performed on late time observations, again with \textsc{ecsnoopy}, with template images taken from SDSS \citep{sdssTrue}, Pan--STARRS1 \citep{Panstarrs} and Liverpool Telescope \citep{LiverpoolTelescope}.
The photometric measurements we obtained are reported in Appendix \ref{AppendixBmeasureddata}.

\begin{figure}
	\includegraphics[width=\columnwidth]{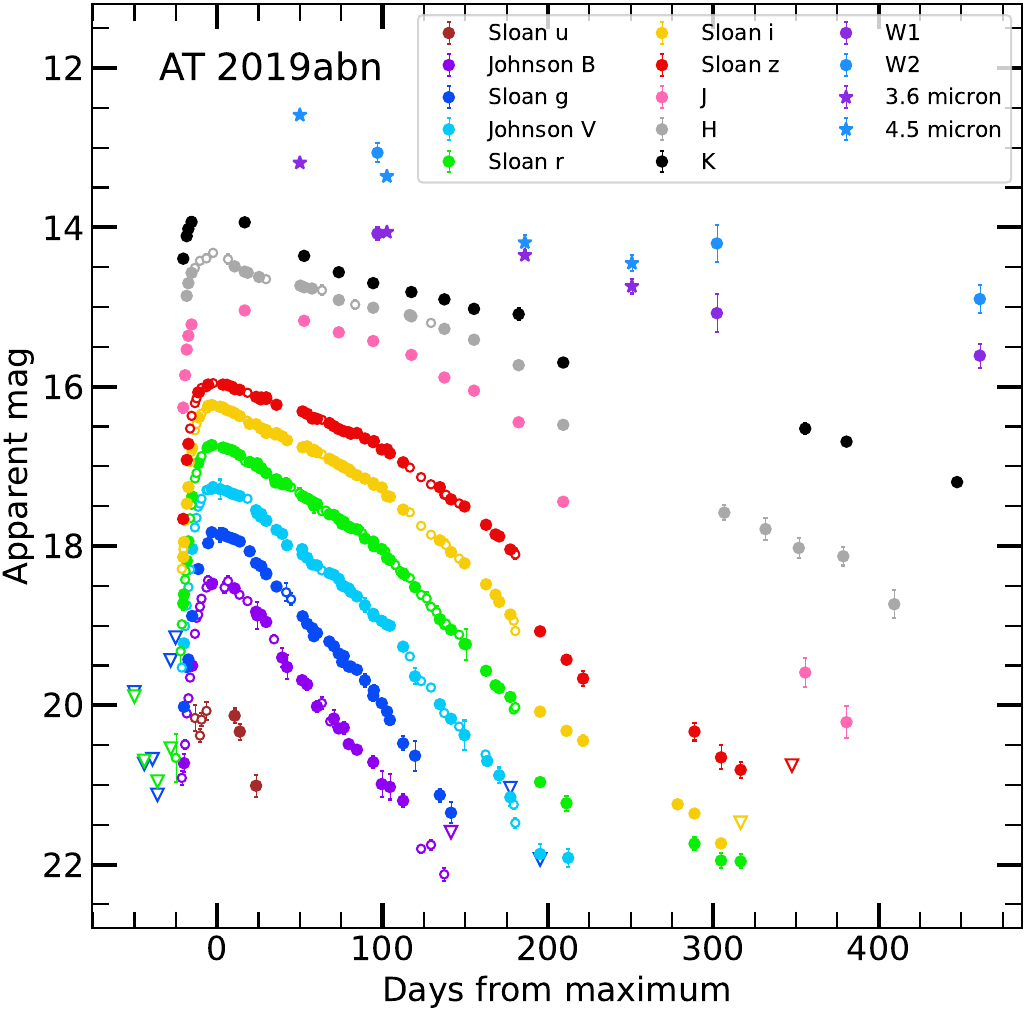}
    \caption{Optical, NIR and MIR light curves of AT 2019abn. Filled circles represent unpublished data, while empty circles represent data points from the literature. Empty triangles represent upper limits.}
    \label{abn_phot}
\end{figure}

\begin{figure}
	\includegraphics[width=\columnwidth]{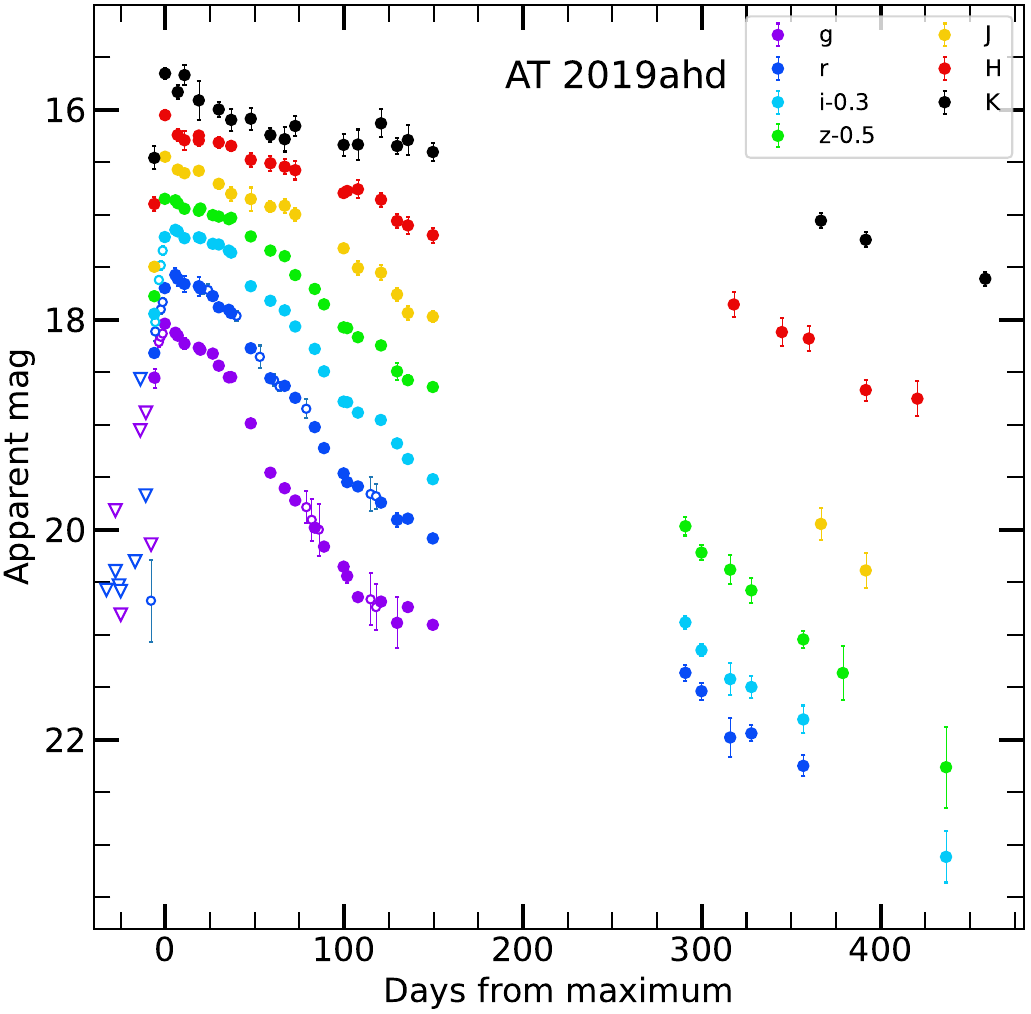}
    \caption{Optical and NIR light curves of AT 2019ahd. Filled circles represent unpublished data, while empty circles represent ZTF data points. Empty triangles represent upper magnitude limits. Magnitude shifts on $i$ and $z$ bands have been applied for clarity.}
    \label{ahd_phot}
\end{figure}

\section{Photometric follow-up} \label{Photometry_sect}

\subsection{AT 2019abn} \label{subsec_photomAT2019abn}
AT 2019abn was discovered on 2019 January 22.6 UT by the Zwicky Transient Facility (ZTF, \citealt{ZTF}) on a spiral arm of Messier 51 (M51) at the coordinates RA$=$ $13^{h}29^{m}42^{s}.41$, Dec $=$ +47\textdegree 11' 16''.6. The discovery and early observations are discussed by \cite{Jencson2019abn}, while the evolution of the transient up until 200 days from the discovery is covered by \cite{Williams2019abn}. In this paper we provide additional optical data, especially at later stages of evolution, while also publishing original NIR and MIR observations obtained with IO:I and NOTCam, which put constraints on a critical section of the Spectral Energy Distribution (SED). 
By measuring the magnitude of the standard stars used as reference by \cite{Williams2019abn} we integrate their dataset with our observations by applying the following magnitude corrections to their observations (likely due to a different choice of reference stars): $\Delta B$ = +0.07 mag, $\Delta r$ = +0.04 mag and $\Delta i$ = +0.05 mag. Discrepancies in $V$, $z$ and NIR bands were within photometric uncertainty, and no correction was needed. Similarly, we incorporate the observations performed by \cite{Jencson2019abn} in our dataset after applying the following corrections: $\Delta J$ = -0.13 mag, $\Delta H$ = -0.05 mag and $\Delta K$ = -0.06 mag. 
We adopt a distance modulus of $\mu$ = 29.67 $\pm$ 0.02 mag to M51, obtained through the method of the tip of the red giant branch \citep{Distance22019abn, Distance12019abn}. The Galactic absorption in the direction of M51 is A$_{V}$ = 0.096 $\pm$ 0.006 mag, from \cite{SchlaflyNEDReddening2011}, under the assumption that $R_{V}$=3.1 \citep{Cardelli}. The local absorption is more challenging to estimate and will be discussed, for all objects, in Section \ref{reddeningEstimate}. 

\begin{figure*}
	\includegraphics[width=\textwidth]{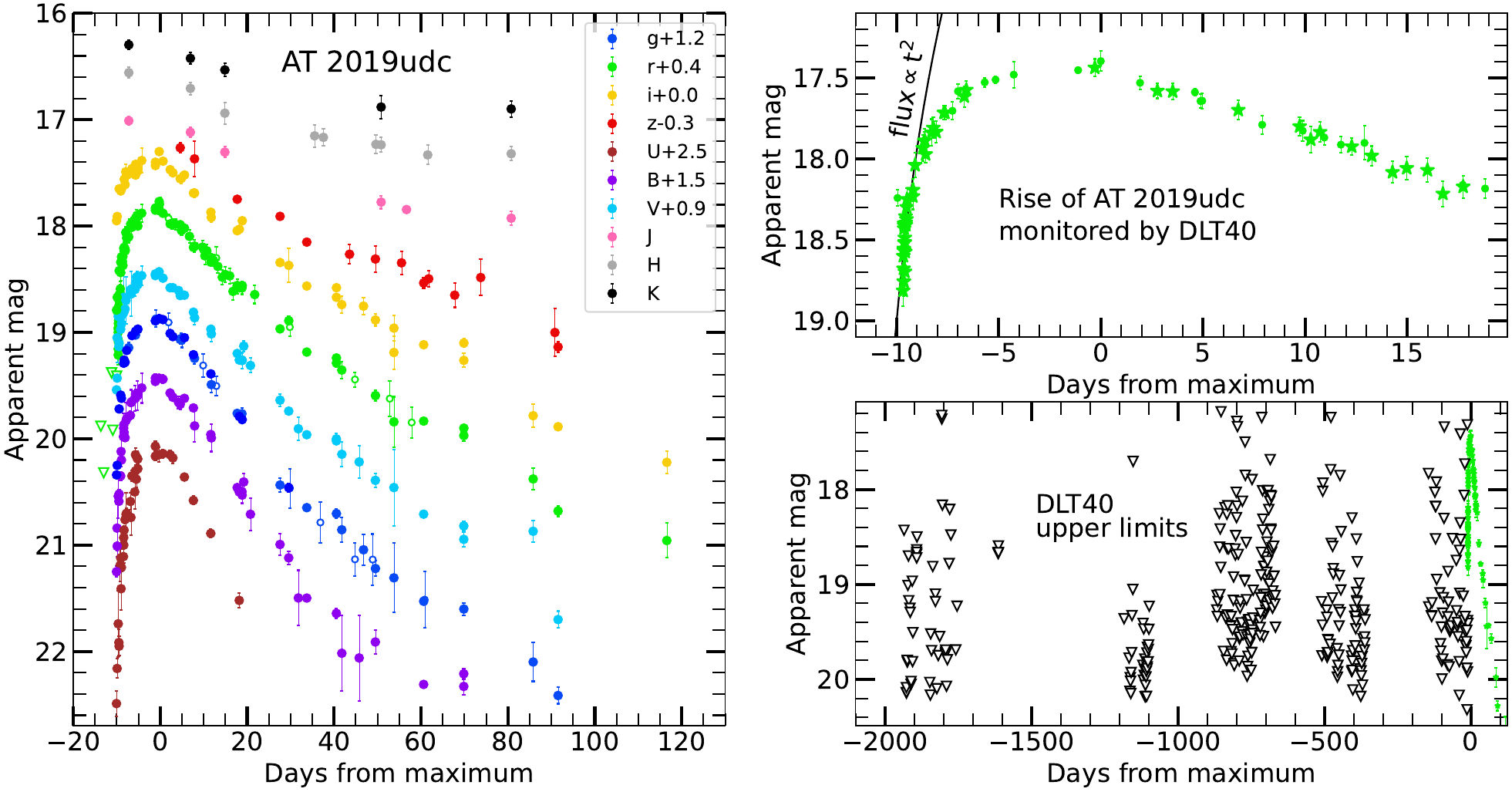}
    \caption{On the right panel we present optical and NIR light curves of AT 2019udc. Magnitude shifts have been applied for clarity. Filled circles represent unpublished data, while empty circles represent ZTF data points. On the left panels, DLT40 monitoring of AT 2019udc. In particular, on the upper left panel is shown the high cadence follow up of the rise of AT 2019udc: DLT40 data points are shown as stars and are integrated with observations obtained through other facilities, represented as circles. On the lower left, years of upper limits collected by DLT40 are displayed as black triangles, with the detections of the transient appearing as green symbols on the right edge of the figure.}
    \label{udc_phot}
\end{figure*}

The light curves of AT 2019abn are shown in Figure \ref{abn_phot}.
Thanks to the early discovery, it is possible to follow the evolution of AT 2019abn from very early stages. The rise in luminosity is observed in multiple bands, and lasts roughly 20 days before the transient reaches a peak apparent magnitude of m$_{r}$ = 16.73 $\pm$ 0.01 mag on MJD = 58527.3. Thanks to this excellent coverage of the rise, it is possible to estimate the rising rates ($\gamma_{1}$) for the observed bands. To do so, we perform a linear regression of the measured magnitudes in each band between 25 and 11 days before maximum: such rising (and declining) rates are useful to quantitatively compare the behaviour of each band at different phases, as well as the differences between each transient (see \citealt{Yongzhi2021ILRT} for similar measurements on other ILRTs). For sake of simplicity, we separate each light curve in multiple linear segments, and we display their declining (or rising) rates in Table \ref{abnrates} and following tables. As a reference, here in the text we report the decline rates measured on the $r$ band in units of [mag/100 days].
At 11 days before maximum, AT 2019abn starts departing from its initial linear increase in magnitude ($\gamma_{1}$=-21$\pm$1), forming a broad peak also described by \cite{Williams2019abn}. The post maximum luminosity evolution is slow ($\gamma_{2}$=1.37$\pm$0.01) especially in the red bands, while the decline rate becomes more steep after 110 days ($\gamma_{3}$=2.49$\pm$0.04). Between 180 and 195 days after maximum, there is a sudden drop in luminosity of $\sim$ 0.9 magnitudes in all observed optical and NIR bands. After this abrupt change, the light curves settle on slow decline rates ($\gamma_{4}$=0.80$\pm$0.06). The MIR sampling of AT 2019abn, obtained through the Spitzer and WISE (survey NEOWISE) space telescopes, is unprecedented for an ILRT. After the first data point, at 50 days after maximum, the MIR light curves show at first a decline faster compared to the optical bands, but after $\sim$100 days the decline becomes more shallow. Interestingly, the luminosity drop at $\sim$180 days is not as evident in the MIR bands.



\subsection{AT 2019ahd} \label{subsec_photomAT2019ahd}
The discovery of AT 2019ahd was reported by the ATLAS survey \citep{2019ahdDiscovery,ATLAS_discovery_Smartt} on 2019 January 29.0 UT. The coordinate of the transient are RA = $10^{h} 51^{m} 11^{s}.737$ Dec = +05\textdegree 50' 31''.03, which is 2''.6 north and 6''.9 south of the centre of its host, the spiral galaxy NGC 3423. As the distance modulus of the host galaxy, we chose to adopt an average of the different independent values reported on the NASA/IPAC Extra-galactic Database (NED) obtaining a distance modulus $\mu$ = 30.22 $\pm$ 0.14 mag, where the error comes from the standard deviation of the measurements \citep{2019ahd_dist1,2019ahd_dist2,2019ahd_dist3,2019ahd_dist4}. We assumed a cosmology where $H_{0}$ = 73 km s$^{-1}$ Mpc$^{-1}$, $\Omega_{\Lambda}$ = 0.73 and $\Omega_{M}$ = 0.27 \citep{SpergelCosmology}, which will be used throughout this whole work. The Galactic absorption in the direction of NGC 3423 is A$_{V}$ = 0.079 $\pm$ 0.003 mag \citep{SchlaflyNEDReddening2011}. The object was initially classified as a Luminous Blue Variable (LBV, \citealt{AHDclass}) due to its narrow hydrogen features and relatively red spectrum. The following photometric and spectroscopic evolution of the transient, in particular the prevalence of calcium features (Ca II H\&K, [Ca II] and Ca NIR triplet; see Paper II), proved that it is an ILRT instead. 
The majority of the follow-up performed for this object is obtained through the GROND telescope, which yielded a remarkably homogeneous data set (Figure \ref{ahd_phot}). The brightest magnitude is well constrained at m$_{r}$ = 17.57 $\pm$ 0.07 mag, reached on MJD = 58525.0. Similar to AT 2019abn, AT 2019ahd displays a slow decline just after peak luminosity ($\gamma_{1}$=1.12$\pm$0.09), but this ''pseudo-plateau'' only lasts for 45 days, followed by a steeper decline ($\gamma_{2}$=2.31$\pm$0.10) that ends 105 days after maximum. The subsequent decline rate is again slower ($\gamma_{3}$=1.05$\pm$0.03), in particular in the NIR bands (Table \ref{ahdudcrates}).

\begin{figure}
	\includegraphics[width=\columnwidth]{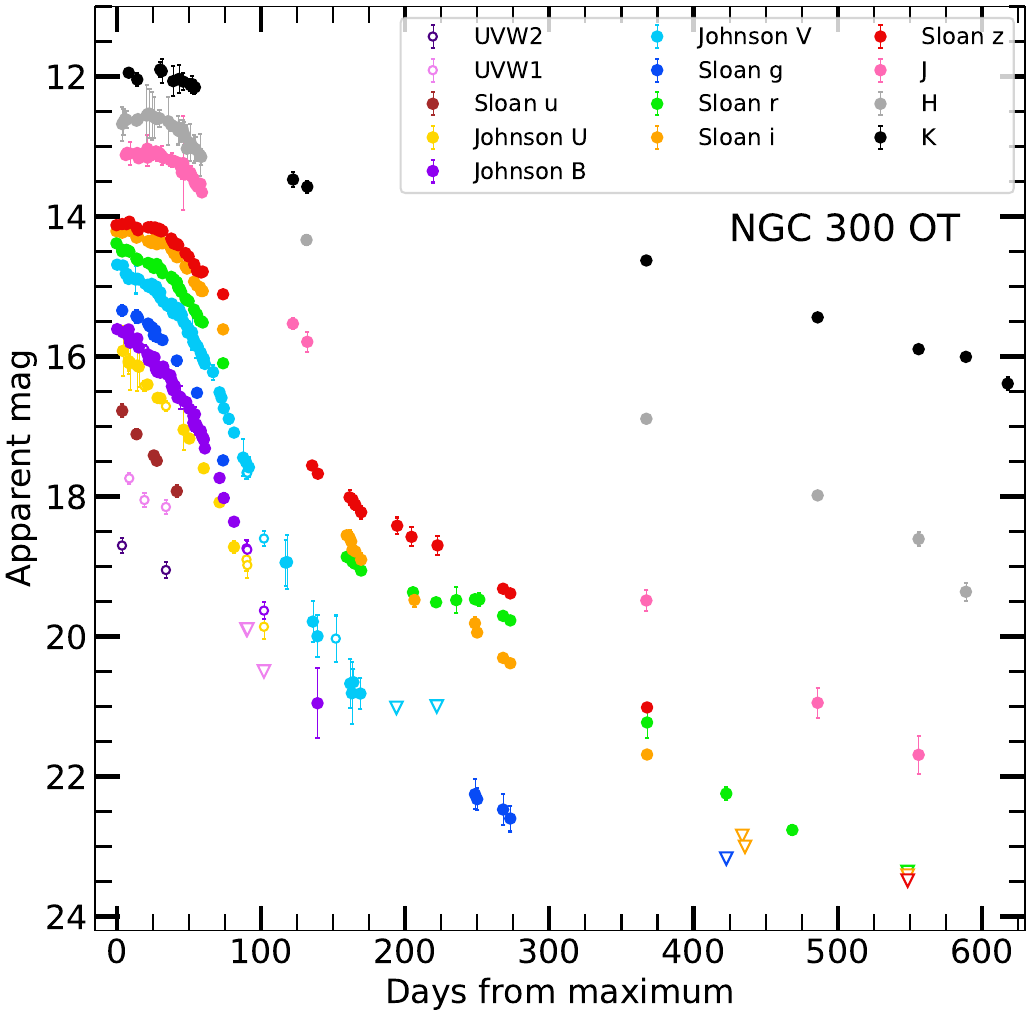}
    \caption{UV, optical and NIR light curves of NGC 300 OT. Empty triangles represent upper limits. Filled circles represent unpublished data, while empty circles represent publicly available SWIFT data.}
    \label{NGC300_phot}
\end{figure}

\subsection{AT 2019udc} \label{subsec_photomAT2019udc}
The discovery of AT 2019udc was reported by the survey DLT40 \citep{DLT40Leo} on 2019 November 4.1 UT. The transient lies on a spiral arm of the galaxy NGC 0718, at RA = $01^{h} 53^{m} 11^{s}.190$ Dec = +04\textdegree 11' 46''.96. We adopt a kinematic measure of distance for NGC 0718, that provides $\mu$ = 31.49 $\pm$ 0.15 mag, obtained through the redshift of the galaxy with respect to 3K CMB \citep{CMB3K}. The Galactic absorption towards NGC 0718 is A$_{V}$ = 0.100 $\pm $ 0.001 mag. Similar to AT 2019ahd, also AT 2019udc was originally classified as an LBV due to its spectral features \citep{UDCclass} but its evolution proves that it is actually an ILRT.
AT 2019udc is the most distant object studied in this sample. The follow-up campaign was stopped 120 days after maximum, due to solar conjunction. The GSP multi-band monitoring yielded a high cadence follow up of the rise and peak of AT 2019udc, crucially contributing to the collection of a high quality data set for this object (Figure \ref{udc_phot}, left panel). During the first two days after discovery (-10 to -8 days from maximum), the rise in magnitudes can be approximated as linear, with an increase rate up to three time faster than the rise of AT 2019abn, with bluer bands showing a systematically faster evolution ($\gamma_{1}$=-47.5$\pm$4.8). After this phase the rise quickly slows down, departing from a linear behaviour and reaching peak luminosity in about 10 days (m$_{r}$=17.40 $\pm$ 0.06, MJD = 58801.28). Immediately after maximum, AT 2019udc displays a fast linear decline ($\gamma_{2}$=3.93$\pm$0.09), again around three times faster than the decline of AT 2019abn just after maximum. The luminosity drop settles on a gentler slope in all bands at around 35 days after maximum ($\gamma_{3}$=2.22$\pm$0.12). The resulting light curve is notably fast evolving for an ILRT: although peculiar compared to the other objects in the sample, these features are not unique among ILRTs, as shown in Section \ref{TemplateLC_sect}.

Thanks to the prompt discovery and high cadence monitoring of the DLT40 survey, the luminosity rise of AT 2019udc is exceptionally well sampled, as presented in the upper right panel of Figure \ref{udc_phot} (DLT40 data are unfiltered observations scaled to sloan $r$ band observations).
The rise immediately after the discovery is remarkably fast, 0.8 magnitudes in less than one day, followed by a more gentle brightening of 0.6 mag in 2.5 days. At this point, 3 days after the first DLT40 detection, AT 2019udc is already close to maximum luminosity, and its magnitude will only marginally change (within $\pm$0.15 mag) in the following two weeks. We try to fit the rise to maximum of AT 2019udc with a simple fireball model, where the flux increases as $t^{2}$ (as detailed by e.g. \citealt{Nugent2011fe}). However, as shown by the black solid line in the upper right panel of Figure \ref{udc_phot}, the observed light curve quickly departs from the fireball model extrapolated from steep rise of the first two days. One possible explanation for this discrepancy is that the early light curves of ILRTs are not dominated by $\textsuperscript{56}$Ni decay, unlike the light curves of SNe Ia (for which the fireball model is a reasonable approximation).
DLT40 also provided several years of upper limits to the optical luminosity of the progenitor of AT 2019udc, which are displayed in the lower right panel of Figure \ref{udc_phot}. The absence of outbursts and the overall non-detection in the optical domain of the progenitor of AT 2019udc is well in line with the expectations for ILRTs, whose precursors have been identified as dust enshrouded stars, heavily extincted in the optical wavelengths but luminous in the MIR \cite{ThompsonProgenitor2009}.


\subsection{NGC 300 2008 OT-1} \label{subsec_photomNGC300OT}
NGC 300 2008 OT-1 (hereafter NGC 300 OT) was discovered on 2008 May 14 during the SN search program at the Bronberg Observatory \citep{MonardDiscovery}. The event, located in the nearby NGC 300 at RA = $00^{h} 54^{m} 34^{s}.51$ Dec = -37\textdegree 38' 31''.4, was extensively studied in the subsequent years \citep{Bond2009,Berger2009,Hump2011,Adams2016} and became a prototype for the class of ILRTs together with SN 2008S \citep{Botticella2008S}. Here we present additional optical, NIR and MIR data, adding this object to our sample of ILRTs. 
For the distance of NGC 300, we adopt the results published by \cite{Gogarten2010}, where a distance modulus $\mu$ = 26.43 ± 0.09 mag is obtained through the Red Clump method. The Galactic absorption towards NGC 300 is A$_{V}$ = 0.034 $\pm$ 0.001 mag \citep{SchlaflyNEDReddening2011}.
NGC 300 OT is the closest ILRT ever observed, making it an extremely valuable which was monitored in detail for an extended period of time. In Figure \ref{NGC300_phot} we present the original optical and NIR photometric data we collected for this transient. The object was behind the sun during its rise and peak luminosity, so we lack the first part of its evolution. Our observed maximum is m$_r$ = 14.38$\pm$0.02 mag on MJD = 54602.38. We note that there is a serendipitous optical detection of NGC 300 OT on MJD = 54580.65 when the transient is rising (m$_{R}$=16.30$\pm$0.04 mag, \citealt{Hump2011}), providing a rough estimate for the onset epoch of the event.
The first 35 days display a slow decline ($\gamma_{1}$=1.11$\pm$0.05), especially in the red bands. From 35 to 75 days, the transient falls from this "pseudo-plateau", and its luminosity starts to fade faster ($\gamma_{2}$=3.32$\pm$0.08). Between 75 and 120 days the decline in luminosity is particularly rapid ($\gamma_{3}$=5.24$\pm$0.06, calculated on the $R$ band due to lack of $r$ band coverage in this phase), comparable to the fast declining phase of AT 2019udc. From 120 and 255 days the fast decline stops, and a slow evolution ensues ($\gamma_{4}$=0.69$\pm$0.07) before the final phase ($\gamma_{5}$=1.56$\pm$0.05) that encompasses from 255 days onwards.
As done for AT 2019abn, we integrate the dataset provided by \cite{Hump2011} with our observations by applying magnitude corrections to their observations, calculated by measuring the magnitude of the reference star chosen in their work: $\Delta B$ = +0.07 mag, $\Delta R$ = +0.05 mag. No correction was needed for $V$ band, $I$ band and NIR data, since they were perfectly matching. We also performed aperture photometry on the public images of NGC 300 OT taken with $Swift$ UVOT which were analysed by \cite{Berger2009}. We find a remarkable agreement in the $U, B, V$ magnitudes reported there, while we measure overall fainter $UVW2$ and $UVW1$ magnitudes, possibly due to a different background selection: in particular, we restricted the aperture down to 3'' to limit possible background contamination. The resulting UV fluxes from our measurements are in line with the behaviour expected from a black body emission, as shown in the top left panel of Figure \ref{sedTotal}.

\begin{figure}
	\includegraphics[width=\columnwidth]{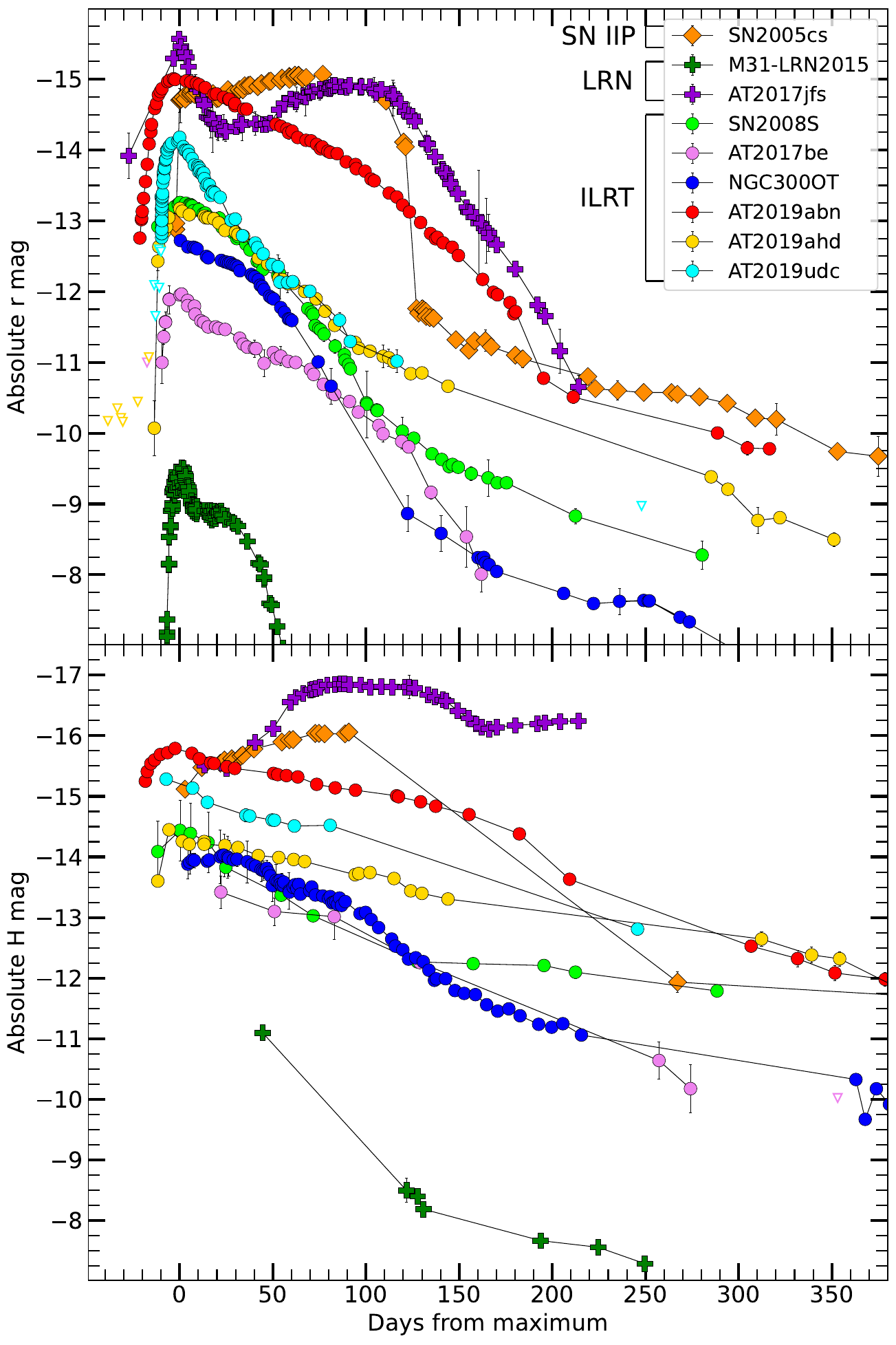}
    \caption{Absolute $r$ band (upper panel) and $H$ band (lower panel) light curve comparison between the ILRTs in our sample and other low-luminosity transients. In particular ILRTs are marked with circles, LRNe with plus signs and the SN IIP with diamonds.}
    \label{Comparison}
\end{figure}

\subsection{Comparison with other transients} \label{Comparison_sect}
In Figure \ref{Comparison}, we compare the absolute $r$ and $H$ band evolution of our sample of ILRTs along with that of other transients of similar luminosity.
The first considerations can be made observing the sample of ILRTs, with the addition of the well studied SN 2008S \citep{Botticella2008S} and AT 2017be \citep{Yongzhi2017beC}. We remark the large spread in the optical peak magnitudes, which span from --12 mag for AT 2017be to --15 mag for AT 2019abn. We note that AT 2019abn (M$_{r}$ = --15.00$\pm$0.08 at maximum) is, to date, the brightest ILRT ever observed, with AT 2019udc (M$_{r}$ = --14.17$\pm$0.16 at maximum) also falling on the brighter end of the ILRTs luminosity distribution. 
AT 2019ahd and NGC 300 OT, instead, display fainter peak magnitudes, --13.17$\pm$0.16 mag and --12.69$\pm$0.11 mag respectively. Among the least luminous ILRT we find AT 2017be, which is barely brighter than M$_{r}$ = --12 mag.
There are also significant differences in light curve shapes within the class. AT 2019abn and AT 2019udc represent the two extreme cases, with the former being characterised by a long phase of slow decline after peak, almost a pseudo-plateau in the $r$ band (1.37 $\pm$ 0.01 mag/100 days), while AT 2019udc undergoes a decline that is three times faster (3.93 $\pm$ 0.09 mag/100 days), after peak luminosity.
This variability is less pronounced in the NIR domain, where both the decline rates and the peak magnitudes span a smaller range of values. Regardless of their differences, ILRTs tend to settle on a linear decline at late times, which is compatible with the expected luminosity decline sustained by $\textsuperscript{56}Ni$ radioactive decay. This can be visually evaluated by comparing the late time behaviour of ILRTs with that of SN 2005cs, one of the prototypes of low-luminosity SNe IIP \citep{Pasto2005cs}, whose late decline is known to be powered by $\textsuperscript{56}Ni$ decay. We also remark that SN 2005cs is fainter than AT 2019abn at peak luminosity, revealing an overlap between the brightest ILRTs and the faintest core collapse SNe.
To summarize the features described above, ILRTs are characterized by their single-peaked, monotonically declining light curves which terminate in a linear decline at late phases.
A clearly different light curve shape is instead associated to Luminous Red Novae (LRNe, \citealt{PastoLRNeGeneral}), another class of transients populating the luminosity "Gap" which separates classical novae from standard SNe.
LRNe are non-degenerate stellar mergers, and typically display double peaked light curves, as shown for by AT 2017jfs \citep{Pasto2017jfsLRN} and to a lesser degree M31-LRN-2015 \citep{Williams2015M31LRN}, both reported in Figure \ref{Comparison}. With its peak absolute magnitude fainter than --10 mag, M31-LRN-2015 appears detached from the other transients shown: indeed, LRNe can be much dimmer, with peak absolute magnitude even below $\sim$--4 mag (e.g. V1309Sco \citealt{TylendaV1309Sco}). On the other hand, ILRTs discovered to date have been strictly confined within the luminosity Gap (--10 mag$<$M$_{V}$$<$ --15 mag).

\begin{figure}
	\includegraphics[width=1.0\columnwidth]{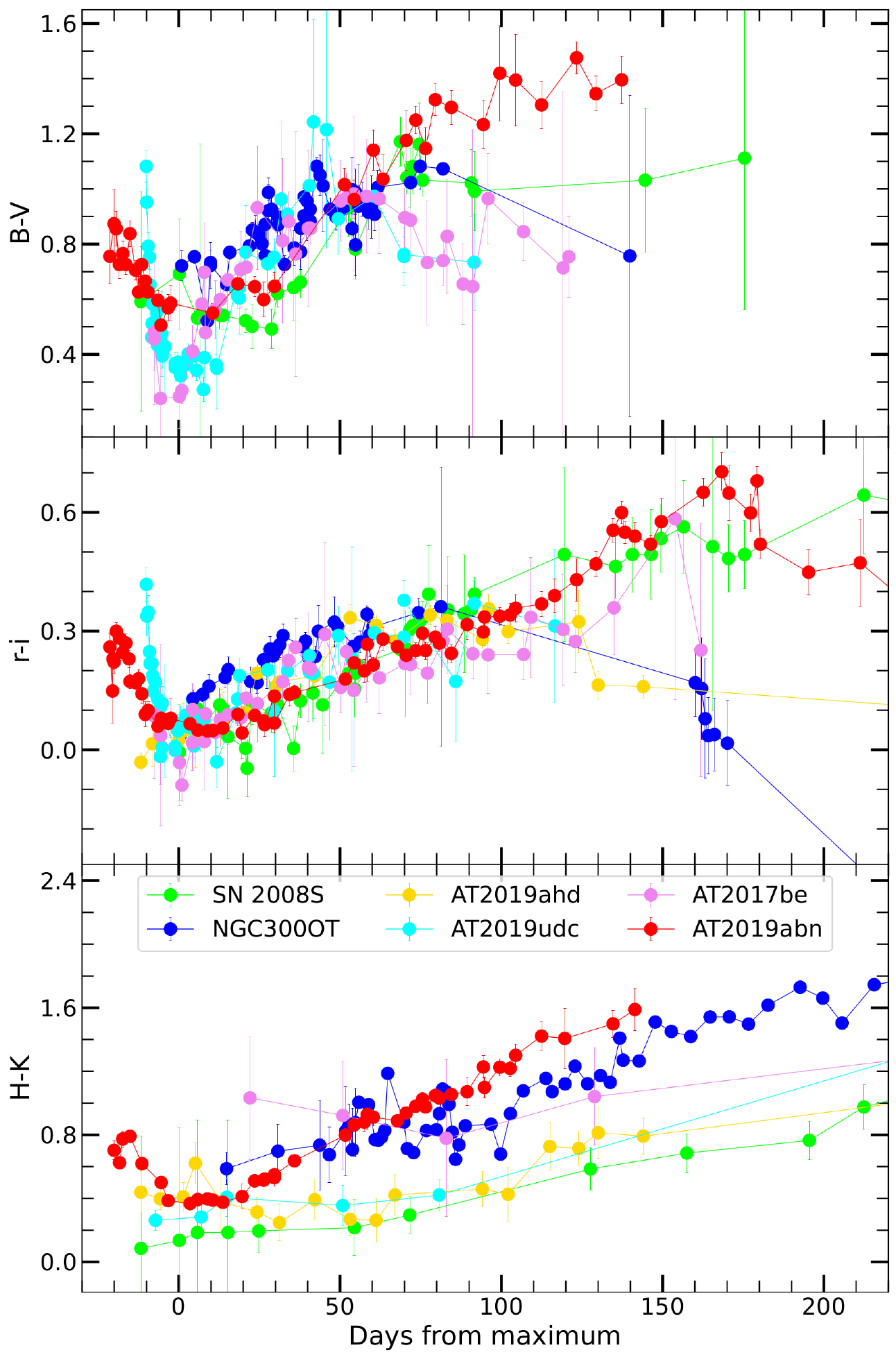}

    \caption{Colour curves for the ILRTs belonging to our sample, with the addition of SN 2008S and AT 2017be. In the top two panels the optical colours are shown. In the bottom panel, it is possible to appreciate the NIR colour evolution.}
    \label{Colours}
\end{figure}

\section{Reddening Estimate} \label{reddeningEstimate}
Estimating the reddening affecting ILRTs is a challenging task, and different approaches can be found in the literature, depending on the available data. A first method consists in using the empirical relation between the Na ID Equivalent Width (EW) and the absorption along the line of sight \citep{Turatto2003,Poznanski2012Na}, as was done by \cite{Yongzhi2017beC}. This method has the advantage of not requiring any assumption on the intrinsic properties of the target, but it needs high S/N spectroscopy in order to be reliable. Furthermore, the EW of Na ID observed may saturate, preventing the application of empirical relations (e.g. \citealt{MaxILRT}). Furtheremore, the EW of the Na ID varies with time in ILRTs (e.g. \citealt{Byrne2023}, and further discussed in Paper II), increasing the uncertainty associated with this method.
A second method is based on the observation that the spectral features of ILRTs resemble those of F-type stars \citep{Hump2011,Jencson2019abn}. 
It is then assumed that all ILRTs reach a temperature of $\sim$7500 K at maximum luminosity, and a reddening correction is applied to the Spectral Energy Distribution (SED) at peak in order to obtain a blackbody continuum corresponding to such temperature. 
Yet another strategy is adopted for SN 2008S by \cite{Botticella2008S}, who estimate the extinction affecting the target based on an IR echo model of the SED at early stages.

In this work we adopt the same procedure carried out by \cite{MaxILRT}, who notes that the observed $V$--$r$ and $r$--$i$ colour curves of several ILRTs have a similar shape. Assuming that ILRTs should display a similar evolution of optical colours, a reddening estimate can be obtained by attempting to match the colour evolution of the bluest objects. While this is likely an approximation, it is not baseless. In fact, these transients present a marked homogeneity in their optical spectra (which will be further discussed in Paper II): since similar conditions in temperature and ionization state are required to produce the same spectral features, we can expect the optical colours of these transients to be at least compatible in their photospheric phases.
Following this reasoning, we consider the bluest object in our sample, AT 2019udc, together with AT 2017be, which shows remarkably similar colours, and we apply a reddening correction to all the other objects in our sample in order to superimpose their $B-V$, $g-r$ and $r-i$ colour curves through a least squares minimization procedure. Only the values measured within 100 days after maximum are considered. The absorption values for each transient inferred in this way are reported in the last column of Table \ref{tab:1}. AT 2019abn is by far the most reddened object in our sample, with an estimated internal absorption of A$_{V}$=2.34 $\pm$ 0.06 mag. AT 2019udc, by construction, is assumed to be reddening-free, since it is among the bluest ILRT observed. 
In Figure \ref{Colours} we report the colour evolution for our sample of ILRTs after applying the reddening correction. The optical colours present an initial decrease (i.e. the objects become bluer) during the rise to maximum. This is particularly clear in AT 2019udc and AT 2019abn, thanks to their high cadence monitoring pre-maximum. After peak luminosity, ILRTs become steadily more red in the following months. This trend may invert as the flux becomes dominated by emission lines rather than continuum emission, as seen for NGC 300 OT $r-i$ after 150 days post maximum.
We note that we are using a conservative estimate for the reddening affecting SN 2008S, for which we adopt just the Milky Way absorption of A$_{V}$=1.13 mag. This is a lower limit to the total reddening along the line of sight: adopting a light echo model, \cite{Botticella2008S} infer an internal absorption of A$_{V}$ $\sim$ 1 mag. This leads to a peak temperature of 8400$\pm$120 K, closer to an A type star, rather than an F type star.
In any case, even with minimal absorption, SN 2008S is among the bluest ILRTs to date: therefore if the colour evolution of ILRTs is indeed perfectly homogeneous, our adopted absorption is likely a lower limit of the actual value. On the other hand, it is reasonable to expect some scatter in the temperature and consequently in the colours of ILRTs, granted that all of them should fulfill the physical requirements for producing the spectrum of an F type star \citep{Hump2011}. In this context, the absorption estimated so far can be seen as a first order correction, crucial for heavily extincted targets such as AT 2019abn.





\begin{figure*}	
	\includegraphics[width=1.0\textwidth]{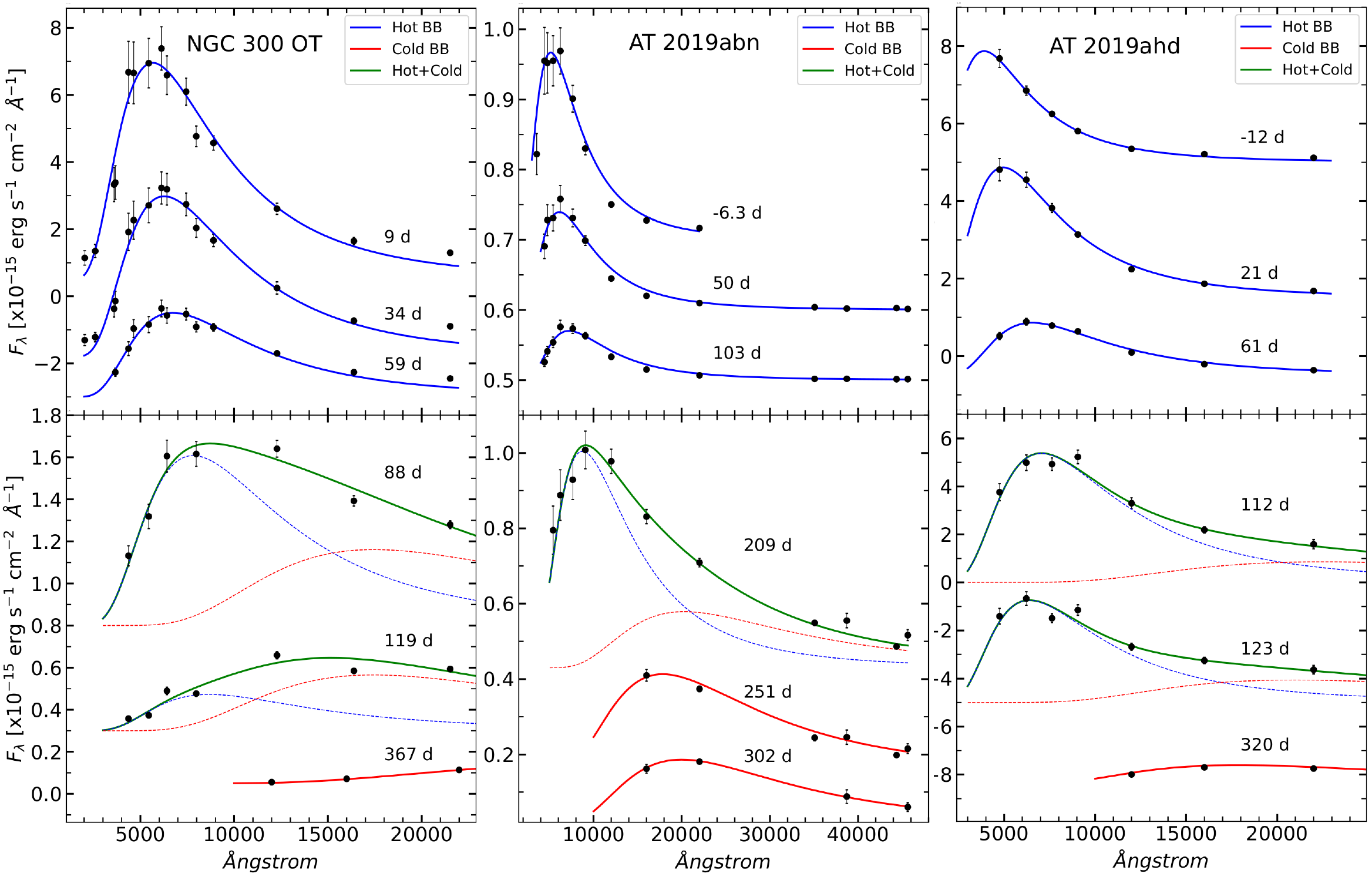}
	
    \caption{From left to right SED evolution of NGC 300 OT, AT 2019abn and AT 2019ahd. In the upper panel, a single black body is sufficient to fit the data, represented in blue. At later phases, shown in the lower panel, a second black body is needed to reproduce the NIR flux excess. Epochs are reported with respect to maximum light. Flux shifts have been applied for clarity.}
    \label{sedTotal}
\end{figure*}

For the sake of completeness we compare the reddening estimates just obtained (which will be adopted for this work) with a straightforward application of the linear relations between Na ID EW and E(B-V) presented by \cite{Turatto2003}. The detailed evolution of the Na ID EW for this sample of ILRTs will be shown and discussed in Paper II: here we simply consider the first measured value for each transient, specifically 0.7$\pm$0.1 {\AA} for NGC 300 OT, 1.1$\pm$0.1 {\AA} for AT 2019abn, 0.9$\pm$0.1 {\AA} for AT 2019ahd and 5.4$\pm$0.2 {\AA} for AT 2019udc. The two different linear relations provided by \cite{Turatto2003} allow to determine both a lower and upper estimates of E(B-V) for each object. The resulting A$_{V}$ range obtained for NGC 300 OT spans between 0.3 and 0.9 mag, in good agreement with our previous estimation of 0.78 mag. In case of the lowest reddening inferred, the peak luminosity of NGC 300 OT in $r$ band would be 0.4 mag fainter, leading to a peak absolute magnitude M$_{r}$= -12.4 mag. For AT 2019ahd we obtain 0.4 mag$<$A$_{V}$$<$1.3 mag, with the lower limit being compatible with our previous measurement of 0.37 mag. If we accepted the larger estimate of A$_{V}$$\sim$1.3 mag, the peak $r$ band absolute magnitude of AT 2019ahd would increase from -13.3 mag to -14.1 mag. For the other two ILRTs in the samples the two methods for estimating A$_{V}$ yield more substantial differences.
According to the empirical relations obtained by \cite{Turatto2003}, the EW of the Na ID observed in the early phases of AT 2019abn would imply an extinction of 0.5 mag$<$A$_{V}$$<$1.6 mag, significantly lower than the value 2.34 mag previously obtained. The peak absolute magnitude of AT 2019abn would decrease from M$_{r}$= -15.1 mag to -14.4 mag, or even -13.5 mag in the case of lowest extinction. On the contrary, AT 2019udc initially displays a remarkably large Na ID EW of 5.4$\pm$0.2 {\AA} (however, we note that the Na ID EW rapidly declines to 1.3 {\AA} in just two days, and even below 1 {\AA} after that phase): this would imply an extinction A$_{V}$ of at least 2.6 mag. The resulting peak magnitude would be at least M$_{r}$= -16.5 mag, firmly within the luminosity realm of full-fledged SNe.
Such critical discrepancies in A$_{V}$ would also lead remarkable variations in temperature, in particular a hotter AT 2019udc and a much cooler AT 2019abn, especially around peak luminosity (compared to the results discussed in the next section and displayed in Figure \ref{ILRThot}). This is an unlikely circumstance, given the already mentioned spectral homogeneity of ILRTs.

\section{SED Evolution} \label{SED_evol_sect}
In order to extract physical quantities of our targets, we perform blackbody fits on the SED at different epochs.
This analysis is carried out through Monte Carlo simulations, using the \textsc{python} tool \textsc{curve\_fit}\footnote{https://docs.scipy.org/doc/scipy/reference/generated/ scipy.optimize.curve\_fit.html} to perform fits on 200 sets of fluxes randomly generated with a Gaussian distribution centered at the measured flux value, and $\sigma$ equal to the error associated to the measurement. Such procedure was already adopted and described in \cite{LRNeHugsI,Valerin2020cxd}. The blackbody fit to the SED of the target yields the estimated temperature, and by integrating over the wavelength we obtain the total flux emitted. 
Adopting the distances discussed in Section \ref{Photometry_sect} and assuming spherical symmetry, we calculate the bolometric luminosity of the source. 
Finally, the radius is estimated through the Stefan--Boltzmann law. This whole procedure is repeated for each epoch with suitable photometric coverage, in order to study the evolution of the inferred physical parameters with time.
During the first phase of evolution all the objects are well fitted by a single black body, associated to the photosphere, throughout the optical and NIR domains. We refer to this component as the ''hot'' blackbody, in contrast with the ''cold'' blackbody which emerges at later stages. The phase in which only the hot blackbody is visible has a variable duration: up to 180 days for AT 2019abn, 110 days for AT 2019ahd and only 80 days for NGC 300 OT (Figure \ref{sedTotal}). The SED of AT 2019udc is well fit by a single black body at all epochs observed, but it is possible that the cold black body emerged after the end of our follow-up campaign.

\begin{figure}
	
	\includegraphics[width=0.9\columnwidth]{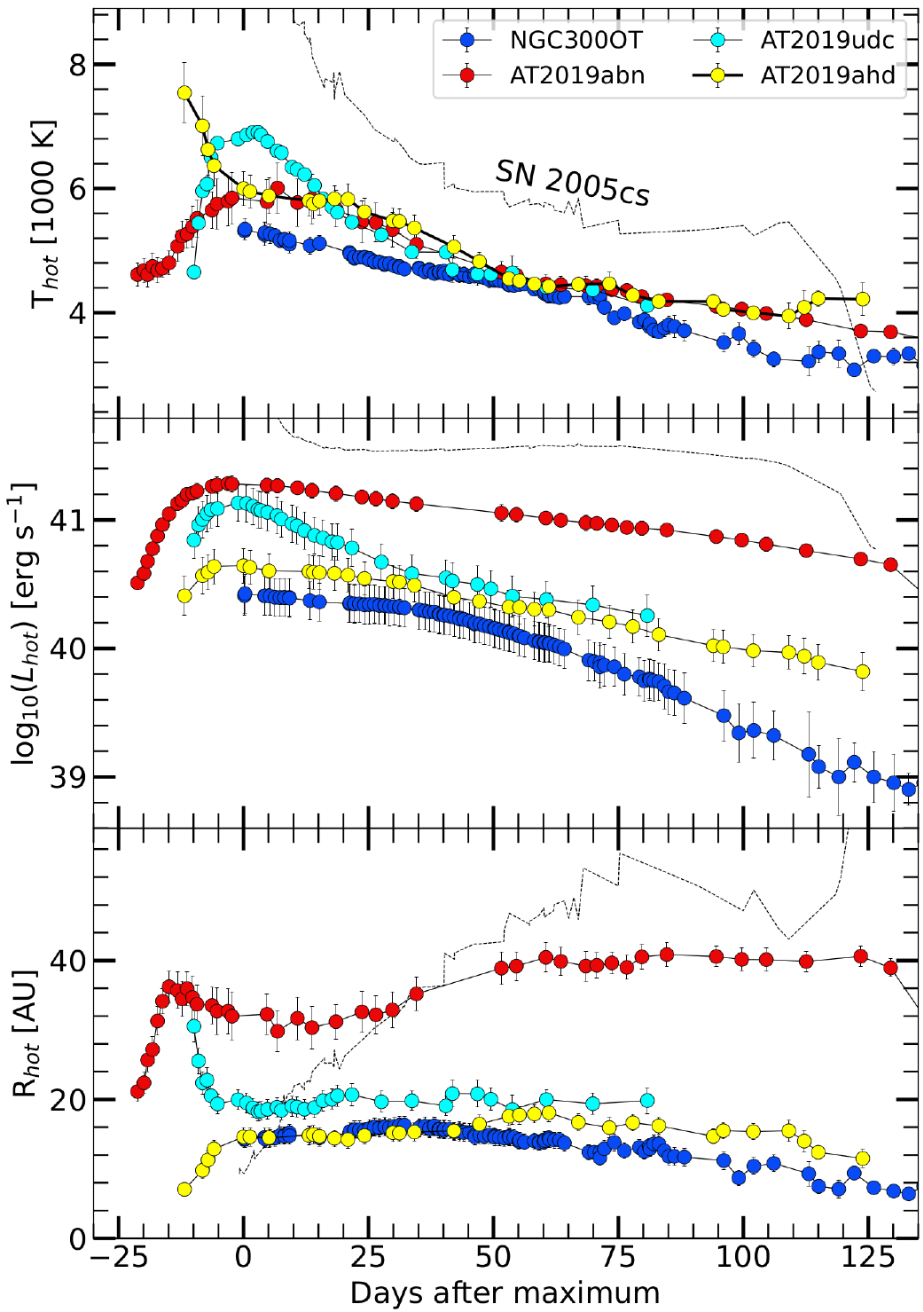}
	
    \caption{Temperature, luminosity and radius evolution of the hot black body of the four ILRTs in our sample. The same values measured for SN 2005cs are shown as a dotted line for comparison. }
    \label{ILRThot}
\end{figure}

The values of temperature, luminosity and radius obtained for this hot blackbody are displayed in Figure \ref{ILRThot}. The temperature evolution of our sample is especially homogeneous between 25 and 75 days after maximum, due to our reddening estimate through colour curves superposition (Section \ref{reddeningEstimate}). However we still find interesting differences between the various targets in the pre--maximum phases. AT 2019abn displays an almost constant temperature of $\sim$4600 K for several days before slowly reaching the peak temperature of $\sim$5900K in $\sim$20 days. AT 2019udc qualitatively follows the same behaviour, but in a shorter timescale (less than 10 days) and reaching 7000 K at peak.
This initial increase in temperature could be explained with the injection of additional energy provided by the interaction between ejecta and CSM, as happens for interacting SNe (e.g. \citealt{DessartBluerInteraction}).
AT 2019ahd, on the other hand, starts from a temperature of $\sim$7500 K, before quickly cooling to 6000 K at peak luminosity. From this point onward, AT 2019ahd closely follows the behaviour of AT 2019abn. In this respect, the early temperature evolution of AT 2019ahd is reminiscent of that of SN 2005cs (displayed in Figure \ref{ILRThot} for comparison) and LL SN IIP in general, where the temperature quickly declines during the rapid expansion of the ejecta. 
NGC 300 OT shows a very simple, monotonic temperature evolution, but in this case we miss the pre--maximum photometric coverage.

\begin{figure}
	
	\includegraphics[width=0.93\columnwidth]{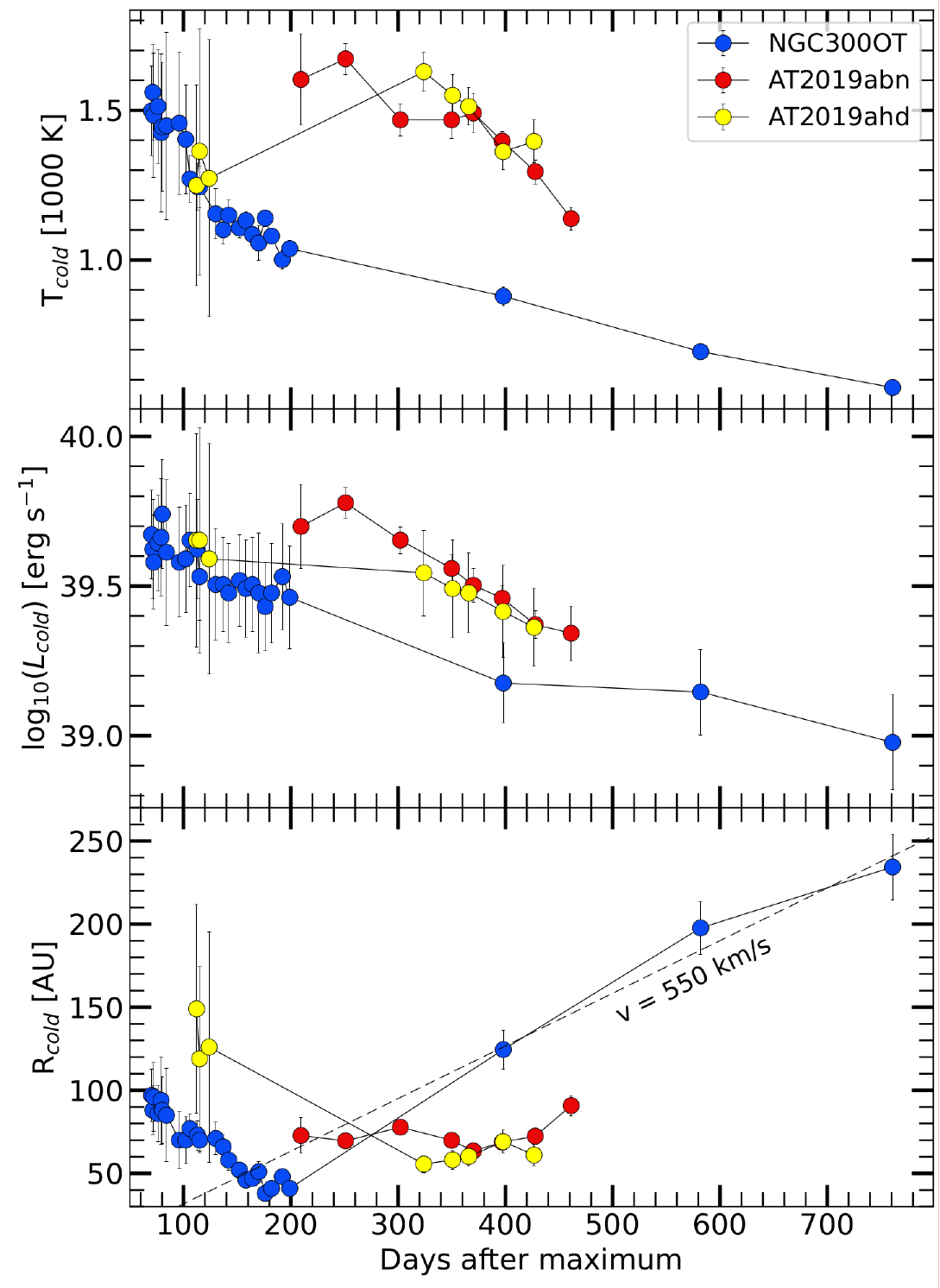}
	
    \caption{Temperature, luminosity and radius evolution of the cold black body of the three ILRTs in our sample that present a NIR excess in their SED. In the third panel, the dashed line represents the position of material expanding with velocity of 550 km s$^{-1}$.}
    \label{ILRTcold}
\end{figure}

In the middle panel of Figure \ref{ILRThot} we present the bolometric luminosity obtained for the ''hot'' blackbody. The bolometric luminosity behaves in a similar way to what is described in Sect \ref{Comparison_sect}, with AT 2019abn being the brightest ILRT, showing a peak luminosity of 1.9$\pm$ 0.3 $\times$ 10$^{41}$ erg s$^{-1}$. As previously pointed out, its decline rate is slower compared to other objects of the same class, but it is not flat like the plateau of SNe IIP. AT 2019udc is characterized by a marked peak (1.4$\pm$0.5 $\times$ 10$^{41}$ erg s$^{-1}$) followed by a fast decline. AT 2019ahd and NGC 300 OT display more modest peak luminosities, respectively of 4.4$\pm$1.1 and 2.7$\pm$1.0 $\times$ 10$^{40}$ erg s$^{-1}$.
As for the evolution of the radius of the emitting source, shown in the bottom panel of Figure \ref{ILRThot}, it appears that all ILRTs in our sample follow a similar behaviour. After maximum, the transients show a roughly constant radius with time, although with different values. AT 2019abn again stands out from the group, showing a blackbody radius that at first quickly increases from 21$\pm$2 to 36$\pm$3 AU from discovery to peak luminosity, and then remains at 30 to 40 AU in the following 125 days. These values are roughly two to three times those obtained for the other three targets. AT 2019ahd shows qualitatively the same behaviour, with a radius growing from 7$\pm$1 to 15$\pm$1 AU from discovery to peak magnitude, and a subsequent slow evolution ranging from 14$\pm$1 to 18$\pm$1 AU within 110 days after maximum. AT 2019udc on the other hand does not show an increase in the radius during the pre--maximum phase, but rather a marginal decrease, from 23$\pm$3 to 19$\pm$2 AU. The following slow evolution of the blackbody radius between 16$\pm$2 and 21$\pm$2 AU over the course of 90 days is reminiscent of those of the other two ILRTs already presented. Finally, for NGC 300 OT we do not have pre--maximum data, but the evolution of the blackbody radius after the observed maximum is again slow, spanning from 11$\pm$1 to 16$\pm$1 AU in 90 days. This tendency of the hot blackbody radius to linger around a constant value is in stark contrast with the monotonic increase in radius for SNe IIP and Low Luminosity SNe IIP during the first 80 days, as shown by the radial evolution of SN 2005cs.

\begin{figure*}
	\includegraphics[width=\textwidth]{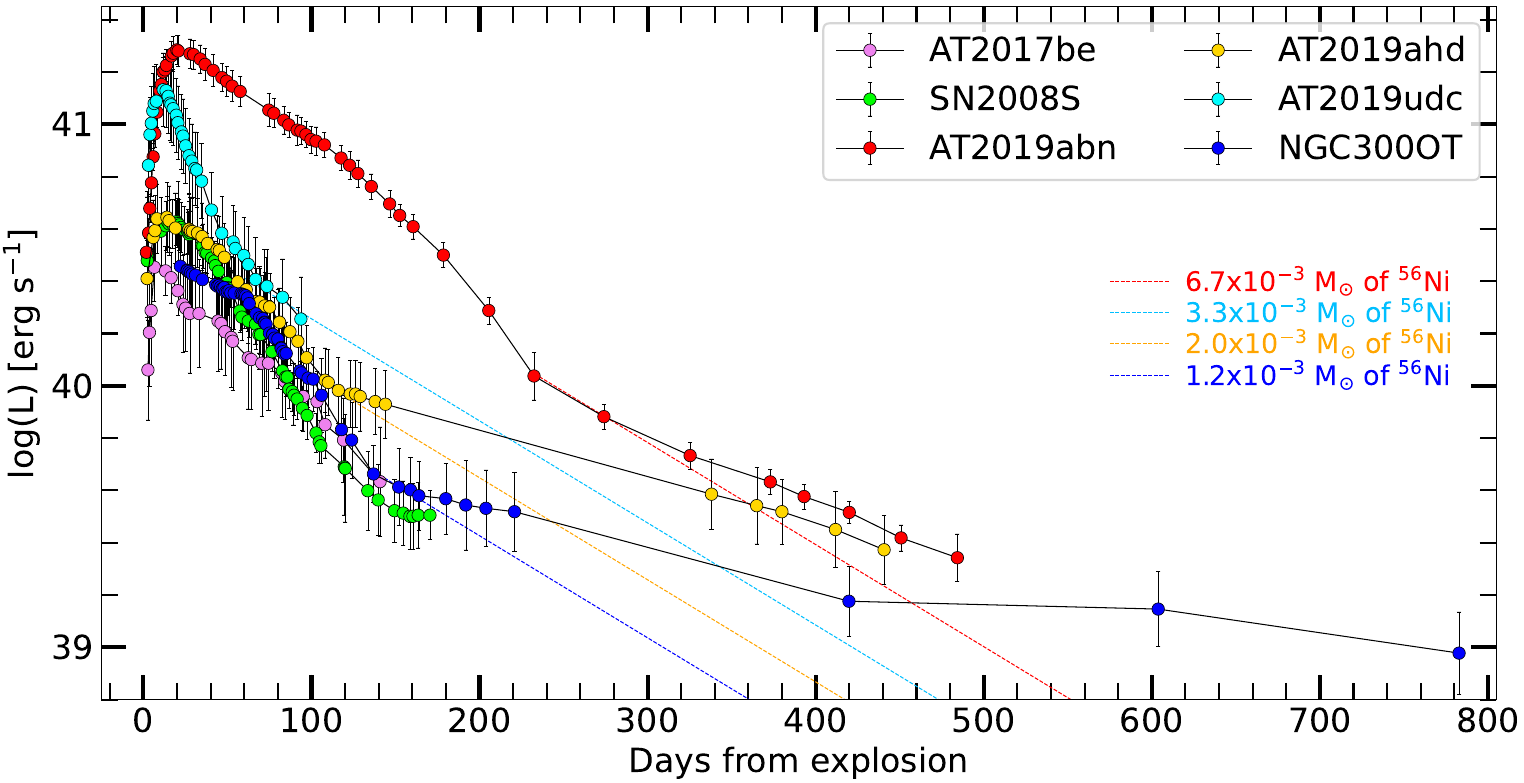}
    \caption{Bolometric light curves of our sample of ILRTs, with the addition of SN 2008S and AT 2017be. Coloured dotted lines show the \textsuperscript{56}Ni decline rates for the upper limits estimated for each object in our sample.}
    \label{bolometric}
\end{figure*}

So far we described the physical properties of the hot blackbody, which is associated to the photosphere of the transient and is usually well visible for $\sim$100-150 days after maximum.
At later phases however, as the gas cools down and expands, an excess in the NIR flux is detected (extending to the MIR, when such measures are available). Such feature can be associated with the formation of dust, which contributes to the SED with a second emission component characterised by a low temperature compared to the hot continuum observed in the first months of evolution (T $\lesssim$ 1500 K, \citealt{Botticella2008S,Yongzhi2017beC}).
In the bottom panels of Figure \ref{sedTotal} are shown blackbody fits performed to the SED of our targets at epochs when two components coexist. At the latest phases, as the objects shift from the photospheric to the nebular phase, the optical flux is entirely supported by emission lines (H$\alpha$, H$\beta$, [CaII], Ca NIR triplet), and there is no more evidence of a hot spectral continuum. At these epochs we still identify a contribution from the "cold" black body associated with dust emission. This feature is especially clear in AT 2019abn and NGC 300 OT (at 761 days after maximum, see Section \ref{NGC300OTdust}) thanks to the MIR measurements. 
In Figure \ref{ILRTcold} we present the physical parameters obtained for the ``cold" blackbody component associated with dust emission. In principle this cold blackbody could be the result of a NIR light echo (e.g. \citealt{LightEcho}). However, the timing of the appearance of this NIR excess (80 days for NGC 300 OT being the shortest) would imply that the dust causing the light echo is located at least $\sim$6900 AU away from the transient. \cite{Botticella2008S} show that the flash from SN 2008S is able to create a dust-free cavity of just $\sim$2000 AU: it appears unlikely that the energy radiated by our ILRTs can bring such a distant body of dust close to its sublimation temperature ($\sim$1400 K for silicates and $\sim$1800 K for graphite, e.g. \citealt{DustSubTemp}). Therefore, we favour the interpretation that this cold blackbody component is emitted by newly condensed dust rather than by a light echo.
AT 2019abn and AT 2019ahd show very similar cold components, with temperatures cooling from $\sim$1600 to $\sim$1200 K over a timescale of hundreds of days. The luminosity of both components evolve slowly, starting from $\sim$3 $\times$ 10$^{39}$ erg s$^{-1}$. The radius of the emitting source is around 60 AU for AT 2019ahd and 75 AU for AT 2019abn.
The most relevant difference between the two transients is that AT 2019ahd develops the "cold" component sooner, at around 110 days. We stress that the blackbody fits at these epochs are affected by large errors, due to the uncertainty in disentangling the hot and cold components in the SED.

This is true also for NGC 300 OT, but in this case we have a higher cadence of NIR observations (also accounting for previously published observations), allowing us to better constrain the evolution of the cold component. A first, fast decline in temperature is observed between 71 days and 130 days after maximum, when the dust cools from 1550$\pm$170 K to 1150$\pm$90 K. The subsequent temperature decline is slower, reaching 570$\pm$10 K at 761 days. Such late estimate is possible thanks to publicly available WISE data. The other two late measurements at 398 and 582 days were performed on MIR data taken by \cite{OhsawaNGC300}.
The luminosity of this "cold" component fades from 5.5$\pm$1.9 $\times$ 10$^{39}$ erg s$^{-1}$, just after its emergence, to 9.5$\pm$0.9 $\times$ 10$^{38}$ erg s$^{-1}$ at 761 days. The radius inferred for the dust emission at first shrinks from $\sim$100 AU to $\sim$40 AU, but around 180 days the trend inverts and the radius grows to 235$\pm$20 AU at 761 days.
The initial shrinking of the emitting source can be qualitatively explained as follows: we are observing dust formation far from the star, with dust condensing at progressively smaller radii as the transient becomes dimmer and the temperature decreases. Indeed, we know that the progenitor star was enshrouded in dust \citep{ThompsonProgenitor2009} before the transient event caused its sublimation, and we may be witnessing the rebirth of this dusty cocoon.
On the other hand, the increase in radius of emitting dust observed after 200 days can be found also in SNe, where the dust expands together with the ejecta (e.g. \citealt{Wesson2015SN1987Adust}). As a simple comparison, in the bottom panel of Figure \ref{ILRTcold} we mark with a dashed line the position of matter expanding with a velocity of 550 km s$^{-1}$, and we note that the three latest measures of radius for NGC 300 OT are compatible with such trend. As will be shown in Paper II, the velocity of 550 km s$^{-1}$ is also well in line with the full width at half maximum velocity estimated from the emission lines of NGC 300 OT, corroborating the idea that the dust is moving jointly with the expanding gas seen during the photospheric phase. As previous observations displayed, part of the dusty cocoon surrounding ILRTs may survive the explosion, avoiding sublimation thanks to a large distance from the star \citep{Botticella2008S,Prieto2009pPNe}. However, in the present work we lack the mid and far-infrared coverage at early phases necessary to detect the emission from such dust component.

In order to obtain the bolometric luminosity of ILRTs, we combine the contribution of the ``hot", photospheric component, and the "cold", dusty component. At late epochs, a blackbody continuum is no longer discernible in the optical domain, with only emission lines dominating the spectra. Therefore, to measure the bolometric luminosity in those cases we integrate the fluxes in the optical domain using the trapezoidal rule. On the other hand, at longer wavelengths it is still possible to perform a blackbody fit and integrate the flux even beyond the observed spectral region. In fact, while the optical domain only displays emission lines at late time, there still is still evidence of thermal continuum in the NIR and MIR domain.
The bolometric luminosity obtained for our sample of ILRTs is reported in Figure \ref{bolometric}.
We analysed an unprecedented amount of NIR and MIR data for ILRTs, which allow us to better infer the luminosity contribution at longer wavelengths, crucial at late times.
The dotted lines in Figure \ref{bolometric} show the expected behaviour of a light curve powered exclusively by \textsuperscript{56}Co decay: it is evident that AT 2019abn, AT 2019ahd and NGC 300 OT are characterised by a late time decline shallower than what is expected if only \textsuperscript{56}Co decay is supporting their late time luminosity.
Of course a more shallow slope in the light curve does not exclude the presence of \textsuperscript{56}Co, but rather it requires an additional mechanism to explain the late time luminosity of ILRTs. In order to estimate an upper limit to the \textsuperscript{56}Ni mass powering the late linear decline of ILRTs, we employ the analytical formulation provided by \cite{Hamuy2003} for the luminosity of \textsuperscript{56}Ni and \textsuperscript{56}Co decay. By considering the first data points available during each linear decline, we obtain upper limits to the \textsuperscript{56}Ni mass of 6.7x10$^{-3}$ M$_{\odot}$ for AT 2019abn, 2.0x10$^{-3}$ M$_{\odot}$ for AT 2019ahd and 1.2x10$^{-3}$ M$_{\odot}$ for NGC 300 OT. AT 2019udc is the only object that shows a linear decline compatible with the one expected from \textsuperscript{56}Co decay, resulting in an estimate of 3.3$\pm$0.3 x10$^{-3}$ M$_{\odot}$ of \textsuperscript{56}Ni. 



\begin{table*}
\centering
\begin{adjustbox}{tabular=cccccc, center}
\\ \hline
   
  Model & L [10$^{38}$ erg s$^{-1}$] & T [K] & R [AU] & $\tau_{5}$ & M$_{d}$ [10$^{-4}$ M$_{\odot}$] \\
  
 \hline
 \hline
 \textbf{Graphite 0.1 $\mu m$} & 9.1 (1.4)   & 576 (5)  & 228  (18) & 5.9  (0.7) & 1.9 (0.5) \\
 \textbf{Graphite 1.0 $\mu m$} & 9.1 (1.4)   & 579 (5)  & 225  (18) & 4.9  (0.6) & 0.45 (0.11) \\
 \textbf{Silicate 0.1 $\mu m$} & 9.2 (1.4)   & 585 (6)  & 222  (17) & 19.4  (2.4) & 19 (5) \\
 \textbf{Silicate 1.0 $\mu m$} & 9.2 (1.4)   & 586 (6)  & 222  (18) & 16.5  (2.0) & 9.5 (2.4) \\

 \hline

\end{adjustbox}

\caption{Parameters obtained for the different SED fit models of NGC 300 OT at 761 d. Errors are reported in brackets.}
\label{GraSilTable}
\end{table*}

\begin{figure*}
\centering
  \makebox[1.7\columnwidth][c]{\includegraphics[width=0.95\textwidth]{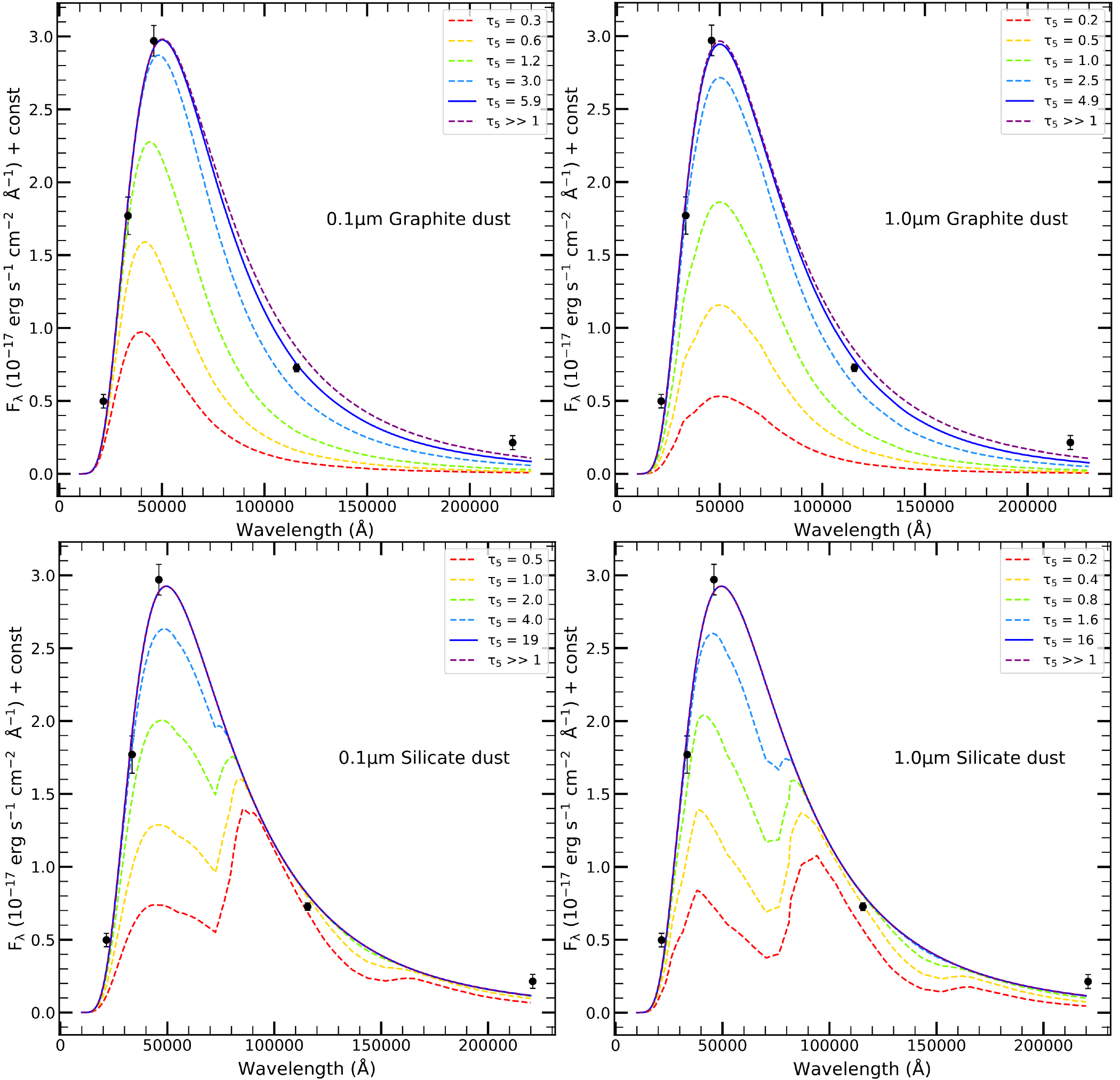}}%
  \caption{Fits to the late time SED of NGC 300 OT. In the different panels are reported the fits for the different dust compositions and grain sizes. The best fit to the data is shown as a solid blue line, while the dashed lines show the effect of changes in the optical depth while keeping temperature and radius of the source fixed.}
  \label{SED_dust_OPTdept}
\end{figure*}

We can also provide values of the total energy radiated by our ILRTs during the time they were monitored: NGC 300 OT emitted 2.5$\times$10$^{47}$ erg over the course of 760 days, with 1.5$\times$10$^{47}$ erg released in the first 120 days after maximum. Similarly, AT 2019ahd radiated 4.0$\times$10$^{47}$ erg in 440 days, 2.3$\times$10$^{47}$ erg in the first 100 days, while AT 2019udc emitted 4.0$\times$10$^{47}$ erg in about 100 days. Finally, AT 2019abn radiated 1.7$\times$10$^{48}$ erg in 480 days, the vast majority (1.6$\times$10$^{48}$ erg) emitted in the first 240 days. Aside from the total energy released by these transients, the decline rate of their bolometric luminosity at late time appears particularly interesting.
\section{Late time mid-infrared monitoring of ILRTs} \label{LATETIME}

\subsection{Dust evolution and composition} \label{NGC300OTdust}

\begin{figure}
  \includegraphics[width=1\columnwidth]{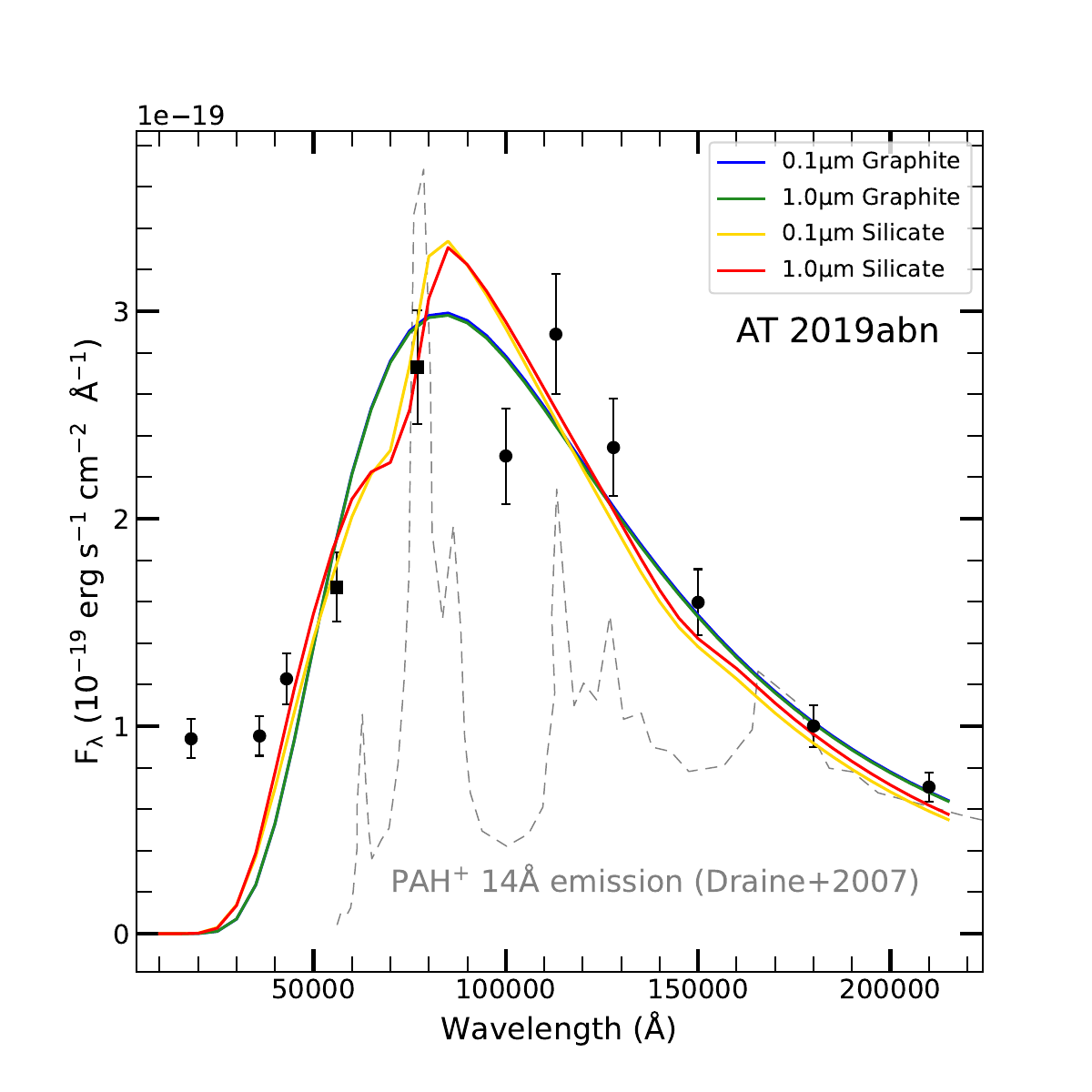}%
  \caption{SED of AT 2019abn 1956 days after discovery. Circles represent flux measurements obtained with JWST on 2024-05-31, while squares represent data extrapolated from the measurements on 2023-06-08. Solid lines trace the best fits for dust with different composition and grain size. The gray dashed line displays a scaled emission spectrum of PAH$^{+}$ with size 14 \AA \ \protect \citep{Draine2007_PAH}.}
  \label{JWST_sed_19abn}
\end{figure}

\cite{OhsawaNGC300} monitored the evolution of NGC 300 OT in the IR domain (2-5 $\mu m$) at 398 and 582 days, finding that the hot dust component had progressively cooled down to 810 K and ultimately to 680 K. The probed wavelength range did not provide information on the cool, pre-existing dust component, which by that time could only contribute $\sim$1\% of the observed flux. \cite{OhsawaNGC300} also set lower limits for the optical depth of the dust: $\tau_{\nu} >$ 12 at 398 days and $\tau_{\nu} >$ 6 at 582 days, effectively stating that the dust remains optically thick even at very late phases. In this condition, the lower limit to the mass of emitting dust was estimated to be $\sim$10$^{-5}$ M$_{\odot}$.

Thanks to the WISE data collected on 2010-06-17 in the $W1$, $W2$, $W3$ and $W4$ filters, we are able to expand the analysis on the late time evolution of the dust in NGC 300 OT, 761 days after maximum. To correct these values for absorption, we extend the reddening law to 22 $\mu m$ by employing the tabulated values and prescriptions found in \cite{Rieke1985_1-13micron} and \cite{Draine1989_upto30micron}.
The late time K band coverage with SOFI obtained 853 days after maximum allows us to linearly extrapolate the K-band flux at 761 days in order to have a more detailed SED of the transient.
We fit our flux measures through the same procedure detailed in Section \ref{SED_evol_sect}. Following in the footsteps of \cite{OhsawaNGC300}, instead of a simple Planck function we use $f(\nu) \propto (1-e^{-\tau_{\nu}}) B_{v}(T_{d})$ as a fitting function, where $\tau_{\nu}$ is the optical depth, $B_{\nu}$ is the Planck function and $T_{d}$ is the dust temperature. The multiplicative factor in front of the Planck function aims to account for the different opacity of dust at different wavelengths.
It is relevant to point out that this model is isothermal, while a cloud of optically thick dust would present a temperature gradient, so this is already an approximation.

To better differentiate the properties of the dust depending on the size of the dust grains and their chemical composition, we perform the fits using four different sets of opacity k$_{\nu}$, tabulated by \cite{Fox2010_TabulatedOpacity}. Two sets are opacity values for graphite dust, with grain size of 0.1 and 1 $\mu m$ respectively, while the other two sets of opacity assume the same grain size, but present a silicate composition. Since the values presented by \cite{Fox2010_TabulatedOpacity} only cover up to 13 $\mu m$ while our data reach 22 $\mu m$ (thanks to the W4 channel), we have to extrapolate the behaviour of k$_{\nu}$ in that region. For the graphite dust we follow the prescription from \cite{Draine2016_graphite}, using a $\lambda^{-2}$ scaling. The silicate opacity is instead characterised by an additional bump in opacity around 19 $\mu m$: to reproduce it, we adopt the opacity behaviour reported by \cite{Draine1984}. Beyond 9 $\mu m$, graphite dust tends to have a lower opacity compared to graphite dust (see Figure \ref{GrainOpacity} in Appendix). Silicate dust stands out because of the "bumps" in opacity in the MIR domain, in contrasts with the smooth behaviour of the graphite opacity.

\begin{table*}
\centering
\begin{adjustbox}{tabular=cccccc, center}
\\ \hline
   
  Model & L [10$^{38}$ erg s$^{-1}$] & T [K] & R [AU] & $\tau_{5}$ & M$_{d}$ [10$^{-4}$ M$_{\odot}$] \\
  
 \hline
 \hline
 \textbf{Graphite 0.1 $\mu m$} & 2.8 (0.1)   & 347 (9)   & 350  (15) & 96  (33) & 71 (24) \\
 \textbf{Graphite 1.0 $\mu m$} & 2.8 (0.1)   & 347 (10)  & 349  (17) & 198  (87) & 46 (20) \\
 \textbf{Silicate 0.1 $\mu m$} & 2.8 (0.1)   & 379 (15)  & 290  (18) & 1.2  (0.6) & 1.8 (0.8) \\
 \textbf{Silicate 1.0 $\mu m$} & 2.8 (0.1)   & 372 (10)  & 303  (13) & 1.6  (0.9) & 1.6 (0.9) \\

 \hline

\end{adjustbox}

\caption{Parameters obtained for the different SED fit models of AT 2019abn at 1956 days. Errors are reported in brackets.}
\label{GraSilTable_19abn}
\end{table*}


\begin{figure*}
  \includegraphics[width=1\textwidth]{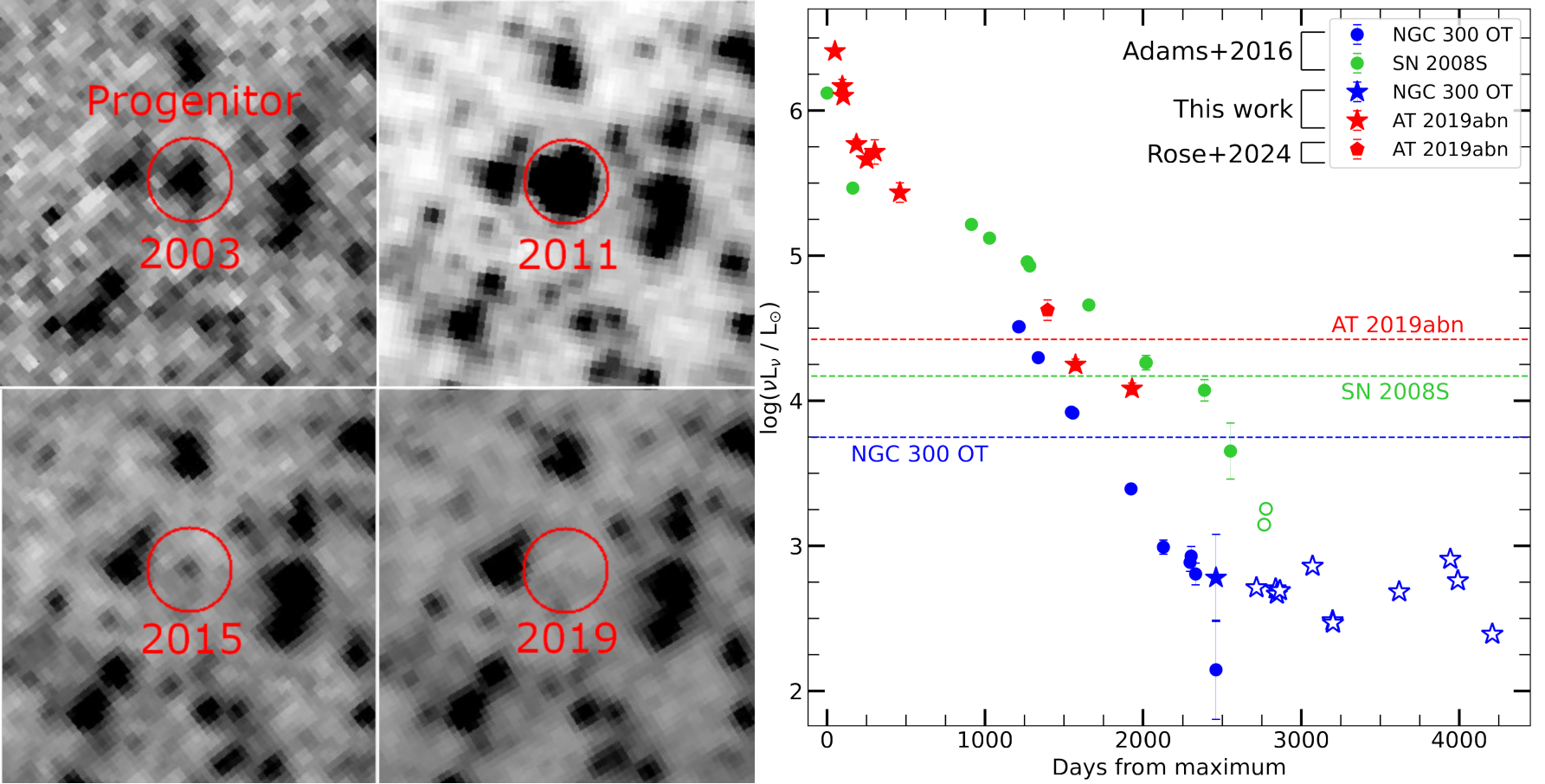}%
  \caption{Left panel: images of NGC 300 OT in the [4.5] $\mu m$ channel of Spitzer from the progenitor until the disappearance of the transient. Right panel: updated version of the figure shown in \protect \cite{Adams2016}, with their measurements in the [4.5] $\mu m$ channel shown with circles. Additional data points for NGC 300 OT and AT 2019abn are shown with stars (upper limits are represented as empty symbols). We also show the JWST data point of AT 2019abn obtained by \protect \cite{Rose_19abn_}. The progenitor luminosity of each transient is shown as a horizontal dashed line.}
  \label{FaintingNGC300OT}
\end{figure*}

In Figure \ref{SED_dust_OPTdept} are reported the results of the SED fitting procedure for the different dust compositions and grain sizes considered. The parameters obtained are reported in Table \ref{GraSilTable}. The parameters of the best fitting black body are similar regardless of dust particles size. There is, instead, a small variation on whether we adopt graphite dust or silicate dust. The dust temperature obtained for graphite is slightly lower than the temperature obtained for silicate dust, with a radius that is consequently larger.
All things considered, the differences between the black body parameters of silicate dust and graphite dust are negligible: the real difference lies in the optical depth inferred. We consider the optical depth at 5 $\mu m$ ($\tau_{5}$), to perform a comparison between the different models. As shown by the dotted lines in Figure \ref{SED_dust_OPTdept}, at low optical depth ($\tau_{5} < $1), the models deviate significantly from the Planck function: silicate dust shows a double peaked emission, while the emission of graphite dust with grains of 1.0 $\mu m$ is characterised by a "shoulder" at around 3.2 $\mu m$. Graphite dust with grains of 0.1 $\mu m$ do not show such clear features, but the SED model is systematically too narrow to properly fit the observed data in the optically thin case. Indeed, all the best fit yield optically thick dust at all wavelengths considered. However, in the case of silicate dust the best fit is obtained with an optical depth of $\tau_{5}$=19.4 for 0.1 $\mu m$ grains and $\tau_{5}$=16.5 for 1.0 $\mu m$ grains, basically returning to a Planck function. On the other hand, the optical depth inferred for graphite dust are $\tau_{5}$=5.9 for 0.1 $\mu m$ grains and $\tau_{5}$=4.9 for 1.0 $\mu m$ grains: optically thick, but still distinguishable from a pure Planck function. We notice that none of the models can accurately fit the point at 22 $\mu m$: we speculate that this infrared excess is produced by the cold dust component that further cooled down since its detection by \cite{Prieto2009pPNe}. 
Finally, through our measures of the optical depth, it is possible to obtain an estimate of the dust mass using the equation (as done by \citealt{Griffin2022dust}):


\begin{equation}
\hspace{2.5cm} M_{dust} = \frac{4 \pi R_{dust}^{2} \tau_{\nu}}{3 k_{\nu}}.
\end{equation}


Here we are assuming a constant density profile in the dust cloud, which is of course an additional approximation. Depending on the composition and grain size considered, the inferred dust mass spans from 4.5$\times$10$^{-5}$ M$_{\odot}$ up to 1.9$\times$10$^{-3}$ M$_{\odot}$ (see Table \ref{GraSilTable}). It is interesting to note that such values, obtained at $\sim$800 days after explosion, are perfectly in line with the amount of dust found in the ejecta of core-collapse SNe \footnote{https://nebulousresearch.org/dustmasses/} (e.g. \citealt{2004etDustMass,2010jlDustMass}).
We remark that this fitting procedure was performed on the dust which was identified by \cite{Prieto2009pPNe} 93 days after discovery at a temperature of 1510 K (and named "warm" dust), and whose cooling is tracked in the upper panel of Figure \ref{ILRTcold}. The initial temperature of this dust component is close to the dust sublimation temperature: this behaviour is compatible with newly formed dust which condensed few months after the peak luminosity of NGC 300 OT, and by the time of our fit has cooled down to $\sim$580 K. Similarly, also the third blackbody component with a temperature of 485 K detected by \cite{Prieto2009pPNe} at 93 days (identified as circumstellar dust that survived the explosion) must have cooled down, and the peak of its emission has shifted outside the monitored domain. As mentioned above, this component may be the origin of the exceeding flux observed at 22 $\mu m$.

We replicate this SED analysis on AT 2019abn, thanks to publicly available images of M51 taken by the JWST on 2022-12-13 (proposal 1240, PI M.E. Ressler), 2023-06-08, 2023-06-14 (proposal 1783 PI A. Adamo) and 2024-05-31 (3435 PI K.M. Sandstrom). The object is still visible even five years after its peak luminosity: we report the magnitudes obtained with PSF fitting through \textsc{ecsnoopy} in Table \ref{JWSTmeasures}.  
The wavelength coverage of the first three epochs is too scarce to provide a well sampled SED, therefore we focus on the measurements obtained on 2024-05-31, 1956 days after discovery. To fill the gap between the F430M and F1000W filters, we infer the behaviour of the F560W and F770W filters observed the previous year (2023-06-08). To do so, we assume a constant colour evolution using the filters F444W (observed on 2023-06-08) and F430M (observed on 2024-05-31) as anchors. The flux in these overlapping filters decreases by a factor of 1.5 between the two epochs, and we assume the same holds true for the F560W and F770W filters. Since cooling objects tend to become redder, this is likely an underestimation of the fluxes in F560W and F770W on 2024-05-31, yet this simple approximation allows us to constrain a crucial spectral region. The resulting SED is displayed in Figure \ref{JWST_sed_19abn}.
Firstly, we point out that the Pa$\alpha$ line is located within the F182M filter, and this could account for the relative brightness of the transient in this band, which is therefore excluded from the following discussion.


In Figure \ref{JWST_sed_19abn} we show as solid lines the various fits to the observed SED of AT 2019abn with dust models characterised by different chemical composition and grain size. Parameters of each fit are shown in Table \ref{GraSilTable_19abn}. The luminosity of the transient is consistently found to be (2.8$\pm$0.1$)\times$10$^{38}$ erg s$^{-1}$ (in Figure \ref{JWST_bolometric} this result is added to the bolometric light curve of AT 2019abn, updating Figure \ref{bolometric}). There is instead a discrepancy in the resulting temperature and radius depending on the dust composition: in the case of graphite, the best fits yield a temperature of $\sim$350 K and a radius of $\sim$350 AU, while for the silicate dust the temperature found is $\sim$375 K, with a blackbody radius of $\sim$300 AU. Grain size only plays a marginal role in shaping these results. Graphite dust models yield high optical depth ($\tau_{5}$$>>$1), leading to infer lower limits to the dust masses in the order of several 10$^{-3}$ M$_{\odot}$. On the other hand, the best fitting silicate dust models are closer to the optically thin regime ($\tau_{5}$$\sim$1), with inferred dust masses of the order of $\sim$2$\times$10$^{-4}$ M$_{\odot}$.
These results, however, need to be interpreted with caution, since even the best fits struggle to accurately reproduce the observed data. Particularly puzzling is the flux ``depression" in the F1000W band compared to the flux values found in the F770W and F1130W bands.
A possible explanation for this departure from a simple blackbody emission may be the presence of polycyclic aromatic hydrocarbon (PAH). In fact, PAH are characterised by emission features around 7 $\mu m$ and 11 $\mu m$. In Figure \ref{JWST_sed_19abn} we display a scaled emission spectrum of ionized PAH with size of 14 \AA \ \citep{Draine2007_PAH} to show the match between the PAH emission features and the peculiarities of the observed SED. The presence of PAH is also corroborated by MIR spectroscopy of AT 2019abn obtained on 2022 \cite{Rose_19abn_}. Accurate PAH modelling is beyond the scope of this study and is deferred to later works. Beyond these considerations on the shape of the SED, JWST observations provide valuable information on the MIR luminosity evolution of AT 2019abn five year after its discovery, as discussed in the following section. 



\subsection{Photometric decline below progenitor luminosity}

\cite{Adams2016}, monitored the two ILRTs NGC 300 OT and SN 2008S with the Spitzer telescope for several years, until they both faded below the luminosity of their respective progenitor.
We update this follow-up campaign by analysing the Spitzer images collected between 2015-10-20 and 2019-11-24 with the Spitzer [4.5] $\mu m$ channel. A visual inspection of the images shows that NGC 300 OT has faded below the detection threshold (Figure \ref{FaintingNGC300OT}, left panel). We measure the flux on all public Spitzer [4.5] $\mu m$ images through PSF fitting employing \textsc{ecsnoopy}. Our results are in good agreement with the values measured with aperture photometry by \cite{Adams2016}, with the exception of the very last detection, on 2015-02-09: we find a value of 600$\pm$300 L$_{\odot}$ against the 140$\pm$110 L$_{\odot}$ obtained with aperture photometry (the difference can be appreciated on the right panel of Figure \ref{FaintingNGC300OT}). The complex background surrounding the target, as seen in the left panel of Figure  \ref{FaintingNGC300OT}, contributes to the uncertainty of this measure. All the images collected from that moment only provide upper limits to the luminosity of NGC 300 OT: in particular, in the right panel of Figure  \ref{FaintingNGC300OT} we show the 1$\sigma$ limits obtained from PSF fitting. We therefore confirm that since 2015 NGC 300 OT has faded below the detection threshold, well below the progenitor luminosity.

We are also able to include AT 2019abn among the ILRTs with MIR coverage spanning several years. The light curve up to $\sim500$ days is provided by the Spitzer and WISE monitoring already shown in Figure \ref{abn_phot}, while the last two epochs are provided by the late JWST observations obtained 1598 and 1956 days after discovery and presented in the previous section. Despite the slight mismatch in the covered wavelength, the JWST filters F444W and F430M are sufficiently close to the [4.5] $\mu m$ and W2 filters to provide a coherent light curve, which is shown with red stars symbols in Figure \ref{FaintingNGC300OT}. The behaviour of AT 2019abn is similar to that of SN 2008S, where the initial decline rate significantly slows down after $\sim200$ days. It is interesting to consider the evolution of the transient compared to its progenitor luminosity in the [4.5] $\mu m$ channel provided by \citealt{Jencson2019abn}, which is marked in Figure \ref{FaintingNGC300OT} as a red horizontal dashed line. It is possible to notice that on 2023-06-08 AT 2019abn has already faded below its progenitor luminosity, with the latest measure obtained on 2024-05-31 being over two times dimmer than the luminosity of the star before the event. All three ILRTs that were monitored for multiple years in the MIR domain eventually became dimmer than their progenitor star, and so far none of them has shown sign of stopping their steady luminosity decline: this fact supports the interpretation that ILRTs are terminal events.


\begin{figure*}
	\includegraphics[width=\textwidth]{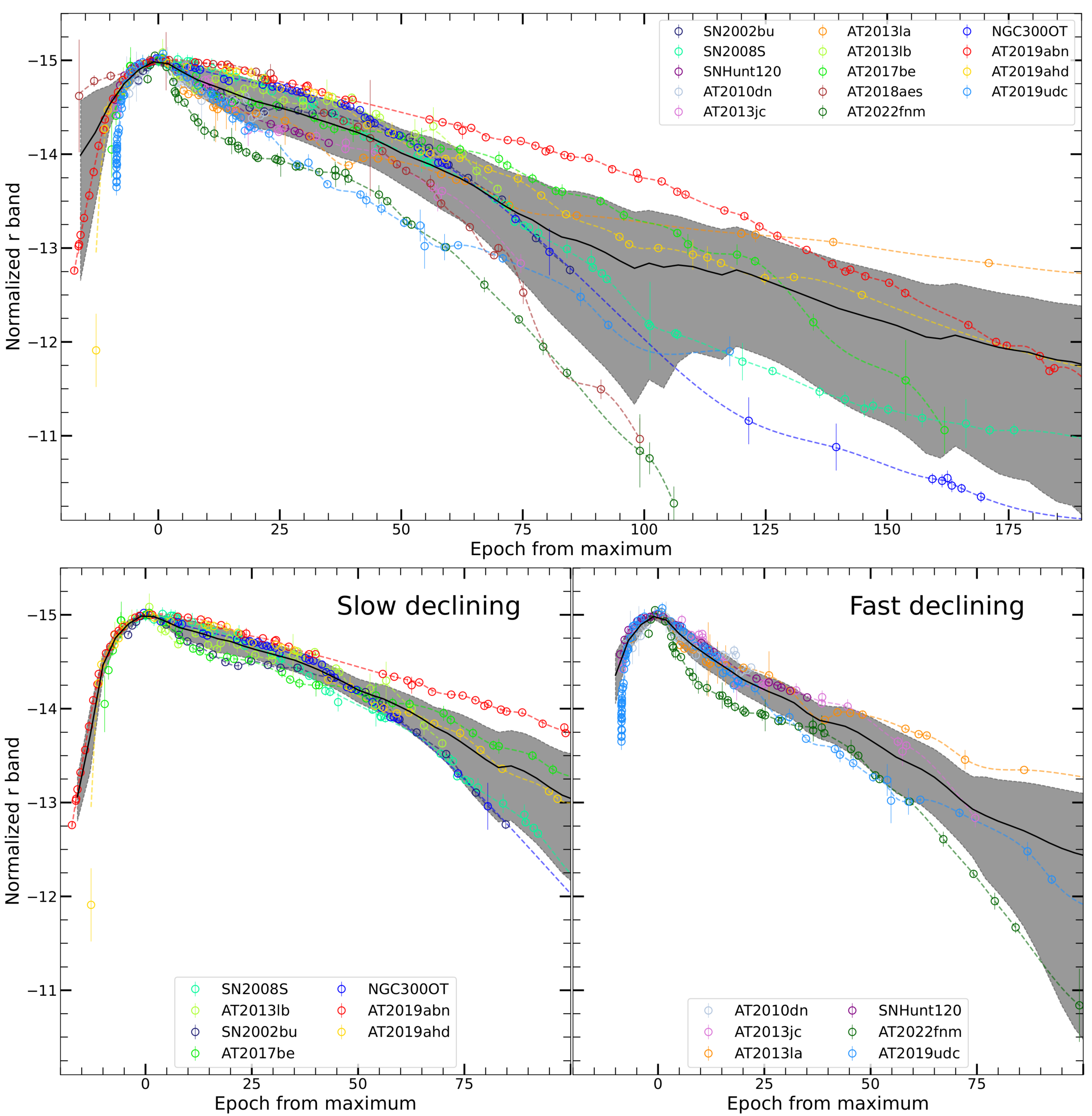}
    \caption{The solid black line represents the ILRTs template light curve, while the gray shaded area shows the 1 $\sigma$ deviation from the mean. The coloured dashed lines display the interpolation performed on each object. In the bottom panels the ILRTs are separated between slow and fast declining objects.}
    \label{Template}
\end{figure*}

\section{Template Light curve} \label{TemplateLC_sect}
In order to better characterize this class of objects, we produce an $r$ band template light curve for ILRTs. 
We are motivated also by the potential usefulness of such templates in our future attempts to classify new transients discovered by the Vera Rubin Observatory (formerly Large Synoptic Survey Telescope, \citealt{2019LSST}) and released by dedicated science brokers such as Lasair \citep{Lasair} and Alerce \citep{ALERCE}. With the foreseen amount of data, it will be impossible to study in detail every single target, and template light curves for different classes of transient could prove to be useful tools for filtering the data stream, in addition to the classification algorithms and tools that are currently being developed (e.g. \citealt{LSST_ml_1,LSST_ml_2}). 
Apart from NGC 300 OT, AT 2019abn, AT 2019ahd and AT 2019udc, which were presented above, we broaden our sample including SN 2002bu \citep{Smith2002bu}, SN 2008S \citep{Botticella2008S}, AT 2017be \citep{Yongzhi2017beC}, SNHunt120 \citep{MaxILRT}, AT 2010dn, AT 2013jc, AT 2013la, AT 2013lb, AT 2018aes \citep{Yongzhi2021ILRT} and AT 2022fnm (Moran et al. 2024, submitted). We chose to focus on the $r$-band, since it is typically well-sampled and allows for a reliable comparison between objects.
The peak absolute magnitude of different ILRTs can span at least three magnitudes (from -12 mag to -15 mag, see Figure \ref{Comparison}): however, here we are interested in recovering just the expected light curve shape of ILRTs. For this reason we normalised the peak magnitude of all objects considered to the same arbitrary value. Both the light curve shape and the absolute magnitude at peak can then be used as filters to identify future ILRTs candidates.

As a first step to produce a light curve template we make use of Gaussian Processes to interpolate the light curves of each object (see e.g. \citealt{Inserra2018GP} for a more detailed description of the procedure).
In particular, we employ the \textsc{Matern} 3/2 kernel found in the \textsc{python} package \textsc{george}  \citep{Ambikasaran2015}, and as for the mean function we adopt equation 1 given by \cite{Bazin2011}.
It is worth noticing that we performed the interpolation procedure in the ''luminosity - time" space.
Given the different sampling of each event, this first step is performed to ensure that each objects holds the same weight in the construction of the template, regardless of how many data points were collected at a given phase.
After interpolating the light curve of each object, we recover the mean value and the standard deviation of the ILRTs luminosity within temporal bins of 3 days, which we empirically choose as a good compromise that yields a rather smooth template which still has a reasonable time resolution. The resulting ''mean light curve" of an ILRT is shown as a black solid line in the upper panel of Figure \ref{Template}, with the grey shaded area representing the 1 $\sigma$ deviation from the mean.



Due to the low number of available ILRTs, the rising part of the template becomes reliable only at about -10 days with respect to maximum luminosity: before that phase the template is dominated by the widely different behaviours of AT 2019abn and AT 2018aes.
The overall template light curve rises of about 0.5 mag from -10 days to 0 days, and fades by 1 mag from maximum to 50 days. As time progresses, the 1 $\sigma$ region gets wider, with objects evolving at different rates. Objects fading below the detection threshold can result in a jagged profile of the template, as happens at about 100 days due to the disappearing of AT 2018aes and AT 2022fnm.
There is a variety of properties in the light curves of the considered ILRTs, but we can roughly divide them in two subclasses, based on their behaviour in the 50 days after maximum. The first group of objects present slowly declining, convex light curves, which on average only fade by 0.8 mag from day 0 to day 50. The second group of objects display fast declining, concave light curves, which on average fade by more than 1.3 mag in the same time span (see Figure \ref{Template}, lower panels). While this distinction is clear in some cases (AT 2019abn and AT 2019ahd are slow decliners, while AT 2019udc and AT 2022fnm are fast decliners), there are some objects with intermediate properties, such as AT 2013la, whose light curve quickly fades in the first two months, only to flatten after 75 d, ultimately becoming even brighter than AT 2019abn after 130 d.
Another outlier is AT 2018aes, which is characterised by the shallowest rise to maximum light, followed by a rather "standard" slow decline for 40 days before abruptly fading by 3.2 mag in 60 days.
The dichotomy between slow and fast declining ILRTs is reminiscent of the division between SNe IIP and IIL (e.g. \citealt{Anderson2014IIPL}), with the progenitors of the slow declining ILRTs retaining larger hydrogen envelopes compared to the progenitors of fast declining ILRTs. Of course, a larger sample of ILRTs is needed to verify if this analogy holds true or if it is a mere coincidence.

\begin{figure*}
\centering
  \includegraphics[width=1.8\columnwidth]{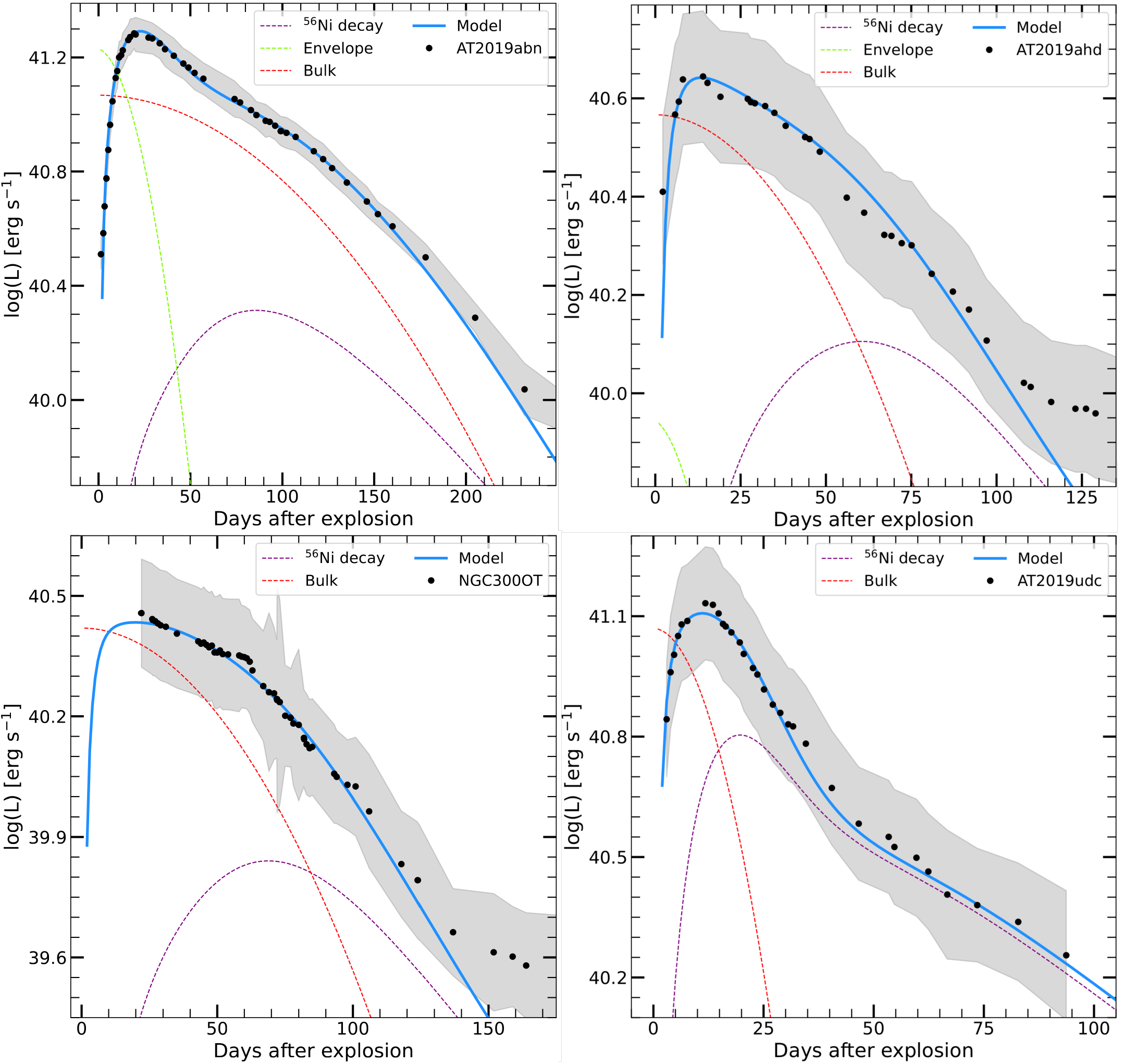}
  \caption{Bolometric light curves of our sample of ILRTs (represented with black circles) with a toy model reproducing their shape. The errors on the data points, dominated by the uncertainty on distance and reddening estimations, are reported as a gray shaded area. The toy models attempting to reproduce the behaviour of each transient is shown as a solid light blue line. The energy sources powering the different phases of the light curves are shown as dashed lines.}
  \label{Models}
\end{figure*}

\section{A toy model for ILRTs light curves} \label{TOYMODEL}

\begin{table*}
\begin{adjustbox}{tabular=ccccccccccc,center}

\\ \hline
  Transient & \textsuperscript{56}Ni & M$_{bulk}$ & V$_{bulk}$ & E$_{bulk}$ & R$_{bulk}$ & M$_{env}$ & V$_{env}$ & E$_{env}$ & R$_{env}$ & t$_{CSM}$ \\ 
   & [M$_{\odot}$] & [M$_{\odot}$] & [km s$^{-1}$] & [erg] &  [cm] & [M$_{\odot}$] & [km s$^{-1}$] & [erg] & [cm] & [days] \\ 
  
 \hline
 \hline

  \textbf{AT 2019abn} & 6.7x10$^{-3}$ & 7.0 & 2000 & 6.5x10$^{49}$ & 2x10$^{13}$ & 1 & 5500 & 2.7x10$^{48}$ & 1x10$^{14}$ & 12 \\
   \textbf{AT 2019ahd} & 1.8x10$^{-3}$ & 3.5 & 4500 & 4.1x10$^{49}$ & 5x10$^{12}$ & 0.5 & 5000 & 7.0x10$^{46}$ & 1x10$^{14}$ & 3 \\
   \textbf{NGC 300 OT} & 1.2x10$^{-3}$ & 4.2 & 3500 & 3.9x10$^{49}$ & 4.5x10$^{12}$ & - & - & - & - & 3 \\
   \textbf{AT 2019udc} & 3.5x10$^{-3}$ & 0.9 & 12000 & 3.7x10$^{49}$ & 1.0x10$^{13}$ & - & - & - & - & 2 \\

 \hline
\end{adjustbox}

\caption{Parameters used in the models displayed in Figure \ref{Models}.}
\label{ModelsParam}
\end{table*}

In this section we present a simple model in the attempt to reproduce the shape of the bolometric light curves of ILRTs in the context of a weak SN explosion. The end goal is to obtain a rough estimate of parameters such as the amount of mass ejected and its velocity. The basic concepts are taken from \cite{Chatzopoulos2012_LCmodel}, which in turn expand the approach introduced by \cite{Arnett1980SneIIP,Arnett1982}. First of all, \textsuperscript{56}Ni radioactive decay is expected to be a relevant power source for SNe, and its luminosity over time is given by the equation:

\begin{multline}
   \hspace{1.5cm} L_{Ni}(t) = \frac{2M_{Ni}}{t_{d}}e^{-[\frac{t^{2}}{t^{2}_{d}} + \frac{2R_{0}t}{vt^{2}_{d}}]}[(\epsilon_{NI}-\epsilon_{Co})\times \\
    \hspace{1.5cm} \times \int^{t}_{0}[\frac{R_{0}}{vt_{d}} +\frac{t'}{t_{d}}]e^{[\frac{t'^{2}}{t^{2}_{d}} + \frac{2R_{0}t'}{vt^{2}_{d}}]}e^{-t'/t_{Ni}}dt' + \\ 
    + \epsilon_{Co}\int^{t}_{0}[\frac{R_{0}}{vt_{d}} +\frac{t'}{t_{d}}]e^{[\frac{t'^{2}}{t^{2}_{d}} + \frac{2R_{0}t'}{vt^{2}_{d}}]}e^{-t'/t_{Co}}dt'](1-e^{-At^{-2}}) ,
\end{multline}

\noindent where M$_{Ni}$ is the mass of synthesised \textsuperscript{56}Ni, t$_{Ni}$ and t$_{Co}$ are the half lives of \textsuperscript{56}Ni and \textsuperscript{56}Co, while $\epsilon_{Ni}$ and $\epsilon_{Co}$ are the energy generation rates of \textsuperscript{56}Ni and \textsuperscript{56}Co released during the radioactive decay. $A$ is a constant that accounts for the opacity of the ejecta to $\gamma$-rays, with large values of $A$ corresponding to complete trapping. The constant values are the same reported in \cite{Chatzopoulos2012_LCmodel}. This equation takes into account the diffusion time through an homologously expanding gas of mass $M$, initial radius $R_{0}$ and characteristic expansion velocity $v$. As discussed in Paper II, the velocity inferred from the width of the emission lines of our sample of ILRTs can reach only up to $\sim$10$^{3}$ km s$^{-1}$. However, the observed emission lines of ILRTs originate in a dense CSM which is obscuring the underlying material: as happens for SNe IIn, fast ejecta may be hiding below this layer of dense gas. Theoretical expectations for ECSNe suggest that their ejecta should be less energetic than a standard SN ($v$ $\lesssim$ 10$^{4}$ km s$^{-1}$).
The characteristic timescale $t_{d}$ of the light curve can be written as a combination of the diffusion timescale $t_{0}$ and the hydrodynamical timescale $t_{h}$:

\begin{equation}
  \hspace{1.5cm}  t_{d} = \sqrt{t_{0} t_{h}} \hspace{1 cm}  t_{0}=\frac{kM}{\beta c R_{0}} \hspace{1 cm} t_{h} = \frac{R_{0}}{v} ,
\end{equation}

\noindent where $\beta$ is a constant linked to the density profile of the mass, and $k$ is the opacity of the ejecta, for which we adopt a value of 0.33 cm$^{2}$ g$^{-1}$. \\
As an additional source of luminosity, we consider the radiation emitted by the expanding and cooling gas, which was heated by the blast wave that followed the explosion. This term was already introduced by \cite{Arnett1980SneIIP} to reproduce the slowly declining light curves of SNe IIP. Using the formalism of \cite{Chatzopoulos2012_LCmodel}, such luminosity term is written as:

\begin{equation}
\hspace{2.5cm} L_{blast}(t) = \frac{E_{th}}{t_{0}} e^{-[\frac{t^{2}}{t^{2}_{d}} + \frac{2R_{0}t}{vt^{2}_{d}}]},
\end{equation}

\noindent where E$_{th}$ is the internal energy which was deposited in the ejecta during the explosion. To better reproduce the observed shape of the light curve, we consider two blast terms: one associated with the envelope, less massive and more extended, characterized by a short diffusion time, and another associated with the bulk of the ejecta, more massive and dense, with a much longer diffusion timescale. Dividing the ejecta in two regions is not a novelty: such approach has been successfully used, for example, by \cite{Nagy2016TwoComponents}.
The total time dependent luminosity is therefore the sum of these three contributions:

\begin{equation}
 \hspace{2.0cm}   L_{tot} = L_{Ni} + L_{bulk} + L_{envelope} .
\end{equation}

The final piece of the model is the thick CSM surrounding the transient, which is a key feature observed in ILRTs spectra. 
This element does not provide additional energy, but it simply reprocesses the total luminosity emitted by the ejecta and \textsuperscript{56}Ni, delaying its appearance. To reproduce this effect, we use the fixed photosphere approximation presented by \cite{Chatzopoulos2012_LCmodel}:

\begin{equation}
  \hspace{1.0cm}  L_{final}(t) = \frac{1}{t_{CSM}} e^{-\frac{t}{t_{CSM}}}\int^{t}_{0} e^{\frac{t'}{t_{0}}} L_{tot}(t')dt' ,
\end{equation}

where $t_{CSM}$ is the diffusion time through the thick CSM shell surrounding the transient. Its effect is crucial in the first phases, during which the blast terms L$_{bulk}$ and L$_{envelope}$ start abruptly from the maximum luminosity, without a rising phase. A CSM shell with a diffusion time of few days allow the model to reproduce the rise observed in ILRTs.
In Figure \ref{Models} are shown the models and the bolometric light curves they are trying to emulate. The contribution of each component is shown with different colours: the parameters relative to each of them are reported in Table \ref{ModelsParam}.
One weakness of the approach outlined above is the large number of free parameters, as well as the degeneracy between some of them: similar solutions can be obtained while inputting different parameters. Instead of blindly applying a fitting procedure, we tried to input reasonable parameters in the context of SNe, gradually modifying them to improve the agreement with the data.
An additional fragility of this approach is the assumption of spherical symmetry, which may be a crude approximation in the case of these transients (e.g. \citealt{SokerJetsILRTS}).

AT 2019abn is definitely the most successful example within our sample. The emission from a rather massive envelope, combined with a CSM diffusion time of 10 days, seems to properly reproduce the rise and the maximum phases. The subsequent slow decline can be explained by slow ejecta which trap the photons for almost 200 days. The \textsuperscript{56}Ni contribution is not significant until the late phases, and it is definitely too faint to explain the evolution after 200 days: as mentioned in the previous section, an additional source of energy is likely needed to explain the late time luminosity of ILRTs. \\
The parameters used to model AT 2019ahd do not differ significantly from the ones used for AT 2019abn: they are mainly scaled down in mass and internal energy of the ejecta. The blast term associated with the envelope is not as crucial, but the overall evolution appears to be reproduced sufficiently well, excluding of course the late shallow decline.
Since we miss the first evolutionary phases of NGC 300 OT, we only use a single blast (associated to the bulk of the ejecta) term plus \textsuperscript{56}Ni decay to reproduce the data: the blast term associated with the envelope is relevant during the early phases, so it is difficult to constrain in this case. Apart from this, the parameters used for NGC 300 OT are very similar to the ones used for AT 2019ahd.

The same conclusions cannot be reached for AT 2019udc, which is definitely an outlier in this sample. First of all, the fast decline hints at the presence of a single blast term. Most strikingly, the evolution of AT 2019udc is much faster compared to the other transients considered so far. For this reason the diffusion time within the ejecta must be remarkably lower: this leads to a low ejected mass and a high scale velocity, several times larger compared to the other ILRTs mentioned so far. Finally, it is worth noticing that in the case of AT 2019udc, \textsuperscript{56}Ni decay is a key element to reproduce the observed late time data points. 
All the masses mentioned so far are just relative to the ejected material: a compact remnant is expected to be left behind, in the scenario in which ILRTs arise from a core collapse event. Since the mass of the remnant can be estimated to be 1.3--2.0 M$_{\odot}$ \citep{PumoModel2017}, the inferred mass of our sample of ILRTs right before the explosion adds up to 5.3--6.0 M$_{\odot}$ for AT 2019ahd,  5.5--6.2 M$_{\odot}$ for NGC 300 OT and 9.3--10 M$_{\odot}$ for AT 2019abn. Such values are compatible with the expectations for SAGB stars and consequently ECSN events, especially accounting for the fact that some mass was likely lost during the evolution of the star, therefore leading to a slightly higher ZAMS. AT 2019udc, with its estimated progenitor mass of just 2.2--2.9 M$_{\odot}$ at the time of explosion, sets itself apart from the rest of the sample, although its origin could be tentatively explained as an ECSN arising from a SAGB star that underwent extreme mass loss.
We remark that the toy model presented in this section is a simplified approach to the light curve modelling of ILRTs: more detailed models envisioning hydrodynamical modelling should be used to retrieve more robust estimates of the explosion parameters of these objects. Accounting for the possible luminosity contribution from ejecta-CSM interaction will also be an important step forward: while ILRTs never display spectral signatures typically associated with shocks, there may be hidden CSM interaction powering their light curve (Paper II).
The puzzling results obtained for AT 2019udc may be significantly revised by using a more refined approach.

\section{Summary and Conclusions} \label{Conclusion_sect}
In this work we present data sets for NGC 300 OT, AT 2019abn, AT 2019ahd and AT 2019udc. AT 2019abn and AT 2019udc stand out for their bright peak magnitude (M$_{r}$=--15 mag and --14.5 mag respectively). On the other hand, NGC 300 OT and AT 2019ahd are characterised by a more modest peak magnitude (M$_{r}$= --12.7 mag and --13.2 mag). We detect a NIR excess in the SED of three of our ILRTs, tentatively associated with dust formation, in three out of four targets. The only exception is AT 2019udc, characterised by the fastest evolution timescale among the ILRTs observed. At late phases, the contributions in the NIR and MIR domains to the bolometric light curve cause a decline shallower than the luminosity decline supported by the \textsuperscript{56}Ni decay, hinting at the presence of an additional powering mechanism.

The late time monitoring of NGC 300 OT with WISE allows us to build the SED of the transient 761 days after maximum. From its study, we infer a dust mass of the order of $\times$10$^{-5}$-$\times$10$^{-3}$ M$_{\odot}$, depending on the composition and size of the grains adopted. Furthermore, the evolution of NGC 300 OT has been monitored with Spitzer for several years, showing a steady dimming of the flux at the transient position over 4 years. This strengthens the argument suggested by \cite{Adams2016} that NGC 300 OT was a terminal event. Thanks to JWST observations we were able to show that also AT 2019abn faded below its progenitor luminosity in the MIR domain. Its SED appears to depart from a simple blackbody, possibly due to the presence of PAH emission features.
Finally, a simple model is presented in the context of a SN explosion, with the goal to estimate the parameters characterising the transients. For AT 2019abn, 2019ahd and NGC 300 OT, the low masses ejected (between 4 and 8 M$_{\odot}$) and low velocity of the material (up to 4500 km s$^{-1}$) are compatible with a weak explosion from a low mass star. AT 2019udc is an outlier in this sense, since its very fast decline rate leads to inferring high velocity (12000 km s$^{-1}$) and remarkably low ejected mass (0.9 M$_{\odot}$). 
Overall, due to their low energy and their apparent terminal nature, ILRTs remain solid candidates for being ECSNe.

\section*{Acknowledgements}

We thank the staff of the various observatories where data were obtained for their assistance.
Based on observations made with the Nordic Optical Telescope, owned in collaboration by the University of Turku and Aarhus University, and operated jointly by Aarhus University, the University of Turku and the University of Oslo, representing Denmark, Finland and Norway, the University of Iceland and Stockholm University at the Observatorio del Roque de los Muchachos, La Palma, Spain, of the Instituto de Astrofisica de Canarias.
Observations from the Nordic Optical Telescope were obtained through the NUTS2 collaboration which are supported in part by the Instrument Centre for Danish Astrophysics (IDA). The data presented here were obtained in part with ALFOSC, which is provided by the Instituto de Astrofisica de Andalucia (IAA).
This work makes use of data from the Las Cumbres Observatory network. The LCO team is supported by NSF grants AST--1911225 and AST--1911151, and NASA SWIFT grant 80NSSC19K1639. 
Data were also obtained at the Liverpool Telescope, which is operated on the island of La Palma by Liverpool John Moores University in the Spanish Observatorio del Roque de los Muchachos with financial support from the UK Science and Technology Facilities Council. Part of the observations were collected at Copernico and Schmidt telescopes (Asiago, Italy) of the INAF – Osservatorio Astronomico di Padova.
Based on observations made with the Gran Telescopio Canarias (GTC), installed in the Spanish Observatorio del Roque de los Muchachos of the Instituto de Astrofísica de Canarias, in the island of La Palma
This work has made use of data from the Asteroid Terrestrial-impact Last Alert System (ATLAS) project. 
Time domain research by the University of Arizona team and D.J.S. \ is supported by NSF grants AST-1821987, 1813466, 1908972, \& 2108032, and by the Heising-Simons Foundation under grant \#20201864. 
G.V., N.E.R. and I.S. acknowledge financial support from PRIN-INAF 2022 ``Shedding light on the nature of gap transients: from the observations to the models".
N.E.R. also acknowledge support from the Spanish MICINN grant PID2019-108709GB-I00 and FEDER funds, and from the program Unidad de Excelencia Maria de Maeztu CEX2020-001058-M.
A.R. acknowledges financial support from the GRAWITA Large Program Grant (PI P. D’Avanzo) and the PRIN-INAF 2022 \textit{"Shedding light on the nature of gap transients: from the observations to the models}.
Y.-Z. Cai is supported by the National Natural Science Foundation of China (NSFC, Grant No. 12303054), the Yunnan Fundamental Research Projects (Grant No. 202401AU070063) and the International Centre of Supernovae, Yunnan Key Laboratory (No. 202302AN360001).
T.E.M.B. acknowledges financial support from the  
Spanish Ministerio de Ciencia e Innovacion (MCIN), the Agencia  
Estatal de Investigacion (AEI) 10.13039/501100011033, and the  
European Union Next Generation EU/PRTR funds under the 2021 Juan de la  
Cierva program FJC2021-047124-I and the PID2020-115253GA-I00 HOSTFLOWS  
project, from Centro Superior de Investigaciones Cientificas (CSIC)  
under the PIE project 20215AT016, and the program Unidad de Excelencia  
Maria de Maeztu CEX2020-001058-M.
MF is supported by a Royal Society - Science Foundation Ireland University Research Fellowship.
The Aarhus supernova group is funded by the Independent Research Fund Denmark (IRFD, grant numbers 8021-00170B, 10.46540/2032-00022B).
L.G. acknowledges financial support from the Spanish Ministerio de 
Ciencia e Innovaci\'on (MCIN), the Agencia Estatal de Investigaci\'on 
(AEI) 10.13039/501100011033, and the European Social Fund (ESF).
This work was funded by ANID, Millennium Science Initiative, ICN12\_009
M.N. is supported by the European Research Council (ERC) under the European Union’s Horizon 2020 research and innovation programme (grant agreement No.~948381). Part of the funding for GROND (both hardware as well as personnel) was generously granted from the Leibniz-Prize to Prof. G. Hasinger (DFG grant HA 1850/28-1).
H.K. was funded by the Academy of Finland projects 324504 and 328898. 
R.K. acknowledges support from the Research Council of Finland (340613).

\bibliographystyle{aa}
\bibliography{example}

\section*{AFFILIATIONS}
\textit{$^{1}$INAF -- Osservatorio Astronomico di Padova, Vicolo dell’Osservatorio 5, I-35122 Padova, Italy\\
$^{2}$INAF -- Osservatorio Astronomico di Brera, Via E. Bianchi 46, 23807, Merate (LC), Italy \\
$^{3}$Yunnan Observatories, Chinese Academy of Sciences, Kunming 650216, P.R. China \\
$^{4}$Key Laboratory for the Structure and Evolution of Celestial Objects, Chinese Academy of Sciences, Kunming 650216, P.R. China \\
$^{5}$International Centre of Supernovae, Yunnan Key Laboratory, Kunming 650216, P.R. China \\
$^{6}$ Graduate Institute of Astronomy, National Central University, 300 Jhongda Road, 32001 Jhongli, Taiwan \\
$^{7}$SRON, Netherlands Institute for Space Research, Niels Bohrweg 4, 2333 CA, Leiden, The Netherlands \\
$^{8}$Department of Astrophysics/IMAPP, Radboud University Nĳmegen, P.O. Box 9010, 6500 GL, Nĳmegen, The Netherlands \\
$^{9}$Institute of Space Sciences (ICE, CSIC), Campus UAB, Carrer de Can Magrans s/n, E-08193 Barcelona, Spain \\
$^{10}$School of Physics, O’Brien Centre for Science North, University College Dublin, Belfield, Dublin 4, Ireland \\
$^{11}$The Oskar Klein Centre, Department of Astronomy, Stockholm University, AlbaNova, 10691 Stockholm, Sweden \\
$^{12}$Hiroshima Astrophysical Science Center, Hiroshima University, Higashi-Hiroshima, Japan \\
$^{13}$Department of Physics, Florida State University, 77 Chieftan Way, Tallahassee, FL 32306, USA \\
$^{14}$Las Cumbres Observatory, 6740 Cortona Dr. Suite 102, Goleta, CA, 93117, USA\\
$^{15}$Department of Physics, University of California, Santa Barbara, CA, 93106, USA\\
$^{16}$School of Physics \& Astronomy, Cardiff University, Queens Buildings, The Parade, Cardiff, CF24 3AA, UK \\
$^{17}$INAF, Osservatorio Astronomico di Capodimonte, Salita Moiariello 16, I-80131 Napoli, Italy \\
$^{18}$DARK, Niels Bohr Institute, University of Copenhagen, Jagtvej 128, 2200 Copenhagen, Denmark \\
$^{19}$Caltech, Mail Code 220-6, Pasadena, CA 91125 \\
$^{20}$Tuorla Observatory, Department of Physics and Astronomy, University of Turku, 20014, Turku, Finland \\
$^{21}$Astrophysics Research Institute, Liverpool John Moores University, IC2, Liverpool Science Park, 146 Brownlow Hill, Liverpool L3 5RF, UK \\
$^{22}$Max-Planck-Institut f\"{u}r Astrophysik, Karl-Schwarzschild Str. 1, D-85748 Garching, Germany \\
$^{23}$Aryabhatta Research Institute of observational sciences, Manora Peak, Nainital, 263001, India \\
$^{24}$Instituto de Alta Investigaci\'on, Universidad de Tarapac\'a, Casilla 7D, Arica, Chile \\
$^{25}$School of Physics, Trinity College Dublin, College Green, Dublin 2, Ireland \\
$^{26}$Steward Observatory, University of Arizona, 933 North Cherry Avenue, Tucson, AZ 85721-0065, USA\\
$^{27}$Department of Physics, University of Oxford, Keble Road, Oxford, OX1 3RH \\
$^{28}$Astrophysics Research Centre, School of Mathematics and Physics, Queens University Belfast, Belfast BT7 1NN, UK \\
$^{29}$Department of Physics and Astronomy, Aarhus University, Ny Munkegade 120, DK-8000 Aarhus C, Denmark \\
$^{30}$INAF – Osservatorio Astronomico d’Abruzzo, via M. Maggini snc, Teramo, I-64100, Italy \\
$^{31}$Department of Physics, University of California, Davis, CA 95616, USA \\
$^{32}$ European Southern Observatory, Alonso de C\'ordova 3107, Casilla 19, Santiago, Chile \\
$^{33}$Millennium Institute of Astrophysics, Nuncio Monsenor S\'otero Sanz 100, Providencia, 8320000 Santiago, Chile \\
$^{34}$INAF-Osservatorio Astrofisico di Catania, Via Santa Sofia 78, I-95123 Catania, Italy \\
$^{35}$Instituto de Astrof\'{i}sica, Universidad Andres Bello, Fernandez Concha 700, Las Condes, Santiago RM, Chile \\
$^{36}$ ICRANet, Piazza della Repubblica 10, I-65122 Pescara, Italy \\
$^{37}$Institut für Theoretische Physik, Goethe Universität, Max-von-Laue-Str. 1, 60438 Frankfurt am Main, Germany \\
$^{38}$INFN-TIFPA, Trento Institute for Fundamental Physics and Applications, Via Sommarive 14, I-38123 Trento, Italy \\
$^{39}$Institut d’Estudis Espacials de Catalunya (IEEC), E-08034 
Barcelona, Spain. \\
$^{40}$Center for Astrophysics, Harvard \& Smithsonian, Cambridge, Massachusetts, MA 02138, US \\
$^{41}$The NSF AI Institute for Artificial Intelligence and Fundamental Interactions, USA \\
$^{42}$Finnish Centre for Astronomy with ESO (FINCA), University of Turku, Väisäläntie 20, 21500 Piikkiö, Finland \\
$^{43}$DTU Space, National Space Institute, Technical University of Denmark, Elektrovej 327, 2800Kgs. Lyngby, Denmark \\
$^{44}$Dipartimento di Fisica e Astronomia ``G. Galilei'', Universit\`{a} degli studi di Padova Vicolo dell’Osservatorio 3, I-35122 Padova, Italy \\
$^{45}$IAASARS, National Observatory of Athens, Metaxa \& Vas. Pavlou St., 15236, Penteli, Athens, Greece \\
$^{46}$Department of Astronomy, University of Virginia, Charlottesville, VA 22904, USA \\
$^{47}$Department of Physics and Astronomy, University of North Carolina at Chapel Hill, Chapel Hill, NC 27599, USA \\
$^{48}$Cosmic Dawn Center (DAWN), Denmark \\
$^{49}$Niels Bohr Institute, University of Copenhagen, Jagtvej 128, 2200 København N, Denmark \\
$^{50}$Manipal Centre for Natural Sciences, Manipal Academy of Higher Education, Manipal - 576104, Karnataka, India \\
$^{51}$Indian Institute Of Astrophysics, 100 Feet Rd, Santhosapuram, 2nd Block, Koramangala, Bengaluru, Karnataka 560034, India
}



%
%


\onecolumn
\appendix

\section{Additional tables and figures}

\begin{figure*}
\centering
  \includegraphics[width=0.65\textwidth]{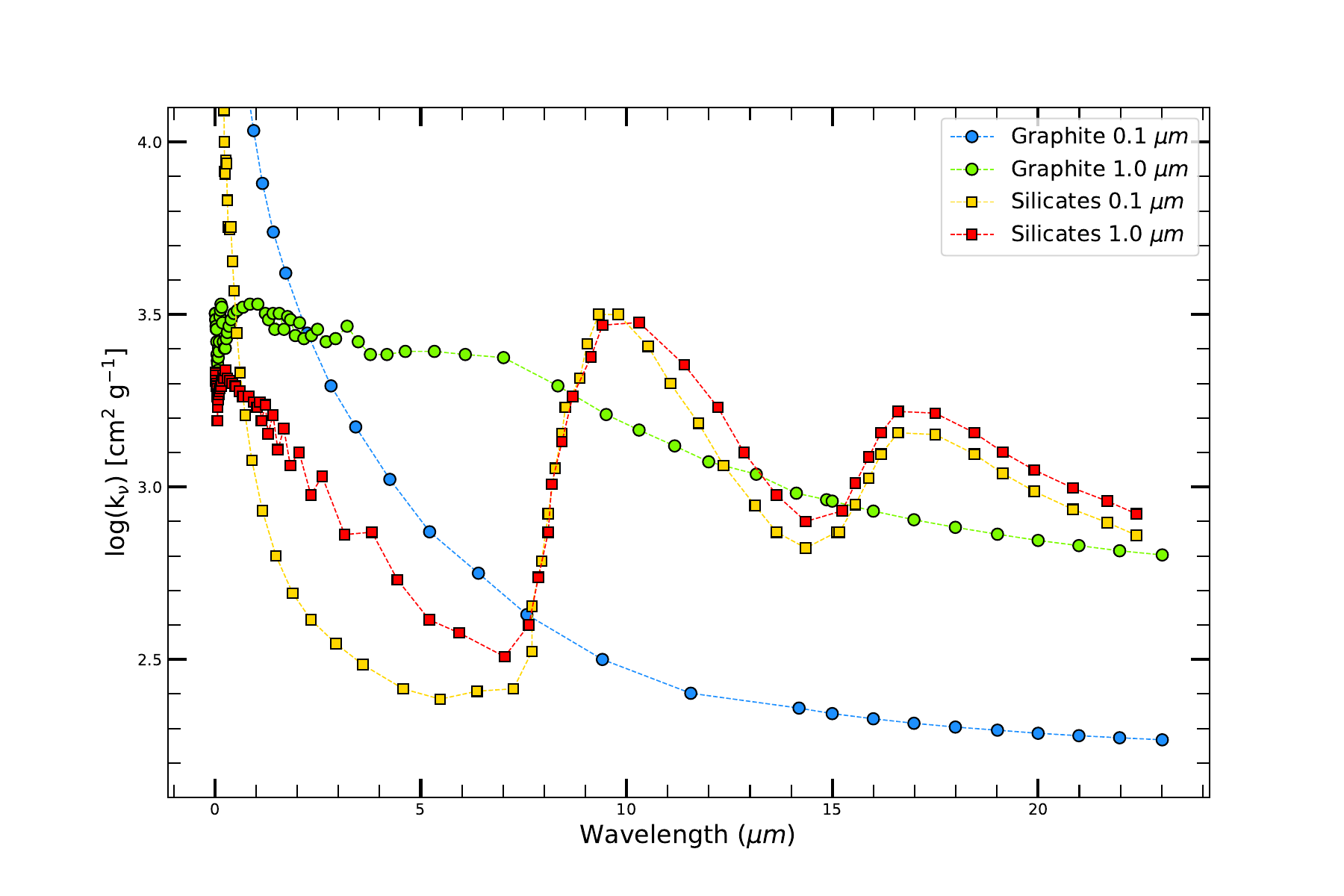}%
  \caption{Opacity of dust with different composition (graphite and silicates) and grain size (0.1 and 1 $\mu m$). Reported values were tabulated by \protect \cite{Fox2010_TabulatedOpacity} and extended to $22$ $\mu m$ as detailed in Section $\ref{NGC300OTdust}$.}
  \label{GrainOpacity}
\vspace{0.3cm}
 \centering
	\includegraphics[width=0.9\textwidth]{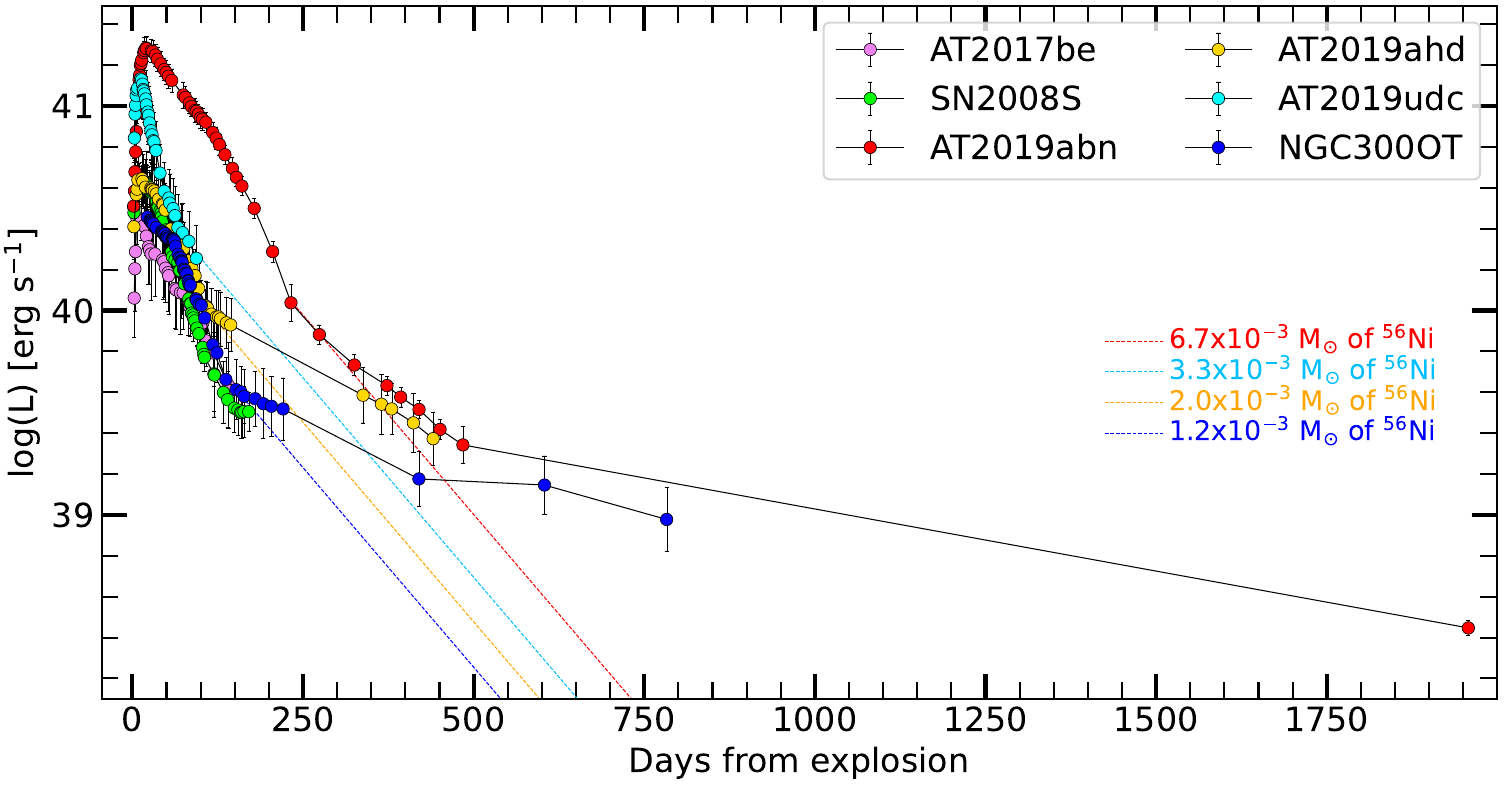}
    \caption{Updated version of Figure \ref{bolometric} including the late time luminosity of AT 2019abn measured on JWST data collected on 2024-05-31.}
    \label{JWST_bolometric}
\end{figure*}



\begin{table*}
\caption{Instruments and facilities used in our follow-up campaigns.}
\label{Instrum}
\centering\resizebox{0.82\textwidth}{!}{
\begin{adjustbox}{tabular= ccccc, center}

  &  &  &  &  \\
		\hline
Code & Diameter [m] & Telescope & Instrument & Site \\
		\hline
PROMPT & 0.41 & PROMPT Telescope & Apogee Alta & Cerro Tololo Inter-American Observatory, Cerro Tololo, Chile\\
REM & 0.6 & REM & ROSS & ESO La Silla Observatory, La Silla, Chile\\
Swope & 1.0 & Swope Telescope & SITe \# 3 & Las Campanas Observatory, Atacama Region, Chile\\
fl03-fl15 & 1.0 & LCO (LSC site) & Sinistro & LCO node at Cerro Tololo Inter-American Observatory, Cerro Tololo, Chile\\
fl05-fl07 & 1.0 & LCO (ELP site) & Sinistro & LCO node at McDonald Observatory, Texas, USA\\
fl06-fl14 & 1.0 & LCO (CPT site) & Sinistro & LCO node at South African Astronomical Observatory, Cape Town, South Africa\\
fl12 & 1.0 & LCO (COJ site) & Sinistro & LCO node at Siding Spring Observatory, New South Wales, Australia\\
ZTF  & 1.22 & Oschin Telescope & ZTF &  Palomar Observatory, United States \\
DFOT  & 1.3 & Devasthal Fast Optical Telescope & DZ436 &  Devasthal, Nainital, India \\
AFOSC & 1.82 & Copernico Telescope & AFOSC & Osservatorio Astronomico di Asiago, Asiago, Italy\\
HCT  & 2.0 & Hymalayan Chandra Telescope & A7iii & Indian Astronomical Observatory, Hanle, India \\
IO:O & 2.0 & Liverpool Telescope & IO:O & Observatorio Roque de Los Muchachos, La Palma, Spain\\
IO:I & 2.0 & Liverpool Telescope & IO:I & Observatorio Roque de Los Muchachos, La Palma, Spain\\
GROND & 2.2 & MPG Telescope & GROND & ESO La Silla Observatory, La Silla, Chile\\
GRONDIR & 2.2 & MPG Telescope & GRONDIR & ESO La Silla Observatory, La Silla, Chile\\
ALFOSC & 2.56 & Nordic Optical Telescope & ALFOSC & Observatorio Roque de Los Muchachos, La Palma, Spain\\
NOTCam & 2.56 & Nordic Optical Telescope & NOTCam & Observatorio Roque de Los Muchachos, La Palma, Spain\\
SOFI & 3.58 & New Technology Telescope & SOFI & ESO La Silla Observatory, La Silla, Chile\\
ACAM & 4.20 & William Hershel Telescope & ACAM & Observatorio Roque de Los Muchachos, La Palma, Spain\\
LIRIS & 4.20 & William Hershel Telescope & LIRIS & Observatorio Roque de Los Muchachos, La Palma, Spain\\
		\hline

\end{adjustbox}}

\end{table*}

\begin{table*}
  \centering
  \begin{adjustbox}{tabular=lcccc,center}
    & &  \textbf{AT 2019abn} & &    \\
  \hline
     & $\gamma_{1}$ [-25 to -11] & $\gamma_{2}$ [0 to 110] & $\gamma_{3}$ [110 to 185] & $\gamma_{4}$ [>185] \\
    \hline
     $B$ & -21.8 $\pm$ 1.0 & 2.72 $\pm$ 0.05 & -- & -- \\
     $V$ & -20.5 $\pm$ 0.9 & 1.70 $\pm$ 0.02 & 2.85 $\pm$ 0.10 & -- \\
     $g$ & -19.7 $\pm$ 1.3 & 2.27 $\pm$ 0.03 & -- & -- \\
     $r$ & -21.0 $\pm$ 1.0 & 1.37 $\pm$ 0.01 & 2.49 $\pm$ 0.04 & 0.80 $\pm$ 0.06 \\
     $i$ & -20.2 $\pm$ 1.2 & 1.04 $\pm$ 0.01 & 2.14 $\pm$ 0.07 & 1.45 $\pm$ 0.05 \\
     $z$ & -16.0 $\pm$ 2.3 & 0.82 $\pm$ 0.01 & 1.67 $\pm$ 0.03 & 1.32 $\pm$ 0.09 \\
     $J$ & -- & 0.84 $\pm$ 0.08 & -- & 1.56 $\pm$ 0.12 \\
     $H$ & -- & 0.65 $\pm$ 0.02 & -- & 1.07 $\pm$ 0.07 \\
     $K$ & -- & 0.67 $\pm$ 0.07 & -- & 0.72 $\pm$ 0.04 \\

    \hline
  \end{adjustbox}
  \caption{Rise and decline rates for the various bands of AT 2019abn. All quantities are reported in [mag/100 days]. For more detailed discussion of these values, see Sect \ref{subsec_photomAT2019abn}.}
  \label{abnrates}
\end{table*}

\begin{table*}
  \centering
\begin{tabular}{lccc}
    &  & \textbf{AT 2019ahd} &    \\
  \hline
     & $\gamma_{1}$ [0 to 45] & $\gamma_{2}$ [45 to 105] & $\gamma_{3}$ [> 105]  \\
    \hline
     $g$  & 1.29 $\pm$ 0.11 & 2.51 $\pm$ 0.12 & -- \\
     $r$  & 1.12 $\pm$ 0.09 & 2.31 $\pm$ 0.10 & 1.05 $\pm$ 0.03 \\
     $i$  & 0.61 $\pm$ 0.06 & 2.18 $\pm$ 0.15 & 1.21 $\pm$ 0.04 \\
     $z$  & 0.51 $\pm$ 0.06 & 1.70 $\pm$ 0.09 & 1.16 $\pm$ 0.04 \\
     $J$  & 0.71 $\pm$ 0.06 & -- & 0.97 $\pm$ 0.03 \\
     $H$  & 0.58 $\pm$ 0.03 & -- & 0.57 $\pm$ 0.04 \\
     $K$  & 0.64 $\pm$ 0.10 & -- & 0.37 $\pm$ 0.03 \\

    \hline
  \end{tabular}
\quad
  \begin{tabular}{lccc}
   &  & \textbf{AT 2019udc} &    \\
  \hline
   & $\gamma_{1}$ [-10 to -8] & $\gamma_{2}$ [0 to 35] & $\gamma_{3}$ [> 35]   \\
    \hline
     $U$ &  -79.3 $\pm$ 8.7 & 8.19 $\pm$ 0.46 &        --       \\
     $B$ &  -62.1 $\pm$ 5.9 & 6.46 $\pm$ 0.27 & 2.69 $\pm$ 0.24 \\
     $V$ &  -30.2 $\pm$ 6.2 & 4.37 $\pm$ 0.15 & 3.19 $\pm$ 0.09 \\
     $g$ &  -57.4 $\pm$ 5.4 & 5.60 $\pm$ 0.12 & 2.88 $\pm$ 0.16 \\
     $r$ &  -47.5 $\pm$ 4.8 & 3.93 $\pm$ 0.09 & 2.22 $\pm$ 0.12 \\
     $i$ &  -17.8 $\pm$ 4.3 & 3.41 $\pm$ 0.17 & 2.12 $\pm$ 0.14  \\
     $z$ & -- & 3.0 $\pm$ 0.2 & 1.71 $\pm$ 0.22 \\
     $H$ & -- & -- & 0.91 $\pm$ 0.15 \\

    \hline
  \end{tabular}
  \caption{Rise and decline rates for various bands of AT 2019ahd and AT 2019udc. All quantities are reported in [mag/100 days]. For more detailed discussion of these values, see Sect \ref{subsec_photomAT2019ahd} and \ref{subsec_photomAT2019udc}.}
  \label{ahdudcrates}
\end{table*}

\begin{table*}
  \centering
  \begin{tabular}{lccccc}
   & & &  \textbf{NGC 300 OT} &  &  \\
  \hline
     & $\gamma_{1}$ [0 to 35] & $\gamma_{2}$ [35 to 75] & $\gamma_{3}$ [75 to 120] & $\gamma_{4}$ [120 to 255] & $\gamma_{5}$ [>255] \\
    \hline
     $U$ & 2.57 $\pm$ 0.16 & 4.20 $\pm$ 0.12 & -- & -- & -- \\
     $B$ & 2.16 $\pm$ 0.08 & 4.06 $\pm$ 0.08 & 6.30 $\pm$ 0.10 & 1.50 $\pm$ 0.08 & -- \\
     $V$ & 1.38 $\pm$ 0.06 & 3.77 $\pm$ 0.07 & 7.06 $\pm$ 0.20 & 1.49 $\pm$ 0.07 & -- \\
     $R$ & 0.81 $\pm$ 0.06 & 3.21 $\pm$ 0.05 & 5.24 $\pm$ 0.06 & 0.95 $\pm$ 0.05 & -- \\ 
     $I$ & 0.49 $\pm$ 0.06 & 2.66 $\pm$ 0.06 & 5.21 $\pm$ 0.10 & 1.35 $\pm$ 0.05 & -- \\ 
     & & & & & \\
     $u$ & 3.38 $\pm$ 0.23 & -- & -- & -- & -- \\
     $g$ & 1.73 $\pm$ 0.11 & 4.14 $\pm$ 0.34 & -- & -- & -- \\
     $r$ & 1.11 $\pm$ 0.05 & 3.32 $\pm$ 0.08 & -- & 0.69 $\pm$ 0.07 & 1.56 $\pm$ 0.05 \\
     $i$ & 0.64 $\pm$ 0.07 & 3.09 $\pm$ 0.09 & -- & 1.54 $\pm$ 0.09 & 1.38 $\pm$ 0.10 \\
     $z$ & 0.33 $\pm$ 0.07 & 2.24 $\pm$ 0.07 & -- & 1.29 $\pm$ 0.09 & 1.71 $\pm$ 0.10 \\
     & & & & & \\
     $J$ & 0.03 $\pm$ 0.10 & 1.90 $\pm$ 0.09 & 2.75 $\pm$ 0.09 & 1.94 $\pm$ 0.12 & 1.01 $\pm$ 0.23 \\
     $H$ & -0.28 $\pm$ 0.09 & 1.46 $\pm$ 0.14 & 2.00 $\pm$ 0.13 & 1.40 $\pm$ 0.08 & 1.21 $\pm$ 0.09 \\
     $K$ & -0.33 $\pm$ 0.30 & 0.98 $\pm$ 0.27 & 1.50 $\pm$ 0.20 & 0.64 $\pm$ 0.10 & 0.81 $\pm$ 0.06 \\

    \hline
  \end{tabular}
  \caption{Decline rates for various bands of NGC 300 OT. All quantities are reported in [mag/100 days]. For more detailed discussion of these values, see Sect \ref{subsec_photomNGC300OT}.}
  \label{NGC300rates}
\end{table*}

\pagebreak
\newpage
\renewcommand{\arraystretch}{1.3}
\begin{table*}
	\centering
	\caption{Template light curves for ILRTs as described in Section 7.}
	\label{TemplateNumbers}
	\begin{tabular}{llll} 
		\hline
Phase  & Total & Slow declining & Fast declining \\
$[days]$ & $[mag]$ & $[mag]$ & $[mag]$ \\
		\hline
-15  & -14.21 $^{-0.48}_{+0.89}$  &  -13.33 $^{-0.19}_{+0.23}$  &           --                    \\
-12  & -14.31 $^{-0.39}_{+0.62}$  &  -13.96 $^{-0.31}_{+0.43}$  &  	   --                    \\
-9   & -14.60 $^{-0.17}_{+0.20}$  &  -14.57 $^{-0.10}_{+0.11}$  &  -14.52 $^{-0.19}_{+0.23}$	 \\
-6   & -14.81 $^{-0.08}_{+0.08}$  &  -14.82 $^{-0.05}_{+0.05}$  &  -14.80 $^{-0.09}_{+0.10}$	 \\
-3   & -14.94 $^{-0.05}_{+0.05}$  &  -14.94 $^{-0.04}_{+0.05}$  &  -14.93 $^{-0.06}_{+0.06}$	 \\
0    & -14.99 $^{-0.01}_{+0.01}$  &  -14.99 $^{-0.01}_{+0.01}$  &  -14.99 $^{-0.02}_{+0.02}$	 \\
\hline
	\end{tabular}
 	\begin{tabular}{llll} 
		\hline
Phase & Total & Slow declining & Fast declining \\
$[days]$ & $[mag]$ & $[mag]$ & $[mag]$ \\
		\hline
96   & -12.83 $^{-0.59}_{+1.36}$  &  -13.13 $^{-0.45}_{+0.78}$  &  -12.49 $^{-0.64}_{+1.75}$	 \\
99   & -12.80 $^{-0.59}_{+1.39}$  &  -13.06 $^{-0.47}_{+0.85}$  &  -12.45 $^{-0.66}_{+1.92}$	 \\
102  & -12.83 $^{-0.57}_{+1.26}$  &  -13.01 $^{-0.49}_{+0.91}$  &  -12.41 $^{-0.67}_{+2.06}$	 \\
105  & -12.78 $^{-0.57}_{+1.29}$  &  -12.96 $^{-0.49}_{+0.93}$  &  -12.38 $^{-0.68}_{+2.25}$	 \\
108  & -12.84 $^{-0.49}_{+0.92}$  &  -12.88 $^{-0.49}_{+0.92}$  &            --	 		\\
111  & -12.78 $^{-0.49}_{+0.90}$  &  -12.81 $^{-0.49}_{+0.92}$  &            --	 		\\
\hline
	\end{tabular}
\end{table*}




\section{Measured data} \label{AppendixBmeasureddata}

\begin{table*}
\caption{Photometric data collected in Johnson bands for NGC 300 OT (Vega mag).}
\label{tab9}
\centering
 \resizebox{0.9\textwidth}{!}{%
 \begin{tabular}{ cccccccc }
\hline
  Date    &    MJD     &     U (err)       &	  B (err) &   V (err) & R (err) & I (err) &	  Instrument \\
 \hline
 \hline

 20080516  &   54602.89  &   -- (--) &       15.61 (0.05) & 14.69 (0.02) & 14.28 (0.02) &   13.83 (0.03) &	 PROMPT3+5	    \\
 20080520  &   54606.88  &  15.92 (0.36) &  15.65 (0.02) & 14.70 (0.02) & 14.24 (0.01) &   13.85 (0.03) &	 PROMPT2+5     \\
 20080521  &   54607.84	 &  -- (--) &       -- (--) &      -- (--) &      14.29 (60.00) &  13.83 (0.04) &	 REM	    \\
20080522  &   54608.83  &   -- (--) &       -- (--) &      14.82 (0.07) & 14.34 (84.00) &  -- (--) &     	 REM	    \\
 20080523  &   54609.55  &  -- (--) &       -- (--) &      14.82 (0.02) & 14.32 (0.01) &   13.84 (0.02) &	 PROMPT5       \\
 20080524  &   54610.87  &  16.09 (0.12) &  15.72 (0.02) & -- (--) &      -- (--) &        -- (--) &     	 PROMPT3	    \\
 20080524  &   54610.87  &  16.07 (0.19) &  15.61 (0.02) & 14.89 (0.02) & 14.36 (0.01) &   13.85 (0.02) &	 PROMPT2+5	    \\
\hline
 \end{tabular}%
 }
 
\end{table*}


\begin{table*}
\centering
\caption{Photometric data collected in Johnson bands for NGC 300 OT (Vega mag).}
\label{tab10}
 \resizebox{0.85\textwidth}{!}{\begin{tabular}{cccccccc}
 
\hline
  Date    &    MJD     &     U (err)       &	  B (err) &   V (err) & R (err) & I (err) &	  Instrument \\
 \hline
 \hline

20080615  &   54632.92  &   16.60 (0.01) & -- (--) &	  -- (--) &      -- (--) &     	-- (--) &         EFOSC	    \\
 20080616  &   54633.80  &  -- (--) &      16.20 (0.02) & -- (--) &      14.52 (0.02) &	13.95 (0.01) &	    PROMPT2+5	    \\
 20080617  &   54634.80  &  -- (--) &      16.14 (0.03) & 15.22 (0.02) & 14.59 (0.02) &	13.98 (0.01) &	    PROMPT2+5	    \\
 20080620  &   54637.79  &  -- (--) &      16.26 (0.06) & 15.27 (0.03) & 14.53 (0.01) &	13.97 (0.02) &	    PROMPT4	    \\
 20080622  &   54639.79  &  -- (--) &      16.27 (0.05) & 15.29 (0.04) & 14.61 (0.02) &	14.02 (0.04) &	    PROMPT4	    \\
 20080622  &   54639.85  &  -- (--) &      16.31 (0.04) & 15.25 (0.03) & 14.60 (0.02) &	13.99 (0.02) &	    PROMPT2+5	    \\
 20080623  &   54640.78  &  -- (--) &      16.35 (0.06) & 15.25 (0.02) & 14.56 (0.01) &	13.99 (0.02) &	    PROMPT4	    \\
\hline
 \end{tabular}}
 
\end{table*}

\begin{table*}
\centering
\caption{Photometric data collected in sloan bands for NGC 300 OT (AB mag).}
\label{tab11}
 \resizebox{0.85\textwidth}{!}{\begin{tabular}{cccccccc}
 
  \hline
  Date    &    MJD     &     u (err)       &	  g (err) &   r (err) & i (err) & z (err) &	  Instrument  \\
 \hline
 \hline
20080516   &   54602.38   &  -- (--) &      -- (--) &      14.38 (0.02) &  14.21 (0.02) &    14.12 (0.02) &	PROMPT \\
20080520   &   54606.40   &  -- (--) &      -- (--) &      14.47 (0.02) &  14.21 (0.01) &    14.11 (0.03) &	PROMPT \\
20080520   &   54606.40   &  16.78 (0.09) & 15.34 (0.07) & 14.50 (0.04) &  14.23 (0.02) &    -- (--) &     	Swope  \\
20080523   &   54609.39   &  -- (--) &      -- (--) &      14.48 (0.01) &  14.20 (0.01) &    14.11 (0.03) &	PROMPT \\ 
20080524   &   54610.38   &  -- (--) &      -- (--) &      14.49 (0.03) &  -- (--) &         -- (--) &     	PROMPT \\
20080525   &   54611.37   &  -- (--) &      -- (--) &      14.50 (0.02) &  14.20 (0.02) &    14.07 (0.01) &	PROMPT \\
20080530   &   54616.35   &  -- (--) &      15.42 (0.06) & 14.61 (0.02) &  14.29 (0.01) &    14.16 (0.01) &	PROMPT \\
\hline
 \end{tabular}}
 
\end{table*}

\begin{table*}
\centering
\caption{Photometric data collected in sloan bands for NGC 300 OT (AB mag).}
\label{tab12}
 \resizebox{0.85\textwidth}{!}{\begin{tabular}{cccccccc}
 
  \hline
  Date    &    MJD     &     u (err)       &	  g (err) &   r (err) & i (err) & z (err) &	  Instrument \\ \hline
 \hline
 20081003   &   54742.09 &  -- (--) &  -- (--) &       -- (--) &        -- (--) &      	17.67 (0.05) &	  PROMPT \\ 
20081023   &   54762.10 &  -- (--) &  -- (--) &       18.86 (0.06) &   18.55 (0.06) & 	-- (--) &	  PROMPT \\
20081025   &   54764.09 & -- (--) &  -- (--) &       18.88 (0.07) &   18.59 (0.10) & 	18.01 (0.10) &	  PROMPT \\
20081026   &   54765.20 & -- (--) &  -- (--) &       18.86 (0.07) &   18.64 (0.13) & 	-- (--) &	  PROMPT \\
20081027   &   54766.12 & -- (--) &  -- (--) &       18.93 (0.07) &   18.76 (0.07) & 	18.04 (0.09) &	  PROMPT \\ 
20081029   &   54768.06 & -- (--) &  -- (--) &       18.96 (0.05) &   18.79 (0.08) & 	18.12 (0.08) &	  PROMPT \\ 
20081102   &   54772.10 & -- (--) &  -- (--) &       19.06 (0.06) &   18.90 (0.09) & 	18.23 (0.09) &	  PROMPT \\
\hline
 \end{tabular}}
 
\end{table*}

\begin{table*}
\centering
\caption{Photometric data collected in NIR bands for NGC 300 OT (Vega mag).}
\label{tab13}
\begin{tabular}{cccccc}

 \hline
  Date    &    MJD     &     J     (err)       &	  H    (err) &        	 K	 (err) &	  Instrument \\
 \hline
 \hline

20/5/08 &  54606.35  &   -- -- &   12.67 (0.24)  & -- --&  REM  \\
21/5/08 &  54607.34  &   -- -- &   12.64 (0.18)  & -- --&  REM  \\
22/5/08 &  54608.83  &   13.12 (0.03) &  12.61 (0.10)  & -- -- &  REM  \\
23/5/08 &  54609.83  &   13.09 (0.06) &  12.62 (0.12)  & -- -- &  REM  \\
24/05/0 &  54610.91    &   -- -- &    -- --  & 11.94 (0.07) &  REM  \\
25/5/08 &  54611.82  &   13.10 (0.16) & -- --  & -- --  &  REM  \\
30/5/08 &  54616.81  &   13.09 (0.05) & 12.63 (0.05) & 12.04 (0.09)   &  REM  \\
31/5/08 &  54617.80  &   13.16 (0.07) & 12.61 (0.04) & -- --  &  REM  \\
06/6/08 &  54623.78  &   13.03 (0.20) & 12.56 (0.44) & -- --  &  REM  \\
07/6/08 &  54624.80  &   13.10 (0.05) & 12.53 (0.04) & -- --  &  REM  \\
08/6/08 &  54625.79  &   13.11 (0.05) & 12.53 (0.35) & -- -- &  REM  \\
09/6/08 &  54626.78  &   13.12 (0.07) & -- -- & -- -- &  REM  \\
10/6/08 &  54627.78  &   13.11 (0.07) & 12.56 (0.34) & -- -- &  REM  \\

 \hline
 \end{tabular}
 
\end{table*}

\begin{table*}
\centering
\caption{Photometric data collected in NIR bands for NGC 300 OT (Vega mag).}

 \begin{tabular}{cccccc}

 \hline
  Date    &    MJD     &     J     (err)       &	  H    (err) &        	 K	 (err) &	  Instrument \\
 \hline
 \hline
 & & &  & &\\

11/6/08 &  54628.28  &   -- -- & 12.6 (0.3) & -- -- &  REM  \\
12/6/08 &  54629.77  &   13.07 (0.06) & -- -- & -- -- &  REM  \\
13/6/08 &  54630.77  &   13.11 (0.09) & 12.61 (0.06) & -- -- &  REM  \\
14/6/08 &  54631.84  &   13.09 (0.04) & 12.59 (0.05) & -- -- &  REM  \\
15/6/08 &  54632.76  &   13.10 (0.09) & 12.60 (0.12) & 11.90 (0.11)   &  REM  \\
16/6/08 &  54633.76  &   13.13 (0.14) & -- -- & 11.92 (0.17)  &  REM  \\
\hline
 \end{tabular}
 
\end{table*}

\begin{table*}
\centering
\caption{Photometric data collected in NIR bands for NGC 300 OT (Vega mag).}

 \begin{tabular}{cccccc}

 \hline
  Date    &    MJD     &     J     (err)       &	  H    (err) &        	 K	 (err) &	  Instrument \\
 \hline
 \hline
 & & &  & &\\

11/6/08 &  54628.28  &   -- -- & 12.58 (0.29) & -- -- &  REM  \\
12/6/08 &  54629.77  &   13.07 (0.06) & -- -- & -- -- &  REM  \\
13/6/08 &  54630.77  &   13.11 (0.09) & 12.61 (0.06) & -- -- &  REM  \\
14/6/08 &  54631.84  &   13.09 (0.04) & 12.59 (0.05) & -- -- &  REM  \\
15/6/08 &  54632.76  &   13.10 (0.09) & 12.60 (0.12) & 11.9 (0.1)   &  REM  \\
16/6/08 &  54633.76  &   13.13 (0.14) & -- -- & 11.92 (0.17)  &  REM  \\
\hline
 \end{tabular}
 
\end{table*}

\begin{table*}
\centering
\caption{Photometric data collected in sloan bands for AT 2019abn (AB mag).}
\label{tab14}
 \resizebox{0.68\textwidth}{!}{\begin{tabular}{ccccccc}

 \hline
  Date    &    MJD     &     g     (err)       &	  r    (err) &        	i	 (err) & z   (err)  &	  Instrument \\
 \hline
 \hline
20190127   &   58510.52  & 20.02 (0.06) &  18.61 (0.02) &  17.95 (0.02) &  -- (--) &     	 fa05		  \\
20190127   &   58510.07  & -- (--) &       18.72 (0.08) &  18.14 (0.02) &  17.66 (0.01) &	 ACAM 	  \\
20190129   &   58512.26  & 19.52 (0.03) &  18.19 (0.02) &  17.47 (0.01) &  16.92 (0.02) &	 ACAM 	  \\
20190130   &   58513.20  & 19.43 (0.05) &  17.94 (0.02) &  17.26 (0.01) &  16.72 (0.01) &	 ACAM		  \\
20190201   &   58515.48  & 18.88 (0.02) &  17.38 (0.01) &  16.77 (0.01) &  -- (--) &     	 fa05 	  \\
20190205   &   58519.17  & 18.29 (0.01) &  16.96 (0.02) &  16.38 (0.02) &  16.07 (0.01) &	 ALFOSC	  \\
20190211   &   58525.14  & 17.96 (0.01) &  16.76 (0.01) &  16.24 (0.01) &  15.97 (0.01) &	 ALFOSC	  \\
\hline
 \end{tabular}}
 
\end{table*}

\begin{table*}
\centering
\caption{Photometric data collected in sloan bands for AT 2019abn (AB mag).}
\label{tab15}
 \resizebox{0.68\textwidth}{!}{\begin{tabular}{ccccccc}

 \hline
  Date    &    MJD     &     g     (err)       &	  r    (err) &        	i	 (err) & z   (err)  &	  Instrument \\
 \hline
 \hline
 
 20190803   &   58698.93  &    --    (--)     &  19.75 (0.05)  &  18.61 (0.02)   &  17.86 (0.04)  & ALFOSC	  \\
20190806   &   58701.14  &    --    (--)     &  19.79 (0.04)  &  18.70 (0.06)  &  17.88 (0.02)  & fa07		  \\
20190812   &   58707.89  &    --    (--)    &  19.90 (0.04)  &  18.86 (0.02)   &  18.04 (0.02)  & ALFOSC 	  \\
20190812   &   58707.89  &   $>$21.0    &   --    (--)    &   --    (--)     &  --    (--)   	 & ALFOSC	  \\
20190830   &   58725.86  &    --    (--)     &  20.96 (0.05)  &  20.08 (0.03)   &  19.07 (0.05)  & ALFOSC	  \\
20190830   &   58725.86  &   $>$21.9    &   --    (--)    &   --    (--)     &  --    (--)     & ALFOSC	  \\
20190915   &   58741.84  &     --    (--)    &  21.23 (0.09)  &  20.32 (0.06)   &  19.43 (0.06)  & ALFOSC	  \\
20190925   &   58751.85  &    --    (--)     &   --    (--)    &  20.44 (0.07)   &  19.66 (0.09)  & ALFOSC	  \\
20191123   &   58809.00  &    --    (--)     &   --    (--)    &  21.24 (0.04)   &   --    (--)    & ALFOSC	  \\
20191202   &   58819.22  &    --    (--)     &  21.74 (0.08)  &  21.36 (0.04)   &  20.33 (0.11)  & ALFOSC	  \\
20191218   &   58835.26  &    --    (--)     &  21.95 (0.10)  &  21.73 (0.05)   &  20.65 (0.15)  & ALFOSC	  \\
20191230   &   58847.13  &   --     (--)    &  21.96 (0.09)  &    --    (--)    &  20.81 (0.10)  & ALFOSC 	  \\
20191230   &   58847.13  &   --     (--)     &   --    (--)    &  $>$21.5    &   --    (--)    & ALFOSC	  \\
20200130   &   58878.14  &    --    (--)    &   --    (--)    &   --    (--)     &  $>$20.76   & ALFOSC	  \\

 \hline
 \end{tabular}}
 
\end{table*}

\begin{table*}
\centering
\caption{Photometric data collected in Johnson bands for AT 2019abn (Vega mag).}
\label{tab16}
 \begin{tabular}{ccccc}

 \hline
  Date    &    MJD     &     B     (err)       &	  V    (err) &  	  Instrument \\
 \hline
 \hline
 & & &  &  \\
 20190127  &  58510.52  &    20.73   (0.12)  &  19.22 (0.05)   & 	fa05		\\
20190201  &  58515.47  &    19.50   (0.04)  &  18.04 (0.03)   & 	fa05		\\
20190213  &  58527.29  &    18.48   (0.05)  &  17.28 (0.02)   & 	fa05		\\
20190218  &  58532.27  &      --      (--)   &  17.29 (0.12)   & 	fa05		\\
20190222  &  58536.50  &      --      (--)   &  17.31 (0.02)   & 	fa05 		\\
20190225  &  58539.50  &      --      (--)   &  17.34 (0.02)   & 	fa05 		\\
20190227  &  58541.15  &    18.53   (0.03)  &  17.35 (0.02)   &  	IO:O		\\
20190302  &  58544.26  &      --      (--)   &  17.37 (0.02)   & 	fa05		\\
20190312  &  58554.13  &    18.83   (0.03)  &  17.55 (0.02)   & 	IO:O		\\
20190312  &  58554.84  &     18.87 (0.17)   &  17.59 (0.05)   & Moravian       \\
20190314  &  58556.93  &    18.86   (0.06)  &  17.63 (0.03)   & 	IO:O 		\\
20190315  & 58557.19   &      --      (--)   &  17.60 (0.03)   & 	fa05		\\
20190318  &  58560.22  &      --      (--)   &  17.68 (0.05)   & 	fa05		\\
20190318  & 58560.24   &    18.95   (0.05)  &  17.68 (0.02)   & 	IO:O 		\\
20190324  & 58566.48   &      --      (--)   &  17.80 (0.02)   & 	fa05		\\
20190327  &  58569.93  &     19.40 (0.12)   & 17.85 (0.05)   &  Moravian       \\
20190330  &  58572.84  &     19.52 (0.15)   & 17.99 (0.06)   &  Moravian       \\

 \hline
 \end{tabular}
 
\end{table*}

\begin{table*}
\centering
\caption{Photometric data collected in Johnson bands for AT 2019abn (Vega mag).}

\begin{tabular}{ccccc}

 \hline
  Date    &    MJD     &     B     (err)       &	  V    (err) &  	  Instrument \\
 \hline
 \hline
 20190408  &  58581.98  &    19.69   (0.06)  &  18.04 (0.02)   &  	ALFOSC	\\
20190409  &  58582.25  &      --      (--)   &  18.12 (0.02)   & 	fa05 		\\
20190411  &  58584.98  &    19.74   (0.07)  &  18.15 (0.02)   & 	IO:O		\\
20190415  &  58588.11  &      --      (--)   &  18.23 (0.05)   & 	fa05		\\
20190417  &  58590.94  &    20.01   (0.07)  &  18.24 (0.02)   & 	IO:O		\\
20190425  &  58598.40  &      --      (--)   &  18.34 (0.02)   & 	fa05		\\
20190428  &  58601.14  &    20.17   (0.10)  &  18.36 (0.03)   & 	IO:O		\\
\hline
 \end{tabular}
 
\end{table*}

\begin{table*}
	\centering
	\caption{$JWST$ data of AT 2019abn (AB mag)}
	\label{JWSTmeasures}
 \vspace{0.15cm}
\small 
	\begin{tabular}{cc}
		\hline
    Date   & F560W \\
            \hline
    2022-12-13 & 19.90 (0.05) \\
\hline

	\end{tabular}

\vspace{0.15cm}

	\begin{tabular}{cccc}
		\hline
    Date  & F444W & F560W & F770W \\
            \hline
    2023-06-08 & 21.30$\pm$0.10 & 20.40 (0.10) & 19.20 (0.10) \\
\hline

	\end{tabular}

 \vspace{0.15cm}

	\begin{tabular}{cccc}
		\hline
    Date  & F200W & F300M & F335M \\
            \hline
    2023-06-14 & 23.75 (0.10) & 22.40 (0.10) & 22.05 (0.10) \\
\hline

	\end{tabular}

 \vspace{0.15cm}
	\resizebox{1.0\textwidth}{!}{\begin{tabular}{cccccccccc}
		\hline
  
    Date  & F182M & F360M & F430M & F1000W & F1130W & F1280W & F1500W & F1800W & F2100W \\
            \hline
   2024-05-31 &  22.40 (0.10) & 22.50 (0.10) & 21.80 (0.10) & 19.40 (0.10) & 18.80 (0.10) & 18.70 (0.10) & 18.75 (0.10) & 18.90 (0.10) & 18.95 (0.10) \\
\hline

	\end{tabular}}
 
\end{table*}

\normalsize


\begin{table*}
\centering
\caption{Photometric data collected in NIR bands for AT 2019abn (Vega mag).}

 \begin{tabular}{cccccc}

 \hline
  Date    &    MJD     &     J     (err)       &	  H    (err) &  K (err)  &	  Instrument \\
 \hline
 \hline
20190127 &  58510.11  &   16.26 (0.06) &  -- (--) &       14.39 (0.03) &	  LIRIS    \\
20190128 &  58511.21  &   15.86 (0.02) &  -- (--) &       -- (--) &     	  LIRIS	\\
20190129 &  58512.16  &   15.53 (0.03) &  14.86 (0.04) &  14.11 (0.03) &	  LIRIS	\\
20190130 &  58513.18  &   15.36 (0.03) &  14.70 (0.03) &  14.02 (0.04) &	  LIRIS	\\
20190201 &  58515.12  &   15.22 (0.02) &  14.57 (0.03) &  13.93 (0.06) &	  LIRIS	\\
20190227 &  58541.16  &   -- (--) &       14.49 (0.01) &  -- (--) &     	  IO:I 	\\
20190305 &  58547.24  &   15.04 (0.01) &  14.56 (0.02) &  13.94 (0.01) &	  NOTCAM	\\
\hline
 \end{tabular}
 
\end{table*}

\begin{table*}
\centering
\caption{Photometric data collected in MIR bands for AT 2019abn (Vega mag).}
\label{tab17}
\begin{tabular}{ccccc}

 \hline
  Date    &    MJD     &     W1     (err)       &	  W2    (err) & 	  Instrument \\
 \hline
 \hline
 & & &  &  \\
20190524  & 58627.52 & 14.08 0.08  &  13.06 0.12	& WISE \\	
20191215  & 58832.73 & 15.08 0.24  &  14.20 0.23   & WISE \\ 
20200522  & 58991.71 & 15.61  0.15   &  14.90 0.18     & WISE \\ 

   \hline
 \end{tabular}
 

\vspace{0.3cm}

\begin{tabular}{ccccc}

 \hline
  Date    &    MJD     &     [3.6]     (err)       &	  [4.5]    (err) & 	  Instrument \\
 \hline
 \hline
 & & &  &  \\
20190407  & 58580.58  &  13.19 0.02  & 12.59 0.02  & Spitzer \\			
20190530  & 58633.22  &  14.06 0.04  & 13.36 0.03  & Spitzer \\	
20190821  & 58716.66  &  14.35 0.02  & 14.19 0.09  & Spitzer \\	
20191025  & 58781.37  &  14.74 0.09  & 14.45 0.10   & Spitzer \\	

   \hline
 \end{tabular}
 
\end{table*}


\begin{table*}
\centering
\caption{Photometric data collected in Sloan bands for AT 2019ahd (AB mag).}
\label{tab19}
\begin{tabular}{ccccccc}

 \hline
  Date    &    MJD     &     g     (err)       &	  r    (err) &  i    (err) &  z    (err)  &	  Instrument \\
 \hline
 \hline
20190130 &  58513.25   &  18.55 (0.02) &  18.32 (0.02) &  18.24 (0.01) &  18.27 (0.02) &	  GROND   \\
20190205 &  58519.11   &  18.04 (0.01) &  17.70 (0.02) &  17.51 (0.01) &  17.35 (0.02) &	  ALFOSC  \\
20190211 &  58525.04   &  18.12 (0.03) &  17.57 (0.07) &  17.44 (0.01) &  17.36 (0.01) &	  ALFOSC  \\
20190212 &  58526.35   &  18.15 (0.01) &  17.61 (0.07) &  17.46 (0.01) &  17.39 (0.03) &	  GROND	  \\
20190216 &  58530.13   &  18.23 (0.05) &  17.66 (0.07) &  17.52 (0.01) &  17.44 (0.03) &	  GROND	  \\
20190224 &  58538.12   &  18.27 (0.04) &  17.68 (0.09) &  17.51 (0.02) &  17.46 (0.04) &	  GROND	  \\
20190225 &  58539.05   &  18.28 (0.01) &  17.70 (0.01) &  17.52 (0.01) &  17.44 (0.03) &	  IO:O	  \\
\hline
 \end{tabular}
 
\end{table*}

\begin{table*}
\centering
\caption{Photometric data collected in NIR bands for AT 2019ahd (Vega mag).}
\label{tab20}
\begin{tabular}{cccccc}

 \hline
  Date    &    MJD     &     J     (err)       &	  H    (err) &  K    (err)   &	  Instrument \\
 \hline
 \hline
20190130 &  58513.25  &  17.49 (0.04) &  16.90 (0.07) &	 16.46 (0.11) &	  GRONDIR  \\
20190205 &  58519.27  &  16.45 (0.04) &  16.05 (0.05) &	 15.65 (0.05) &	  GRONDIR  \\
20190212 &  58526.35  &  16.57 (0.03) &  16.24 (0.06) &	 15.83 (0.07) &	  GRONDIR  \\
20190216 &  58530.14  &  16.60 (0.04) &  16.29 (0.09) &	 15.67 (0.10) &	  GRONDIR  \\
20190224 &  58538.15  &  -- (--) &       16.24 (0.04) &	 -- (--) &     	  IO:I     \\
20190224 &  58538.12  &  16.58 (0.03) &  16.29 (0.06) &	 15.91 (0.18) &	  GRONDIR  \\
20190307 &  58549.27  &  16.70 (0.04) &  16.31 (0.05) &	 16.00 (0.07) &	  GRONDIR  \\
\hline
 \end{tabular}
 
\end{table*}

 
 \begin{table*}
\centering
\caption{Photometric data collected in Sloan bands for AT 2019udc (AB mag).}
\label{tab21}
\resizebox{0.8\textwidth}{!}{\begin{tabular}{cccccccc}

 \hline
  Date    &    MJD    & u    (err)   &     g     (err)       &	  r    (err) &  i    (err)   &   z    (err)   &	  Instrument \\
 \hline
 \hline
20191107 &  58794.03 &  -- (--) &	17.94 (0.05) & 	17.70 (0.05) &	17.51 (0.04) & -- (--) &     		  Prompt5	\\
20191108 &  58795.61 &  -- (--) &	17.82 (0.02) & 	17.53 (0.03) &	17.52 (0.02) & -- (--) &		  fa12		\\
20191109 &  58796.14 &  -- (--) &	17.77 (0.02) & 	17.51 (0.02) &	17.48 (0.03) & -- (--) &		  fa15		\\
20191110 &  58797.03 &  -- (--) &	17.85 (0.10) & 	17.48 (0.08) &	17.38 (0.12) & -- (--) &		  Prompt5 	\\
20191113 &  58800.18 &  -- (--) &	17.69 (0.02) & 	17.45 (0.02) &	17.43 (0.02) & -- (--) &		  fa15		\\
20191118 &  58805.89 &  18.78 (0.07) &	17.87 (0.04) & 	17.59 (0.02) &	17.55 (0.02) & 17.56 (0.04) &		  ALFOSC 	\\
20191119 &  58806.17 &  -- (--) &	17.86 (0.05) & 	17.64 (0.04) &	17.55 (0.05) & -- (--) &		  Prompt5 	\\
\hline
 \end{tabular}}
 
\end{table*}

\begin{table*}
\centering
\caption{Photometric data collected in Johnson bands for AT 2019udc (Vega mag).}
\label{tab22}
\begin{tabular}{cccccc}

 \hline
  Date    &    MJD     &     U     (err)       &	  B    (err) &  V    (err)   &	  Instrument \\
 \hline
 \hline
20191108 &   58795.59    & 17.80 (0.06)  & 18.12 (0.03) &  17.64 (0.03)  &    fa12 		\\
20191109 &   58796.13    & 17.78 (0.04)  & 18.09 (0.03) &  17.67 (0.03) &    fa15 		\\
20191109 &   58796.02    &    --    (--)   & 18.10 (0.14) &  17.59 (0.07)  &    Prompt5 		\\
20191110 &   58797.04    &    --    (--)   & 18.02 (0.14) &  17.57 (0.09)  &    Prompt5 		\\
20191113 &   58800.16    & 17.66 (0.04)  & 17.96 (0.03) &  17.57 (0.03)  &    fa15 		\\
20191118 &   58805.89    &    --    (--)   & 18.18 (0.05) &  17.75 (0.03)  &    ALFOSC 	\\
20191122 &   58809.20    &    --    (--)   & 18.38 (0.15) &  17.96 (0.07)  &    Prompt5		\\
\hline
 \end{tabular}
 
\end{table*}

\begin{table*}
\centering
\caption{Photometric data collected in NIR bands for AT 2019udc (Vega mag).}
\label{tab23}
\begin{tabular}{cccccc}

 \hline
  Date    &    MJD     &     J     (err)       &	  H    (err) &  K    (err)   &	  Instrument \\
 \hline
 \hline
 & & &  & &   \\
20191107 &  58794.06  &  17.01 (0.04) &  16.56 (0.05) &  16.30 (0.04) & SOFI		\\
20191121 &  58808.23  &  17.12 (0.05) &  16.71 (0.06) &  16.42 (0.05) & SOFI 		\\
20191129 &  58816.16  &  17.31 (0.05) &  16.94 (0.10) &  16.53 (0.06) & SOFI		\\
20191219 &  58836.81  &    --  (--)  &  17.15 (0.11) &    --  (--)  & IO:I		\\
20191221 &  58838.83  &    --  (--)  &  17.17 (0.08) &    --  (--)  & IO:I		\\
20200102 &  58850.92  &    --  (--)  &  17.23 (0.09) &    --  (--)  & IO:I		\\
20200104 &  58852.12  &  17.78 (0.06) &  17.24 (0.07) &  16.88 (0.11) & SOFI 		\\
20200109 &  58857.96  &  17.85 (0.03) &  --  (--)  &    --  (--)  & NOTCAM		\\
20200114 &  58862.83  &    --  (--)  &  17.33 (0.09) &    --  (--)  & IO:I		\\
20200203 &  58882.04  &  17.93 (0.06) &  17.32 (0.07) &  16.900 (0.08) & SOFI		\\
20200717 &  59047.00  &    --  (--)  &  19.03 (0.11) &    --  (--)  & NOTCAM		\\

   \hline
 \end{tabular}
 
\end{table*}

\begin{table*}
\centering
\caption{Photometric data collected for AT 2019udc by the DLT40 survey (AB mag).}
\label{tab24}
\begin{tabular}{cccc}

 \hline
 MJD    &    mag (err)     &     MJD        &	mag     (err)   \\
 \hline
 \hline
& & & \\
58791.61   &	18.81	(0.09)   &	58792.20   &	18.04	(0.08)    \\
58787.62   &	19.48	(0.10)      &	58792.57   &	17.93	(0.08)   \\
58790.10   &	18.98	(0.10)      &	58792.65   &	17.89	(0.07)   \\
58791.26   &	19.01	(0.10)      &	58792.67   &	17.89	(0.05)   \\
58791.62   &	18.60	(0.09)   &	58792.71   &	17.97	(0.05)   \\
58791.63   &	18.68	(0.09)   &	58793.01   &	17.84	(0.08)   \\
58791.65   &	18.76	(0.08)   &	58793.09    &	17.81	(0.06)   \\
58791.65   &	18.55	(0.09)   &	58793.23   &	17.83	(0.07)   \\
58791.65   &	18.57	(0.08)   &	58793.61   &	17.71	(0.04)   \\
58791.65   &	18.76	(0.09)   &	58793.66   &	17.71	(0.04)   \\
58791.65   &	18.56	(0.09)   &	58794.61   &	17.61	(0.06)   \\
58791.66    &	18.43	(0.08)   &	58794.70   &	17.57	(0.05)   \\
58791.66   &	18.40	(0.08)   &	58800.99   &	17.44	(0.05)   \\
58791.67   &	18.41	(0.08)   &	58804.06   &	17.58	(0.05)   \\
58791.68   &	18.52	(0.08)   &	58804.80   &	17.58	(0.05)   \\
58791.69   &	18.51	(0.07)   &	58808.01   &	17.70	(0.06)   \\
58791.69   &	18.49	(0.09)    &	58811.01   &	17.80	(0.06)   \\
58791.70     &	18.70	(0.08)   &	58811.56   &	17.88	(0.08)    \\
58791.71   &	18.40	(0.08)    &	58812.01   &	17.83	(0.06)   \\
58791.72   &	18.57	(0.08)   &	58813.57   &	17.92	(0.05)   \\
58791.73    &	18.35	(0.07)   &	58814.54   &	17.98	(0.06)   \\
58791.74   &	18.39	(0.08)   &	58815.56   &	18.08	(0.07)   \\
58791.75   &	18.41	(0.07)    &	58816.26   & 	18.06	(0.07)   \\
58791.77   &	18.37	(0.07)   &	58817.27    &	18.07 (0.07)   \\
58791.79   &	18.30	(0.08)   &	58818.03   &	18.21	(0.08)   \\
58791.80   &	18.25	(0.07)   &	58819.02   &	18.17	(0.07)   \\
58792.09   &	18.23	(0.08)   &	58823.02   &	18.24	(0.08)   \\
58792.10   &	18.19	(0.07)   &		    &			 \\
   \hline
 \end{tabular}
 
\end{table*}

\begin{table*}
	\centering
	\caption{$Swift$ measurements for the ILRTs in our sample.}
	\label{SWIFTmeasures}
	\begin{tabular}{cccccccc}
            &            &         & \textbf{NGC 300 OT}	&	        &  	       &  		 \\
		\hline
    MJD	   &	UVW2	&  UVM2	  &  UVW1	&      U        &     B	       &   V             \\
            \hline
54606.22   & 18.70 (0.11) &   -- (--)	  &    -- (--)	&      -- (--)        &       -- (--)      &       -- (--)	 \\
54611.27   &    -- (--)   	&   -- (--)	  & 17.74 (0.08)  &  16.11 (0.07)   &  15.67 (0.06)  &   14.86 (0.05) 	 \\
54621.80   &	-- (--)	&   -- (--)	  & 18.05 (0.10)  &  16.42 (0.07)   &  15.90 (0.06)  &   14.97 (0.05)	 \\
54636.69   & 19.05 (0.12) &   -- (--)     & 18.15 (0.10)  &  16.71 (0.07)   &  16.22 (0.06)  &   15.25 (0.05)	 \\
54692.57   &    -- (--)   	&   -- (--)	  & >19.9    &  18.9 (0.17)   &  18.74 (0.12)  &   17.66 (0.09)	 \\
54693.16   &	-- (--)	&   -- (--)	  & >19.9    &  18.98 (0.18)   &  18.76 (0.12)  &   17.66 (0.09)	 \\
\hline
	\end{tabular}
\end{table*}

\end{document}